\documentclass[twocolumn]{aastex631}
\usepackage{mathtools}

\shorttitle{A Unified Model for the Fan Region and the North Polar Spur}
\shortauthors{West et al.}
\received{June 2021}
\accepted{\today}
\submitjournal{ApJ}

\graphicspath{{./}{figures/}}

\begin{document}

\title{A Unified Model for the Fan Region and the North Polar Spur: A bundle of filaments in the Local Galaxy}

\correspondingauthor{J. L. West et al.}
\email{jennifer.west@dunlap.utoronto.ca}

\author[0000-0001-7722-8458]{J. L. West}
\affil{Dunlap Institute for Astronomy and Astrophysics, University of Toronto, Toronto, ON M5S 3H4, Canada}

\author[0000-0003-1455-2546]{T. L. Landecker}
\affil{National Research Council Canada, Herzberg Research Centre for Astronomy and Astrophysics, Dominion Radio Astrophysical Observatory, PO Box 248, Penticton, V2A 6J9, Canada}

\author[0000-0002-3382-9558]{B. M. Gaensler}
\affil{Dunlap Institute for Astronomy and Astrophysics, University of Toronto, Toronto, ON M5S 3H4, Canada}
\affil{David A. Dunlap Department of Astronomy and Astrophysics, University of Toronto, Toronto, ON M5S 3H4, Canada}

\author[0000-0003-2645-1339]{T. Jaffe}
\affil{NASA Goddard Space Flight Center, Greenbelt, MD 20771, USA}

\author[0000-0001-7301-5666]{A. S. Hill}
\affil{Department of Computer Science, Math, Physics, and Statistics, Irving K. Barber School of Science, University of British Columbia, Kelowna, BC V1V 1V7, Canada}
\affil{National Research Council Canada, Herzberg Research Centre for Astronomy and Astrophysics, Dominion Radio Astrophysical Observatory, PO Box 248, Penticton, V2A 6J9, Canada}

\begin{abstract} 
We present a simple, unified model that can explain two of the brightest, large-scale, diffuse, polarized radio features in the sky, the North Polar Spur (NPS) and the Fan Region, along with several other prominent loops. We suggest that they are long, magnetized, and parallel filamentary structures that surround the Local arm and/or Local Bubble, in which the Sun is embedded. We show this model is consistent with the large number of observational studies on these regions, and is able to resolve an apparent contradiction in the literature that suggests the high latitude portion of the NPS is nearby, while lower latitude portions are more distant. Understanding the contributions of this local emission is critical to developing a complete model of the Galactic magnetic field. These very nearby structures also provide context to help understand similar non-thermal, filamentary structures that are increasingly being observed with modern radio telescopes. 

\end{abstract}


\section{\label{sec:intro}Introduction}

\begin{figure*}[!ht]
\centering 

\includegraphics[width=17cm]{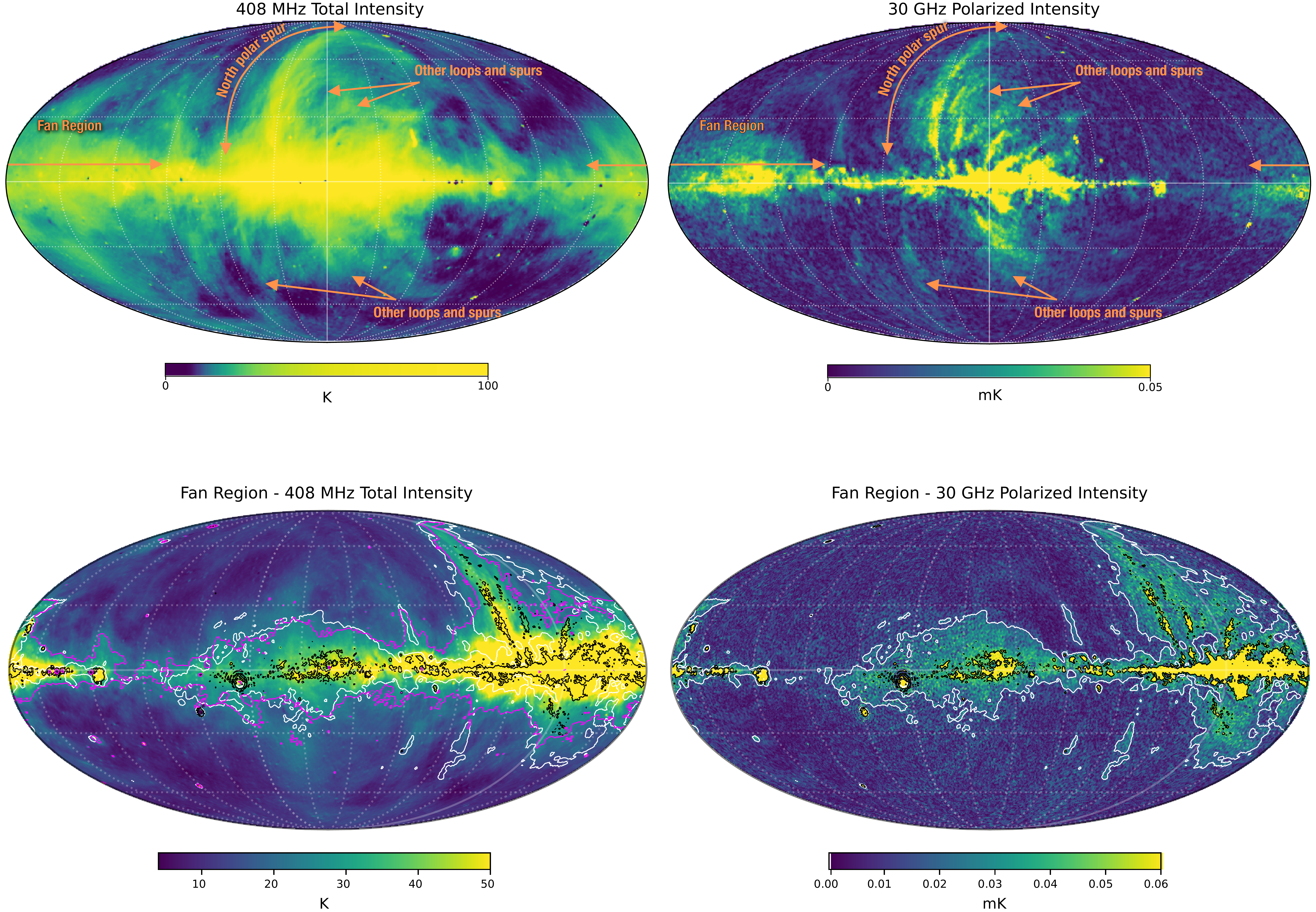}

\caption{ \label{fig:fig1} Top row: 408 MHz Stokes~$I$, total intensity data \citep{1982A&AS...47....1H} (left) and 30~GHz polarized intensity data \citep{2016A&A...594A...1P} (right) shown in Galactic coordinates with the centre of the map at $(l,b)=(0^\circ,0^\circ)$, and with labels identifying the main features discussed in this work. The 30~GHz data have been smoothed to a resolution of $1^\circ$. Bottom row: The same 408 MHz total intensity (left) and 30 GHz polarized intensity data (right) as above, but centred on the approximate centre coordinates of the Fan Region, $(l,b)= (135^\circ, 0^\circ)$. These maps are shown with slightly different brightness scaling, and the 30~GHz map is shown at full resolution ($33'$).  The white contour in these maps shows the faint 30~GHz polarized intensity at a level of 0.015~mK (smoothed to $2^\circ$). The black contour is also 30~GHz polarized intensity, but at a level of 0.05~mK (unsmoothed), which highlights the bright polarized features towards this direction. The magenta contour in the bottom left panel shows the 24~K level of the 408 MHz data for direct comparison with the polarized intensity contour. }
\end{figure*}

The North Polar Spur (NPS) \citep{1958PMag....3..370T, 1960Obs....80..191H} and the Fan Region \citep{1964BAN....17..465B} are, by far, the two brightest and most dramatic extended features in the large-scale radio synchrotron sky across a broad range of frequencies. Since the time of their discoveries and right up to the present day, astronomers have questioned the origin of these two regions, with some arguing that they are local features \citep{1971A&A....14..252B, 1972A&A....21...61S,1974MNRAS.167..593W, 1984A&A...135..238S, 1998LNP...506..229H, 2015MNRAS.452..656V, 1995A&A...294L..25E, 2007ApJ...664..349W, 2020MNRAS.498.5863D}, while others argue that they are distant, Galactic-scale features \citep{1977A&A....60..327S, 2003ApJ...582..246B, 2016MNRAS.459..108S, 2017MNRAS.467.4631H, 2018Galax...6...27K, 2019MNRAS.482.4813S, 2020Natur.588..227P,2020ApJ...904...54L}. Their origins remain an open question. In Fig.~\ref{fig:fig1} we present a labeled map showing the location of these regions using the Stokes~$I$, total intensity all-sky map at 408 MHz from \citet{1982A&AS...47....1H}\footnote{For the 408 MHz data we use the synchrotron emission map provided by the Planck collaboration, which has been corrected for the contribution due to the Cosmic Microwave Background. \url{https://irsa.ipac.caltech.edu/data/Planck/release_2/all-sky-maps/foregrounds.html}} alongside the 30~GHz linear polarized intensity map from Planck \citep{2016A&A...594A...1P}. The resolutions, i.e., the full-width at half maximum of the Gaussian beams of these maps, are $51'$ and $33'$ respectively. 

If these features are indeed local, understanding their structure and morphology is critical since we are embedded among the stars, dust, and gas that comprise them, and all features beyond must be observed through this local veil of material. Features that are extremely nearby can have a very large angular size on the sky, and only with good models can we account for this ``contamination'' when developing large-scale models of the Galactic magnetic field and foreground models for cosmology experiments \citep[e.g.,][]{2019BAAS...51g.188G}. Yet, understanding the local environment is also very difficult because there is superposition of emission at all distances, and the large angular extent means the surface brightness of local emission can be small and can easily be overwhelmed by background sources. 

In addition to the NPS and the Fan Region, there are several other polarized loops and spurs in the large-scale Galactic emission \citep[e.g.,][also Fig.~\ref{fig:fig1}]{planck-et-al-dust}. \citet{2015MNRAS.452..656V} propose a common origin for the NPS and these loops as filaments in the wall of a nearby bubble, but the Fan Region is not part of their analysis (see their Fig.~2). The NPS and the Fan Region share many similar features. Both have high fractional polarization and coherent magnetic fields, revealing the synchrotron origin of this emission, and confirming the presence of relativistic cosmic ray electrons (CREs). Both are clearly seen in dust polarization \citep{2015A&A...576A.104P}, and both have similar angular scale, several tens of degrees long and several degrees wide. Despite the similarities, very few studies have considered that they could have a common origin. A notable exception is \citet{1965AuJPh..18..635M} who suggested they are features of the Local arm in which the Sun is embedded.

The synchrotron nature of these features means that they are intrinsically polarized, making radio polarization observations a powerful tool for their study. A radio telescope observes the linearly polarized components of the complex polarization vector using the Stokes parameters, $Q$ and $U$. The observed intensity of the polarized radiation, PI, is found by

 \begin{equation}
      \text{PI}=\sqrt{(Q^2+U^2)}, 
 \end{equation}

and the observed orientation of a polarized wave, i.e., the polarization angle, $\chi_\text{obs}$ is given by 

 \begin{equation}
      \chi_\text{obs}=\frac{1}{2}\arctan \left( {\frac{U}{Q}} \right). 
 \end{equation}

The telescope observes a wave with polarization angle $\chi_\text{obs}$, different from the angle $\chi_0$ at the point of emission. The plane of polarization has been modified by Faraday rotation in the intervening magneto-ionic medium. ${\chi_\text{obs}}-{\chi_0}$ is dependent on the observation frequency, $\nu$, and on the properties of the medium along the propagation path where

 \begin{equation}
 \label{eq:deltachi}
      \chi_\text{obs} - \chi_0  =  \text{FD}\left( {\frac{c}{\nu}}. \right)^2
 \end{equation}

Here $c$ is the speed of light and FD is the Faraday depth. Thus one can see that for large $\nu$, the difference between $\chi_\text{obs}$ and $\chi_0$ becomes very small. The FD (in discrete form), along a particular line of sight, from some source at a distance $\sum_{i} \Delta R$, is defined by:

\begin{equation}
\centering
\text{FD}=\sum_{i} 0.81{n_{e, i}}{B_{\parallel, i}}\Delta R~{\rm{[rad~}}{{\rm{m}}^{\rm{-2}}}{\rm{]}}.
\end{equation}

Here $n_{e, i}$ is the electron density, and $B_{\parallel, i}$ is the line-of-sight component of the magnetic field at each step, $i$.  The step size along the line of sight is given by $\Delta R$.

The long and narrow morphology of the NPS and the Fan Region is also reminiscent of other long and narrow filamentary structures that are ubiquitous throughout our Galaxy and observed across many different wavelengths and through different emission mechanisms.  At optical wavelengths there are radiative filaments around old supernova remnants (SNRs) \citep[e.g.,][]{2015ApJ...812...37F} and isolated filaments of unknown origin \citep{2001AJ....122.1500M}. Thermal and non-thermal radio filaments are found in SNRs and in the walls of Galactic chimneys and supershells  \citep[e.g., the W4 Chimney, ][]{1996Natur.380..687N}. Long (up to 430~pc), narrow ($\sim0.5$~pc wide) filaments of molecular gas trace spiral structure and may form the ``bones'' or ``skeleton'' of the Galaxy \citep{2014ApJ...797...53G, 2015ApJ...815...23Z}.  In addition, simulations show the formation of filamentary structure in molecular clouds \citep[e.g.,][]{2018MNRAS.481.5275T}. Recently,  \citet{2020Natur.578..237A} discovered a narrow and coherent wave of dense gas in the solar neighborhood that is 2.7~kpc long, and aligned with Galactic longitude, $l=60^\circ\pm10^\circ$. Although molecular gas is quite different from synchrotron emission, this example demonstrates that large-scale filamentary structure can trace spiral structure over large distances.

Across all wavelengths, filamentary structures are increasingly evident as the resolution and sensitivity (both brightness and spatial scale) of observations improve. Radio data have revealed a plethora of non-thermal filaments in the Galactic centre \citep{1984Natur.310..557Y, 2018MNSSA..77..102.}, which \citet{2020PASJ..tmp..161S} proposes may be relics of old SNRs,  and isolated filaments from LOFAR measurements \citep{2015MNRAS.454L..46Z, 2015A&A...583A.137J}. Fine filamentary structure can be seen in the NPS in reprocessed data from the NVSS survey \citep{2009AJ....137..145R}.  In addition, the HI sky has many large-scale filaments, now shown to be strongly aligned with the Galactic magnetic field \citep{2018ApJ...857L..10C, 2019ApJ...887..136C}, and other recent studies are showing alignment of HI filaments with other straight, polarized features (e.g., Campbell et al., submitted; West et al. in prep). These examples probe different phases of the interstellar medium (ISM), which may not sample the same region of space. However, there is increasing evidence that these phases are interconnected.

Recent studies suggest that the Local Bubble dominates the very local Galactic magnetic environment. \citet{2018A&A...611L...5A} find that the Local Bubble magnetic field is directed towards the Galactic latitude and longitude, $(l, b) = (70^\circ\pm11^\circ, +43^\circ \pm 8^\circ)$ in the northern polar cap and $(74^\circ \pm 8^\circ, -14^\circ \pm 18^\circ)$ in the southern cap, and \citet{planck-et-al-dust} find it is directed towards $(l , b ) = (70^\circ , +24^\circ)$ in the south \citep[see Table 3 in][for a summary of these values]{2020A&A...636A..17P}. These values are close to $l = 83^\circ\pm4^\circ$, the local field direction found from starlight polarization \citep{1996ApJ...462..316H}. \citet{2019A&A...631L..11S} find that polarized dust emission is dominated by magnetic structure within 200-300~pc, within the shell of the Local Bubble. These measurements all have a fairly consistent longitude, but the latitudes of these vary greatly.

In this paper, we are inspired by the suggestion of \citet{1965AuJPh..18..635M} that the Fan Region and the NPS are part of the Local arm, and consider this in the context of recent observational evidence. Given the ubiquity of filamentary structure and the alignment of HI filaments with the magnetic field, we consider a model where the Sun is embedded in an environment with filamentary geometry instead of a uniform magnetic field. In this model, the filaments that we observe as the Fan Region and the NPS have a common origin as the elongated fragments of a supershell surrounding the Local Bubble that has been stretched along the Local Arm. We do not endeavor to do a full simulation in this work. Instead, the goal of this paper is to present a new idea using a simple model, and compare it to a wide variety of observational features including morphology, radio brightness, FD, and absolute polarization angle, while at the same time considering ways to resolve a number of observational discrepancies that have been presented in the literature. 

In Sec.~\ref{sec:observations} we give a brief overview of the large amount of relevant literature concerning the distances to these features. In Sec.~\ref{sec:data} we offer a new perspective and compare previously published data at different radio frequencies. We present our model in Sec.~\ref{sec:model} in three parts of increasing complexity: a straight-line model (Sec.~\ref{sec:straightlines}), curved filaments (Sec.~\ref{sec:curvedlines}), and tilted, curved filaments (Sec.~\ref{sec:tiltedlines}), and then discuss implications for the synchrotron intensity (Sec.~\ref{sec:brightness}) and FD (Sec.~\ref{sec:faradayrotation}). In Sec.~\ref{sec:discussion} we discuss the distance (Sec.~\ref{sec:distance}), and possible origins (Sec.~\ref{sec:origin}). Conclusions are presented in Sec.~\ref{sec:conclusions}.

\begin{figure*}[!ht]
\centering 
\begin{minipage}{8.5cm}
\includegraphics[width=8.4cm]{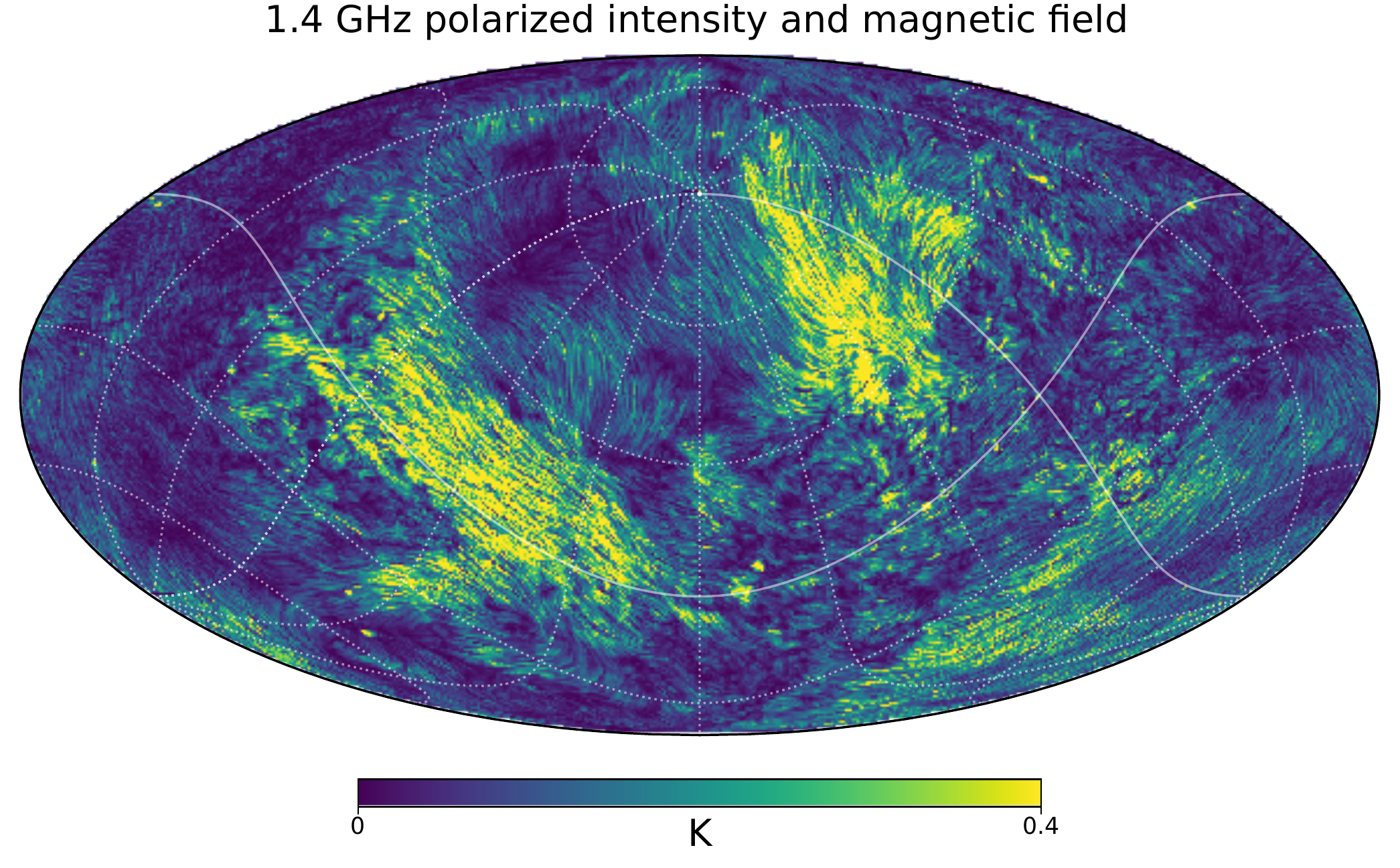}
\end{minipage}
\hfill
\begin{minipage}{8.5cm}
\includegraphics[width=8.4cm]{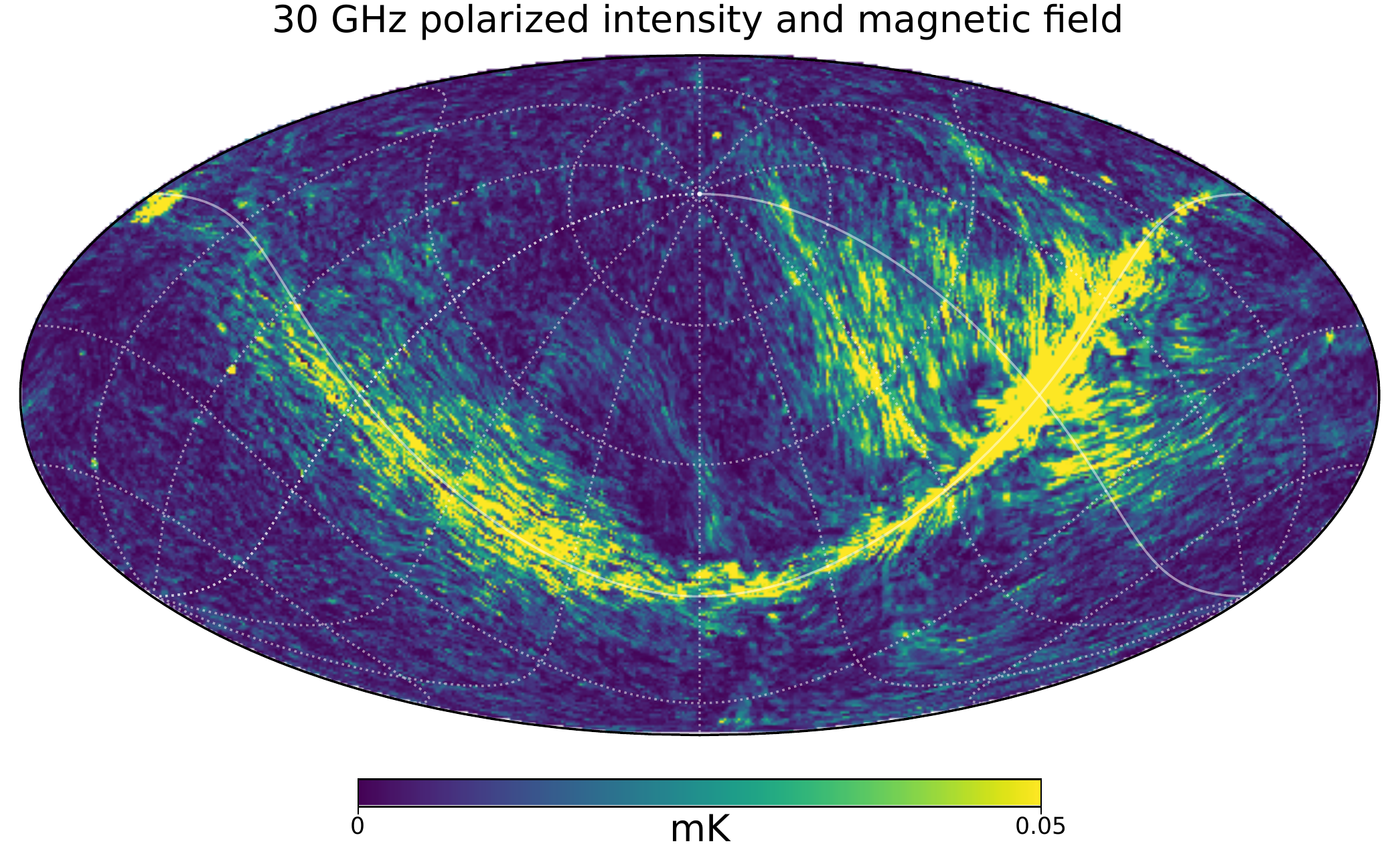}
\end{minipage}
\caption{ \label{fig:reprojection} Left: DRAO/Villa Elisa 1.4~GHz peak polarized intensity map projected to a centre point at $(l,b) = (90^\circ,+45^\circ)$ with line integral convolution of the observed magnetic field lines as measured from the 1.4~GHz data. The grid shows Galactic coordinates with increments of $30^\circ$, with the North Galactic pole near the top at the centre. The Fan Region is the bright extended feature to the lower left of the map centre, and the NPS is located on the upper right of the map centre. Right: 30~GHz polarized intensity data with line integral convolution of the observed magnetic field lines as measured from the 30~GHz data. The projection and grid are the same as the plot on the left.  }
\end{figure*}

\section{\label{sec:observations}Observational Features}

Given the large angular extent and prominence of these features, many observational studies have revealed a wealth of information. However, these studies do not always agree on the conclusions, particularly concerning the distance. There is significant controversy over the NPS in particular, and whether it is very nearby, or at the Galactic centre. Here we provide a summary of the main observational features, which provide important context for the paper.

\subsection{\label{sec:nps}North Polar Spur}

In addition to radio synchrotron, the NPS is observed through many other observational tracers, including neutral hydrogen \citep{1970Natur.225..364B}, dust traced by microwave emission \citep{2015A&A...576A.104P}, tracers from starlight polarization \citep{2014A&A...561A..24B}, thermal emission from X-rays \citep{1972ApJ...172L..67B, 2014A&A...566A..13P, 2020Natur.588..227P}, and Gamma rays \citep{2009arXiv0912.3478C}. Although the emission is coincident across these wide variety of tracers, there is not a consensus as to whether these trace the same structures.

Observations at X-ray wavelengths tend to favor a distant NPS:
\begin{enumerate}
    \item A recent study by \citet{2020Natur.588..227P} presents new, all-sky, X-ray observations from eROSITA. The fourfold symmetry around the Galactic centre and the tantalizing association with the diffuse gamma ray emission called the Fermi bubbles \citep{2010ApJ...724.1044S} lead the authors to conclude that this is a Galactic scale feature possibly associated with outflow from the supermassive black hole Sgr A*. 
    
    \item Other authors \citep[e.g.,][]{1977A&A....60..327S, 2003ApJ...582..246B, 2016MNRAS.459..108S, 2018Galax...6...27K, 2019MNRAS.482.4813S} use magneto-hydrodynamic simulations and observations near the Galactic centre to support the central black hole outflow argument. These works demonstrate the plausibility of creating an NPS-like structure at the Galactic centre that could explain the observed X-ray emission. 
    
    \item  \citet{2014A&A...566A..13P} made a detailed study comparing the local 3D dust distribution with soft X-ray emission. They conclude that the brightest part of the NPS in X-rays (i.e., for $b<8^\circ$) originates from hot gas beyond $\sim200$~pc. 
    
    \item In a follow-up study of X-ray absorption, \citet{2015MNRAS.447.3824S} conclude that the NPS must be behind the Aquila Rift, giving a lower limit on the NPS distance of $1.02\pm0.25$~kpc.  Other studies of X-ray absorption toward the NPS by \citet{2016A&A...595A.131L} and \citet{2020ApJ...904...54L} also favor a distant origin for the X-ray emission in the direction of the NPS.  
    
\end{enumerate}

It must be noted, however, that these analyses provide no solid lower limits beyond a few hundred pc and no real information beyond 4\,kpc.  \citet{2016A&A...595A.131L} find a minimum distance to the near side of the NPS of 260~pc. Though the analyses favor a distant origin for the X-ray emission, distant in this case means beyond most, if not all, of the intervening dust.  That dust is so thinly distributed about the disk ($\approx 100$\,pc) that there is little information in the dust to constrain the distance beyond a kpc even for the lower part ($b\approx 10\degr$) of the NPS.

The main argument presented by \citet{2020Natur.588..227P} is a morphological one. While visually compelling, many other observational studies have presented strong evidence that contradicts this picture to conclude that the NPS must be considerably more local. This evidence includes: 

\begin{enumerate}
    \item Its alignment with polarized emission from nearby dust grains \citep{2019A&A...631L..11S, 2020MNRAS.498.5863D, 2021arXiv210614267P}.
    \item Polarized starlight from 150 - 500 pc away demonstrates the presence of a magnetic field in that distance range that strongly resembles the morphology of the NPS \citep{1996ApJ...462..316H,2011ApJ...728..104S}. 
    \item There is an association with neutral hydrogen that suggests a local structure \citep{1970ApL.....6..215V,1971A&A....14..252B,1971A&A....12..388H,1994AJ....107..287V,2009Ap&SS.323....1W}.
    \item There is no associated structure detected in current FD maps of the sky \citep{2015A&A...575A.118O, 2020A&A...633A.150H, 2021arXiv210201709H}, which primarily use measurements of the Faraday rotation of radio galaxies as probes to trace the full FD out to the edge of the Milky Way Galaxy.This makes it very unlikely that the NPS is a physically large structure. If the NPS is a Galactic-scale feature located near the Galactic center, its thickness would measure hundreds of pc ($6^\circ$ corresponds to $\sim840$~pc at a distance of 8~kpc). Since the NPS is very prominent in polarized intensity, this implies that it contains a coherent magnetic field, which should contribute to a significant Faraday rotation through a structure with such a large path length, even if the electron density is relatively small. 
\end{enumerate}

Taking all of this evidence together would seem to indicate that the NPS itself is local and that the X-ray observations most likely present a confused superposition of local and more distant emission. Suggestions for a local origin of the NPS include a large, old supernova remnant \citep{1971A&A....14..252B, 1972A&A....21...61S} or a portion of the local arm \citep{1965AuJPh..18..635M}. Still others have suggested that it could be a bright filament from an expanding shell with a diameter of 120~pc \citep{1998LNP...506..229H, 2015MNRAS.452..656V} or due to a collision of shells \citep{1995A&A...294L..25E, 2007ApJ...664..349W}.

Although there is a large amount of evidence that suggests that the NPS is local, there is disagreement as to the exact distance, and several studies have pointed out what seems to be contradictory evidence that suggests a portion of the NPS is local ($\sim100$~pc) while other portions must be more distant ($>250$~pc). These include:
\begin{enumerate}
\item \citet{2015ApJ...811...40S} present a FD study of the diffuse NPS emission and conclude that the high latitude portion of the NPS at $b>50^\circ$ is local ($\sim100$~pc), but that the region for which $b<40^\circ$ is likely more distant. This is because the part nearest the Galactic plane shows strong depolarization at lower frequencies ($\sim1~$GHz) \citep{2015ApJ...811...40S}, which implies a foreground depolarizing screen. Conversely the FD from the ISM in front of the NPS for $b> 50^\circ$ is zero, which implies that this part of the NPS is local.%One argument that has been made against the NPS as a local feature is that

\item  \citet{2011ApJ...728..104S} study starlight polarization with a view to test the hypothesis of an annular shell. They find ``very discrepant distances along the annular region: $\approx$100 pc to the left side and 250 pc to the right side, independently confirming the indication from a previous photometric analysis.''

\item  \citet{2020MNRAS.498.5863D} use Gaia DR2 data and interstellar reddening to map the 3D distance. They find that they can ``account for nearly 100 per cent of the total column density of the NPS as lying within 140~pc for latitudes $>26^\circ$ and within 700~pc for latitudes $<11^\circ$.'' 

\item \citet{2021arXiv210614267P} analyze a combination of stellar polarization, polarized synchrotron radio emission, and polarized thermal dust emission to show a significant alignment of the relative angles from these tracers concluding that the NPS is local for $b>30^\circ$. 

\end{enumerate}

It is interesting that these varied studies all come to a similar conclusion that the high latitude portion of the NPS is local, while the lower latitude portion(s) are more distant. The bubble or shell-type models that have been proposed in the literature are not able to reconcile the observations, suggesting very different distances to different parts of the NPS. We interpret this evidence as a key clue in understanding the true nature of the NPS.

\begin{figure}[!ht]
\centering 
\includegraphics[width=8.5cm]{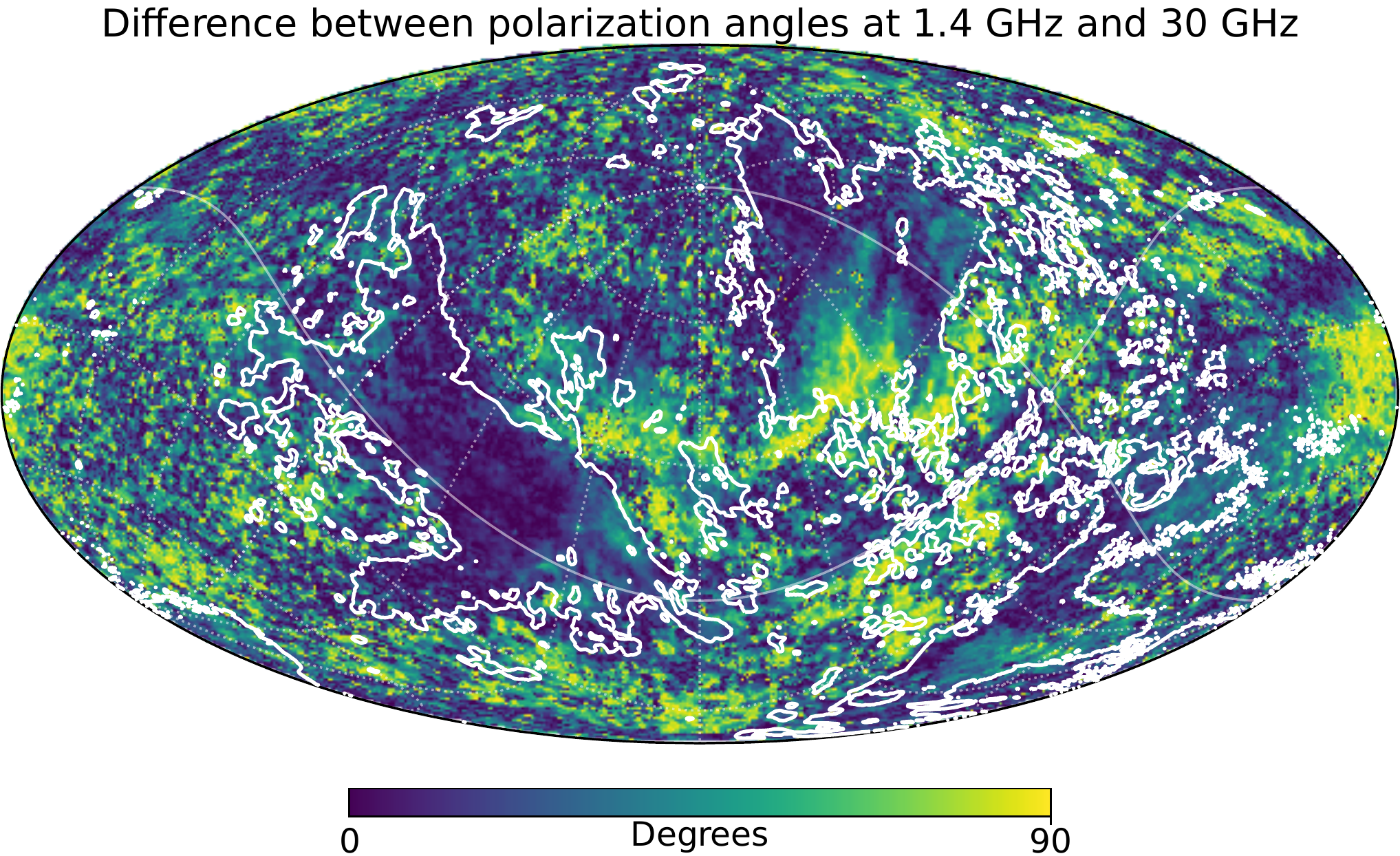}
\caption{ \label{fig:datadiff} Absolute difference in polarization angle between 1.4 and 30 GHz data, shown in the same projection as Fig.~\ref{fig:reprojection}. Contours show the 1.4~GHz polarized intensity for a value of 0.15~K, highlighting that the regions with the highest polarized intensity have some of the smallest differences in polarization angle, with the notable exception of the base of the NPS. }
\end{figure}

\begin{figure*}[!ht]
\centering 
\begin{minipage}{8.5cm}
\includegraphics[width=6.0cm]{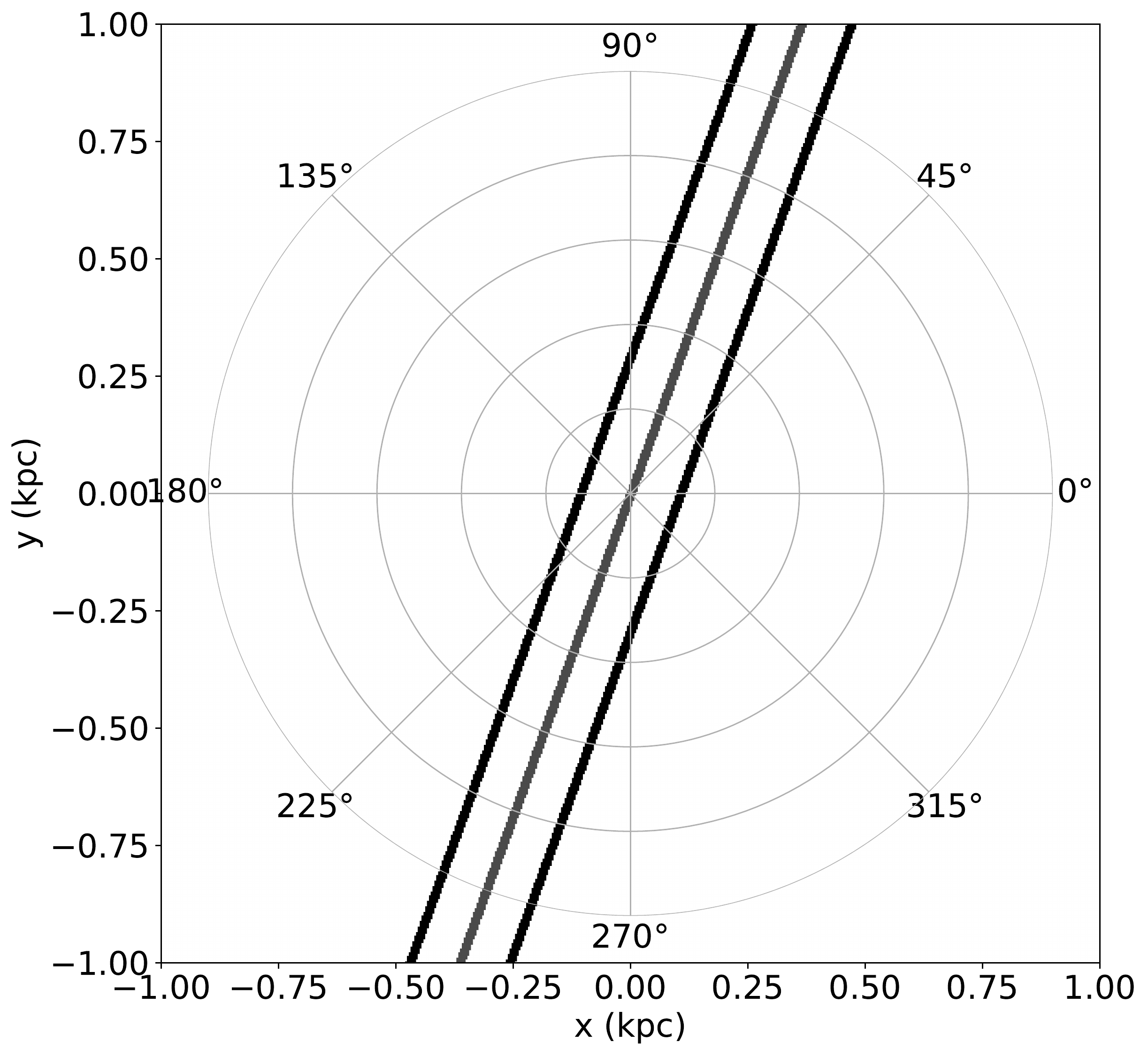}
\end{minipage}
\hfill
\begin{minipage}{8.5cm}
\includegraphics[width=8.4cm]{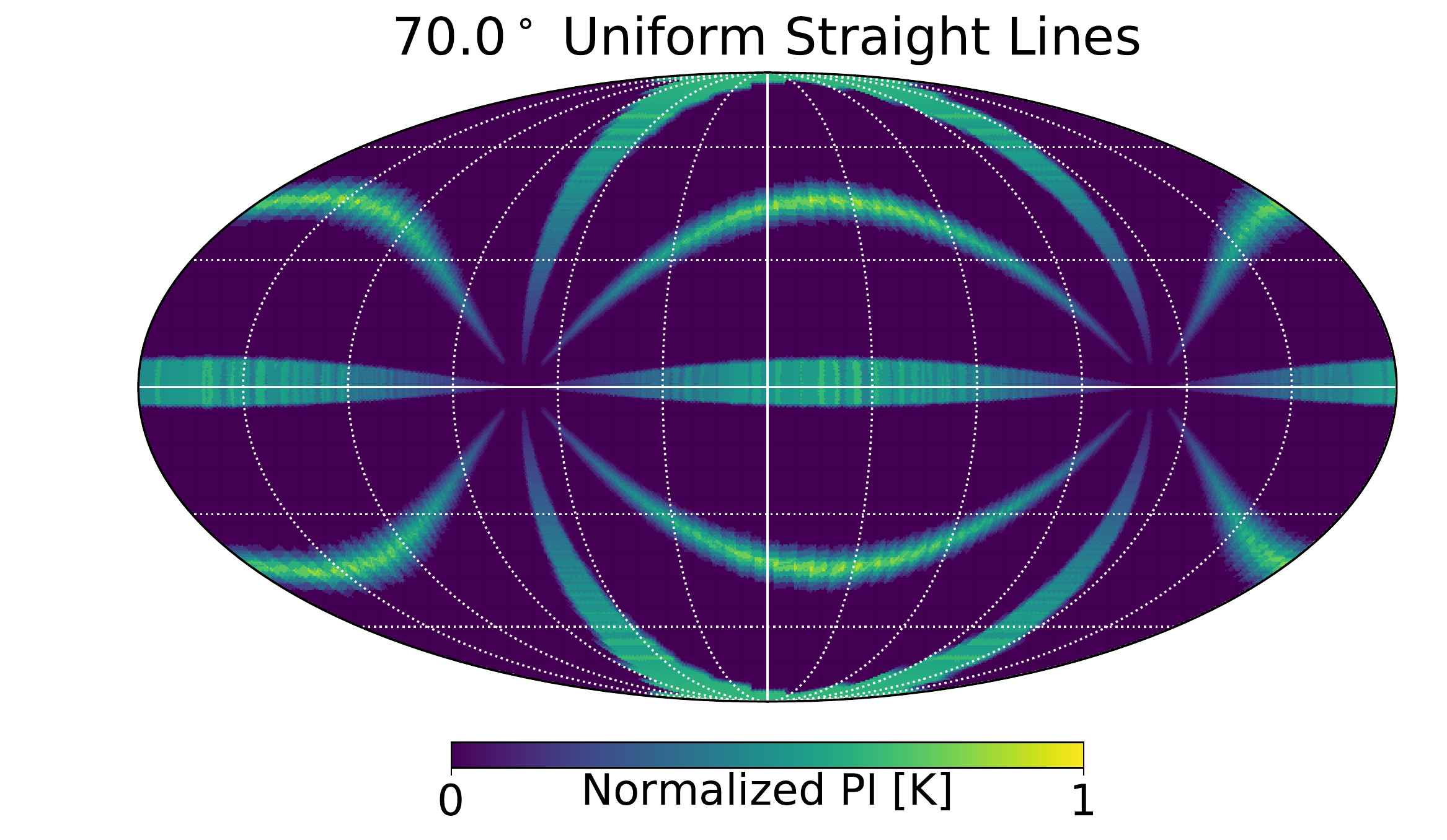}
\end{minipage}
\begin{minipage}{8.5cm}
\includegraphics[width=6.0cm]{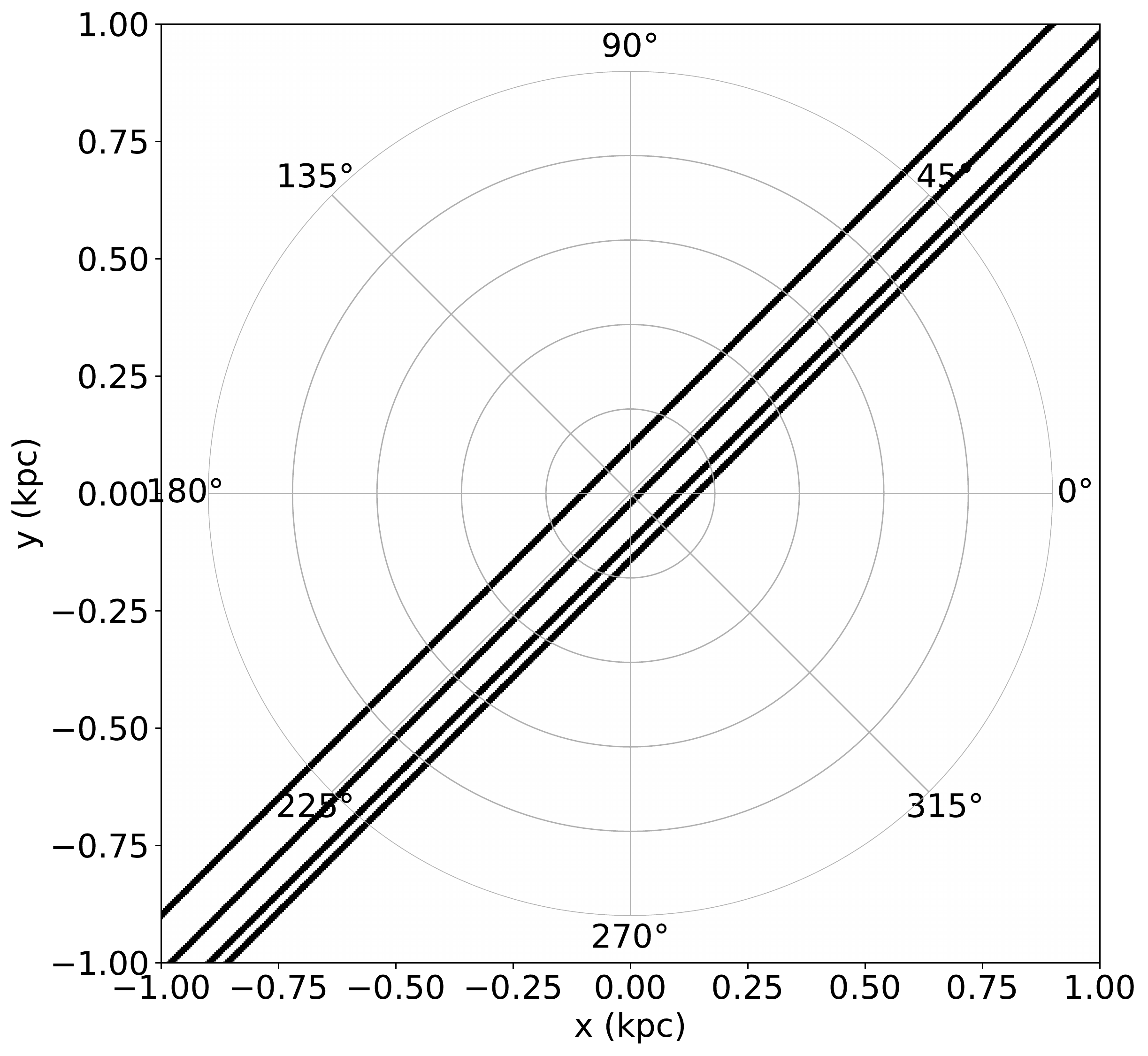}
\end{minipage}
\hfill
\begin{minipage}{8.5cm}
\includegraphics[width=8.4cm]{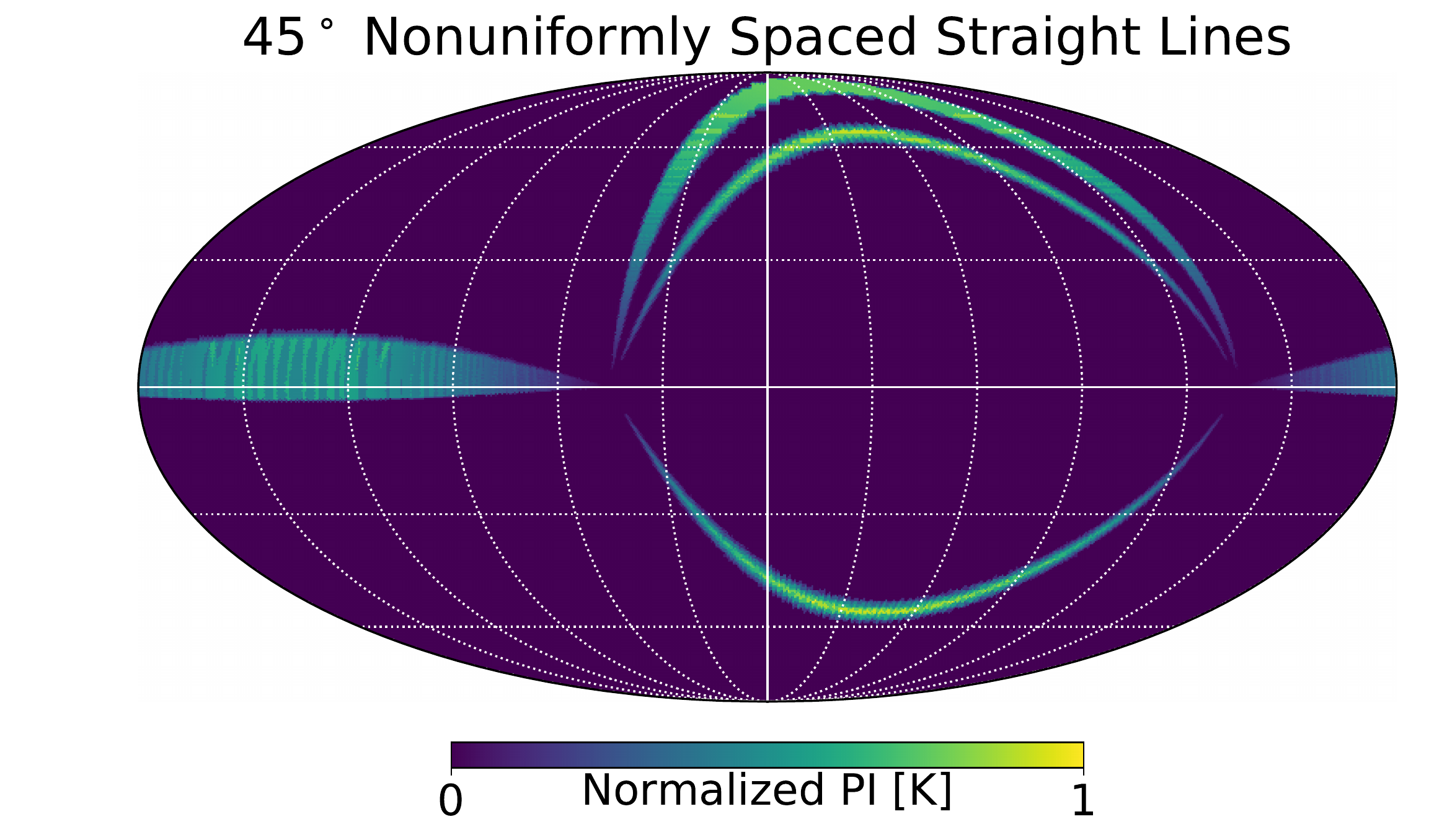}
\end{minipage}

\caption{ \label{fig:uniformfields} Top Left: Top-down view of a selection of straight-line filaments oriented towards $l=70^\circ$. The Sun is located at the centre of this plot and the Galactic centre is to the right. Galactic longitude is labeled on the circular grid. There are a total of eight filaments in this model, arranged uniformly around the Sun: one directly above us, one directly below us, one to each side, and one each in the top-right, top-left, bottom-right, and bottom-left. Only three filaments are visible in this figure since the others are stacked in the z-direction.  Bottom Left: Top-down view of a selection of straight-line filaments oriented towards $l=45^\circ$.   
Right column: Simulated synchrotron emission from these filaments. Galactic longitude circles and lines of Galactic latitude are drawn every $30^\circ$, with $l=0^\circ$, $b=0^\circ$ at the centre. }
\end{figure*}

\subsection{\label{sec:fan}Fan Region}

The term ``Fan Region'' has been used in the literature to refer to a variety of features in diffuse polarized radio emission in the second Galactic quadrant extending several degrees North and somewhat South of the Galactic plane. It was originally called the Fan Region because its electric field vectors, as seen from  polarized radio emission, appear to fan out and away from the Galactic plane \citep{1976A&AS...26..129B}. Some of these features are known discrete objects that are located at a variety of longitudes (and distances) within the broad extended region of diffuse emission. In some cases, the features are only visible at some range of frequencies.  \citet{2017MNRAS.467.4631H} argue that at least $30\%$ of the brightest emission at $\gtrsim 1 \textrm{ GHz}$ — the yellow emission in the lower right panel of Fig.~\ref{fig:fig1} — must originate $\gtrsim 2 \textrm{ kpc}$ away. \citet{2013A&A...549A..56I} examine the Faraday spectrum towards a polarized feature at $(\ell, b) = (137\arcdeg, +7\arcdeg)$ at $\sim 150 \textrm{ MHz}$ and conclude that one of the components of this low-frequency emission is nearby ($\lesssim 100 \textrm{ pc}$) and extended, and is likely part of the Local Bubble wall, while another component is a large, discrete Faraday structure. In the \citet{1976A&AS...26..129B} data, the Fan Region appears to only extend from $120^\circ<l<160^\circ$, though from Fig.~\ref{fig:fig1}, one can see it extends much farther. We attempt to clarify the definition of the Fan Region for the purposes of this work.

Like the NPS, the Fan Region is also very highly polarized (up to 40\%) and extremely bright and obvious in polarized emission. But unlike the NPS, the Fan Region is not nearly as obvious in radio continuum and is mostly unremarkable at other wavelengths, possibly owing to its superposition on the Galactic plane. Some authors have stated that there is no Stokes~$I$ counterpart to the Fan Region \citep[e.g.,][]{2017MNRAS.467.4631H}, but on careful examination of the available data, we think this is a misconception for several reasons. At higher frequencies, Stokes~$I$ emission becomes increasingly mixed with contributions from other emission mechanisms (e.g., thermal Bremsstrahlung) making it more difficult to differentiate the synchrotron contribution, whereas polarized intensity more directly probes the synchrotron emission \citep[see Fig.~51 of][]{2016A&A...594A..10P}. At 30~GHz, the Stokes~$I$ synchrotron component is mixed considerably with other emission mechanisms, making many of the features indiscernible. At lower frequencies the Stokes~$I$ is increasingly dominated by synchrotron, but the polarized intensity has a greater influence from Faraday rotation, whereas Stokes~$I$ is unaffected by Faraday effects. Thus, in the bottom row of Fig.~\ref{fig:fig1}, we choose to compare polarized intensity at 30~GHz to total intensity at 408~MHz. The 30~GHz polarized intensity map is dominated by the synchrotron component, and additionally, at this frequency, is almost completely free from the effects of wavelength dependent Faraday rotation, which can complicate the interpretation (although, as a vector quantity, polarized emission is still subject to geometric depolarization along the line-of-sight).

The bottom row of Fig.~\ref{fig:fig1} highlights two main components of polarized emission towards the approximate centre coordinates of the Fan Region, $(l,b)= (135^\circ, 0^\circ)$. First there are several smaller scale features that are very bright in polarization intensity (black contours), and second, there is a larger and fainter region of polarized emission, which we identify with a white contour. In the bottom-left panel, the 24~K total intensity contour, shown in magenta, demonstrates a clear correspondence to the white, polarized contour. When observing towards the Fan Region with a larger beam (which is particularly true of some historical observations), the bright and faint components become mixed and more difficult to distinguish.

For the purposes of this work we consider the Fan Region to be defined by the white contour in the bottom row of Fig.~\ref{fig:fig1}, i.e., the large and diffuse envelope of emission, but excluding the bright, discrete objects that are identified by the black contours. We make this distinction since the black contours in Fig.~\ref{fig:fig1} include several well known, discrete Galactic synchrotron emitting objects, such as the W3/W4/W5 star-forming complex and superbubble and 22 known supernova remnants in the region $100^\circ<l<180^\circ$ \citep[][]{2019JApA...40...36G}, which include several bright and extended historically known radio sources \citep[e.g., 3C10, 3C58,][]{1962MmRAS..68..163B}. In addition, the magnetic field appears to be coherent over the entire area defined by the white contour, as shown in Fig.~\ref{fig:reprojection}.

An additional complication is that the position of the Fan Region also corresponds to longitudes where we would expect to have excess emission due to the Perseus spiral Arm. Global Galactic magnetic field models cannot reproduce either the Stokes~$I$ or the polarized emission towards the Fan Region. In both cases there is excess emission in the data compared to the model \citep[see Figure 5 of ][]{Collaboration:2016eh}, however we note that the problem is much less severe for Stokes~$I$ than for polarized intensity. 

\begin{figure*}[!ht]
\centering 
\begin{minipage}{18cm}
\includegraphics[width=18cm]{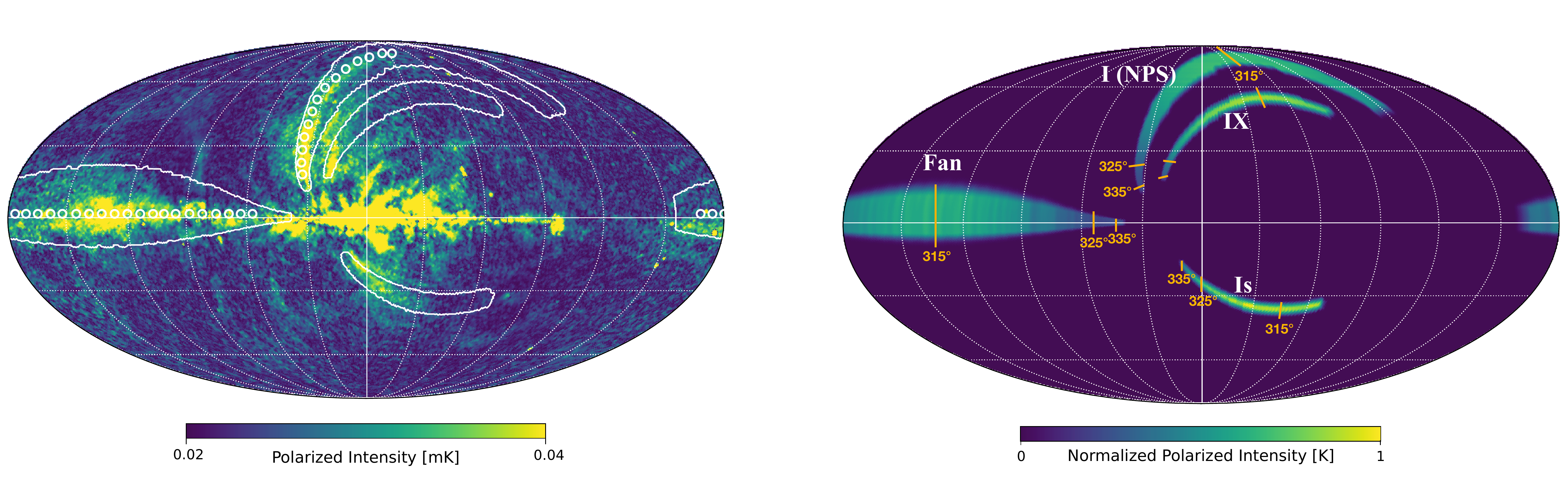}
\end{minipage}

\begin{scriptsize}\caption{\label{fig:planck-pi-overlay}Left: 30~GHz polarized intensity image with contours of the model filaments as they would appear in the sky (see Sec.~\ref{sec:model}). The small circles shown have a diameter of $2^\circ$, and are used to compare the model and the data in Fig.~\ref{fig:brightness}. Right: Simulated synchrotron emission for the parallel filaments as illustrated in Fig.~\ref{fig:schematic} and Fig.~\ref{fig:local-galaxy}, and using the parameters in Table~\ref{tab:modelparams}.  The labels follow the naming convention used by \citet{2015MNRAS.452..656V}. The orange ticks label the values of the variable $\phi$, which is used to define the extent of the filament in the model (see point 2 in Sec.~\ref{sec:curvedlines}). The corresponding points on the top-down view of the filaments are labeled in Fig.~\ref{fig:local-galaxy}.}
\end{scriptsize} 
\end{figure*}

\section{\label{sec:data}A new perspective on existing data}

The idea that the Fan Region and the NPS could both be emission from the Local Arm has a simple premise: synchrotron emission is brightest perpendicular to the field \citep{1965AuJPh..18..635M}. An observer embedded in a uniform magnetic field aligned with the Local Arm should observe two bright patches of synchrotron emission filling a large fraction of the sky at the approximate positions of the NPS and the Fan Region. This can be more easily seen and understood by altering the map projection used to view an image. All-sky maps are most commonly presented in standard map projections (e.g.,  Mollweide) with the Galactic centre at the centre of the map (e.g., as shown in the top panels of Fig.~\ref{fig:fig1}). These projections can distort our perspective and alter how we interpret the data. 

 We compare the linearly polarized radio data at two frequencies: high frequency data at 30~GHz from Planck \citep{2016A&A...594A...1P}, and a combined lower frequency, all-sky map at 1.4 GHz using 26-m Dominion Radio Astrophysical Observatory (DRAO)  \citep{2006A&A...448..411W} and the 30-m Villa Elisa telescope \citep{2008A&A...484..733T} polarization surveys provided by the Centre d'Analyse de Données Etendues (CADE) \citep{2012A&A...543A.103P}. The resolution of these maps is nearly identical ($33'$ and $36'$ respectively). 
 
 In Fig.~\ref{fig:reprojection} we show peak polarized intensity data (i.e., synchrotron emission) from the two data sets. Instead of the standard centre point, we reproject these maps to $(l,b) = (90^\circ,+45^\circ)$, which is roughly mid-way between the Fan Region and the NPS. We also overlay the position angle of the polarized emission, $\chi$, rotated by $90^\circ$, plotted using healpy's line-integral convolution function \citep{Zonca2019}, which represents the measured plane-of-sky magnetic field orientation that has been integrated along the line-of-sight. In this new map projection, we see that the Fan Region and the NPS resemble two parallel features on the sky as we would expect for features that are parallel to the Local Arm. The magnetic field lines are also largely parallel to these features.

The interpretation of these plots must be carefully considered since we observe radiation that is a projection, integrated through the Galaxy, and which also undergoes Faraday rotation as it traverses the magneto-ionic interstellar medium. The amount of rotation is $\propto\lambda^2$, where $\lambda$ is the wavelength of observation. Thus, radiation observed in the 1.4~GHz data will undergo about 400 times more Faraday rotation than that observed in the 30~GHz data, which should undergo negligible rotation.

In Fig.~\ref{fig:datadiff}, we show the acute difference in the polarization angles, $\Delta\chi$, between the two data sets using:
\begin{equation}
\begin{aligned}
    \Delta\chi = 90^\circ - | 90^\circ - |\chi_{1.4~\text{GHz}} - \chi_{30~\text{GHz}}||.
    \end{aligned}
\end{equation}

The magnetic field lines in the two maps are remarkably similar given the differences we might expect to see at the two frequencies due to Faraday rotation. Here the regions of bright polarized intensity show some of the smallest differences between the angles (i.e., dark regions in Fig.~\ref{fig:datadiff}), which supports the idea that these bright features are local. In addition to observation frequency, the amount of Faraday rotation also depends on the path length traversed by the photons, as well as the electron density and line-of-sight magnetic field strength. If the path is very long, then we have a greater potential for Faraday rotation, and we would expect to see larger differences between the polarization angles of the 1.4~GHz and 30~GHz data for the more distant emission. On the other hand, for nearby emission, the path length is short, then the amount of Faraday rotation should be smaller, and we would expect greater agreement between the two maps.

This is another way of showing that the Faraday rotation towards these bright polarized regions is small, which has been shown by other authors \citep[][]{1984A&A...135..238S,2015ApJ...811...40S, 2021AJ....162...35W}. Small FD in the direction of the Fan Region is not unexpected since, from both modelling and observations, we expect that the Galactic magnetic field in this direction is dominated by the plane-of-sky magnetic field ($B_{\perp}$), whereas the line-of-sight field ($B_{\parallel}$) is small. The main takeaway here is that these two regions of similarly low FD, and high polarized intensity, have magnetic field directions that, on the sky, appear largely parallel to each other.

\begin{figure}[!ht]
\centering \includegraphics[width=7cm]{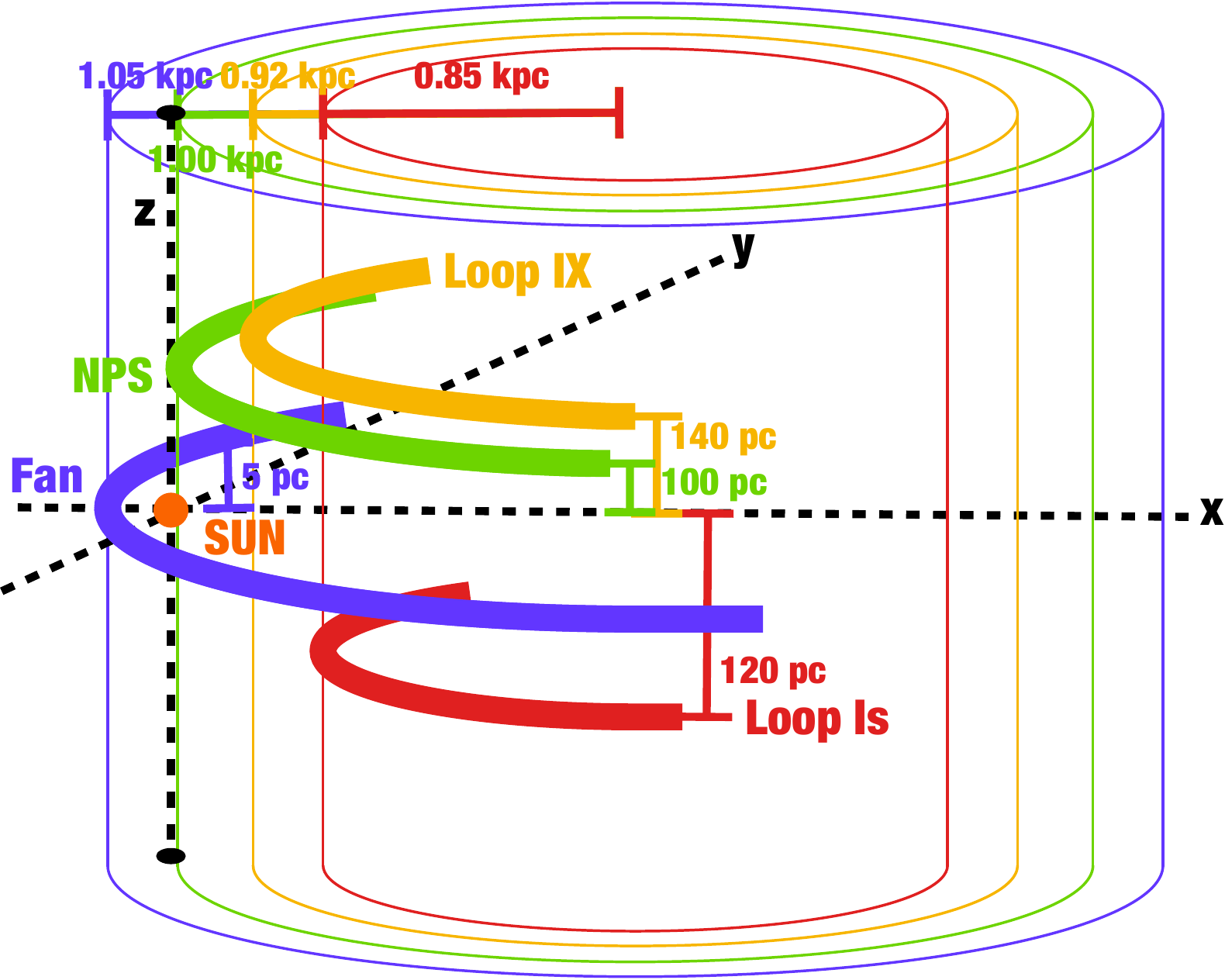}
\begin{scriptsize}\caption{\label{fig:schematic}The arrangement of filaments, constructed on nested cylinders. The diagram is not to scale, but placement of the filaments is correct with respect to the Sun's position and relative to each other. The x-, y-, and z- axes are labeled, with x-axis directed towards $l=0^\circ$ and the y-axis towards $l=90^\circ$. 
}
\end{scriptsize} 
\end{figure}

\section{\label{sec:model}Model}

We use the Hammurabi code \citep{Waelkens:2009bn} to simulate observations of the local environment around the Sun ($\pm1$~kpc). Hammurabi is a synchrotron emission and Faraday rotation modelling code that uses a grid of 3D magnetic field, thermal electron distribution, and CRE distribution as inputs. Our grid uses $512^3$ elements, giving a resolution of $\sim4$~pc per element, with the Sun located at the centre, at the origin of our coordinate system. We use a constant value for the thermal electron and CRE densities. An enhancement of CREs along a filament-like geometry would enhance the synchrotron emission, but it seems probable that there would be a corresponding enhancement of magnetic field. We assume a simple scenario  where the magnetic field has a filamentary geometry.  The internal magnetic field is aligned parallel to each filament, and the field is zero elsewhere. We use a 3D-grid where the $x$- and $y$-axes are in the Galactic plane, the $z$-axis is perpendicular to the plane, the Sun is at the origin, and the Galactic centre is towards positive-$x$ (i.e., $x>0$ is towards $l=0^\circ$, and $y>0$ is towards $l=90^\circ$, see Fig.~\ref{fig:uniformfields}).

The filaments of the NPS in the 30~GHz data are about $6^\circ$ wide. At a distance of 100~pc, they would be $\sim10$~pc wide, and so initially (in Sec.~\ref{sec:straightlines}) we set the width of each filament to $w=10$~pc. In Sec.~\ref{sec:curvedlines} and Sec.~\ref{sec:tiltedlines}, we change this to $w=20$~pc to compensate for the fact that a long filament will appear narrower at a farther distance.

\subsection{\label{sec:brightness}Brightness}

We compute the model synchrotron intensity as

\begin{equation}
\centering
\text{I}(\nu, p)=K\sum_{i} {J_{\text{CRE},i}}{B_{\perp, i}^{{(p+1)}/2}}\Delta R~{\text{[K]}},
\end{equation}

where for $i$ elements along a line-of sight, $\nu$ is the observing frequency, $p$ is the power-law index of the electron energy spectrum, $K$ is a normalizing constant, $J_{\text{CRE}}$ is the spatially dependent CRE density, $B_{\perp}$ is the magnetic field component in the plane of the sky, and $\Delta R$ is the step size along the line-of-sight. 

We assume $p=3$, which has been shown to be approximately correct over much of the Galaxy across a wide frequency range, from 408~MHz to 23~GHz \citep{2011A&A...534A..54S}, and which gives $\text{I}(\nu)\propto B^2$. However other authors have found a range of spectral indicies with generally less steep (i.e., smaller) values at lower frequencies (down to $p\approx2.5$)  \citep[e.g.][]{1988A&A...196..211R,2011A&A...534A..54S,2019MNRAS.485.2844D,2016A&A...594A..25P}. 

Synchrotron intensity depends on the (plane-of-sky) magnetic field strength, the CRE density, and the path length, and by adjusting these quantities, it is possible to model an arbitrary intensity. Here we examine whether the proposed model can plausibly explain the observed radio brightness of the NPS and the Fan Region. 

In Fig.~\ref{fig:fig1} we show that the Fan and NPS are both enveloped by a contour with a level of 24~K for the 408~MHz total intensity data. The CRE density is very uncertain, but previous work has used values in the range from  $0.25<J_{\text{CRE}}< 0.4$~GeV$^2$m$^2$s$^{-1}$sr$^{-1}$ \citep[][]{2010MNRAS.401.1013J}. Assuming these values for $J_{\text{CRE}}$, and using a typical Galactic magnetic field strength of $6~\mu$G, and also assuming the width of our proposed filament is 20~pc, we find a synchrotron intensity of approximately $0.2$~K to $0.3$~K, which is a factor of 80-120 smaller than we observe. 

However, if these filaments are a result of multiple supernova explosions, we expect that the gas would be compressed, resulting in a stronger magnetic field and increased CRE density. The theoretical compression factor in a young supernova remnant (i.e., a strong shock) is at least a factor of 4. For older, radiative filaments the compression factor is thought to be much higher \citep{2017hsn..book.1981R}, and for multiple explosions the factor could be higher yet. Using a compression factor of 4 as a lower limit, we find the magnetic field strength is $24~\mu$G and a CRE density of $1.0<J_{\text{CRE}}< 1.6$~GeV$^2$m$^2$s$^{-1}$sr$^{-1}$. Thus, the resulting increase in the synchrotron intensity is a factor of at least 64. Using these reasonable assumptions about the magnetic field and CRE density, we find an intensity for the model filament of the same order of magnitude as for the observations. 

At 30~GHz there are additional uncertainties that make a comparison more difficult. For example, at 30~GHz, we measure a 0.015~mK polarized intensity contour level as enclosing approximately the same region as the 24~K total intensity level at 408~MHz, as shown in Fig.~\ref{fig:fig1}. However, the total synchrotron component in Stokes~$I$ at 30~GHz is uncertain due to the contributions from other emission mechanisms, calibration errors in the data, and uncertainties in the fractional contributions to this brightness from other sources (i.e., we do not know how much of this emission is due to the background Galactic or extragalactic contributions). The polarized fraction and spectral index are also both uncertain, making an extrapolation from low to high frequency very unreliable. We can do a rough order of magnitude estimate: assuming a polarized fraction of 40\%, we find that a spectral index of $p=2.63$, gives the measured 30~GHz polarized flux of 0.015~mK. Using $p=3$, we find a 30~GHz polarized brightness that is 3-4 times smaller than the measurement.

Given the degeneracy between $J_{\text{CRE}}$ and $B_{\perp}$, and the uncertainties in these quantities, as well as $p$, we use the normalized brightness for the models that we show in the next few sections that focus on morphology.

\begin{figure*}[!ht]
\centering 
\begin{minipage}{5.7cm}
\includegraphics[width=5.6cm]{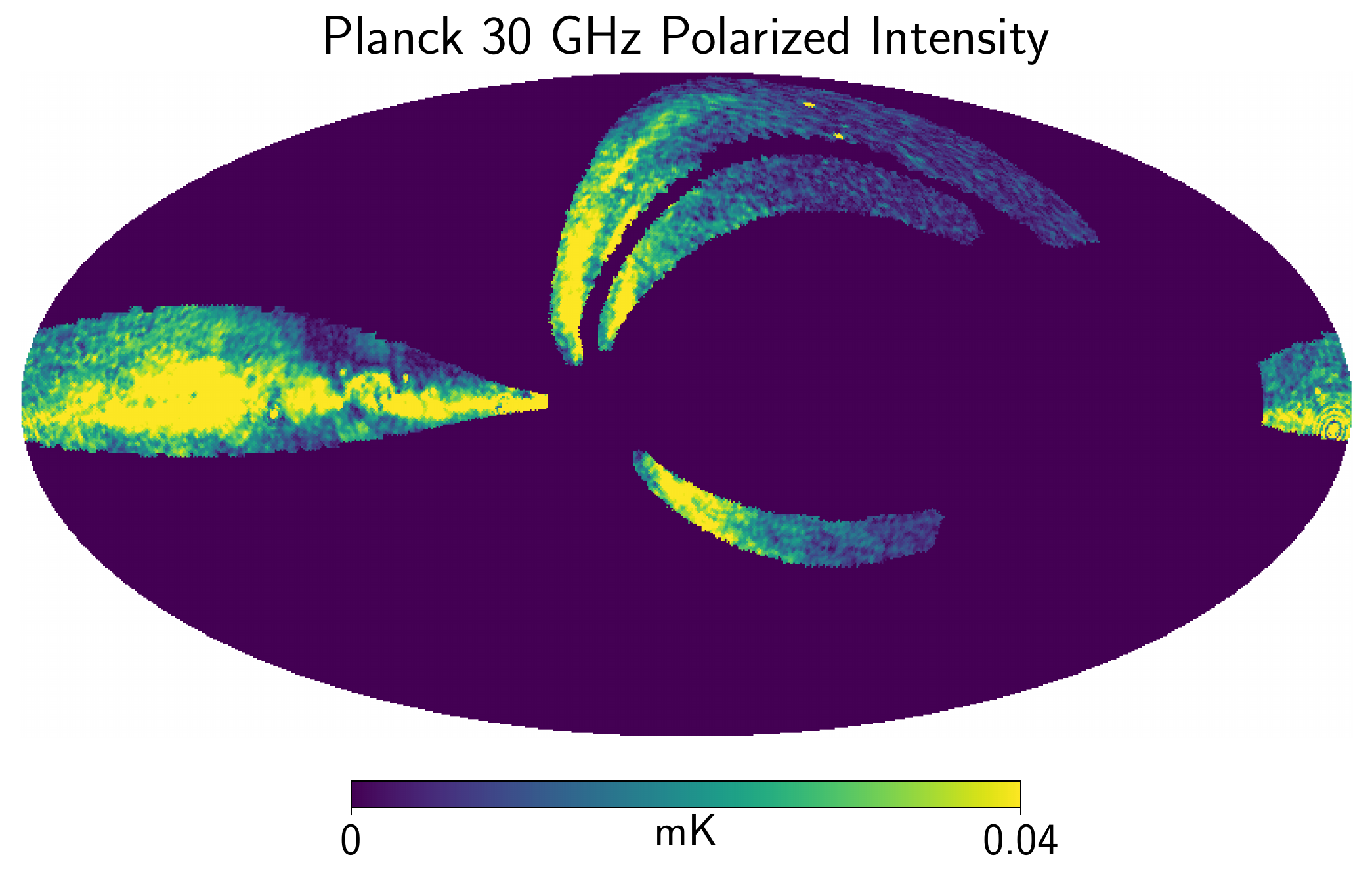}
\end{minipage}
\hfill
\begin{minipage}{5.7cm}
\includegraphics[width=5.6cm]{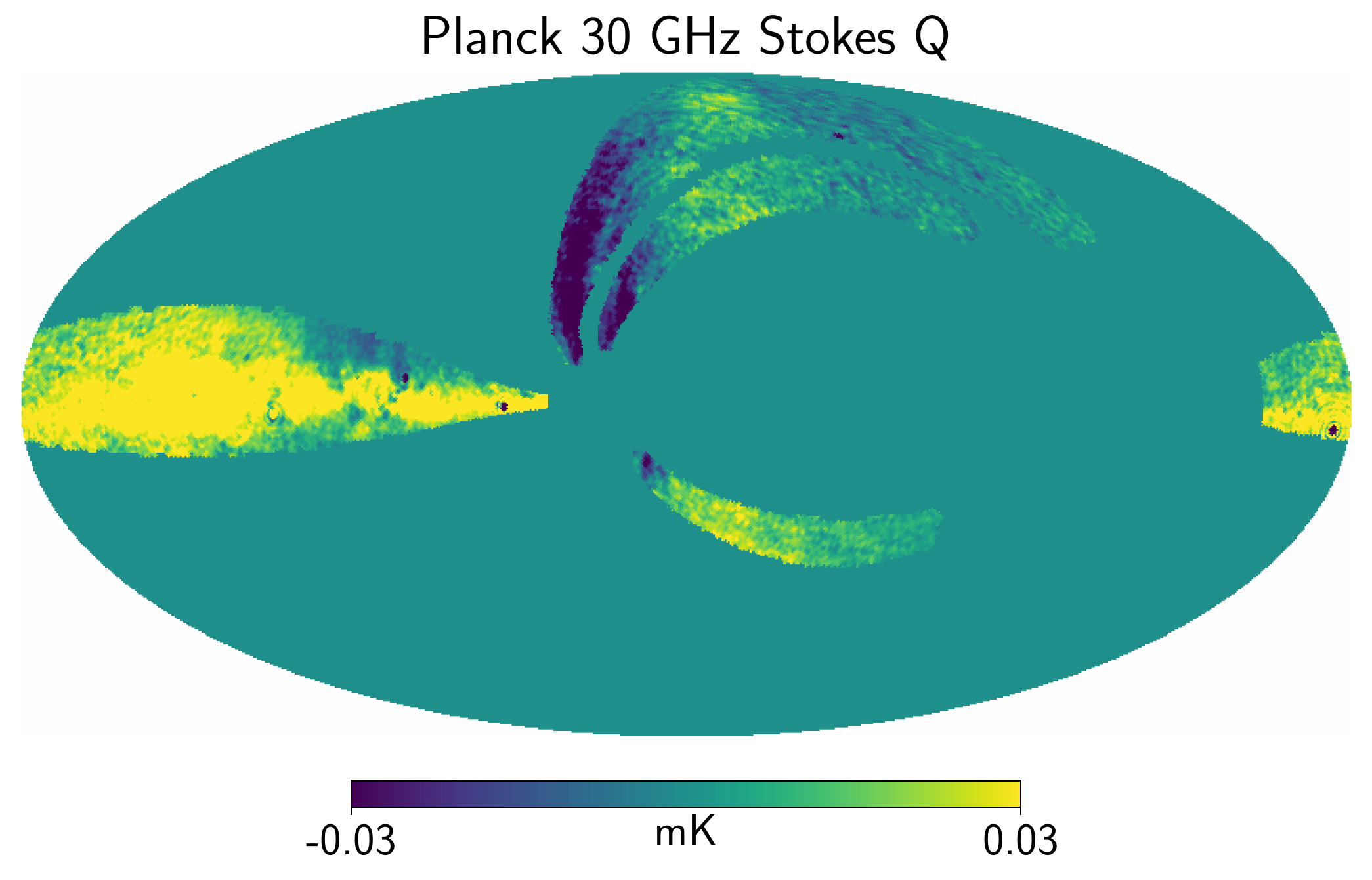}
\end{minipage}
\hfill
\begin{minipage}{5.7cm}
\includegraphics[width=5.6cm]{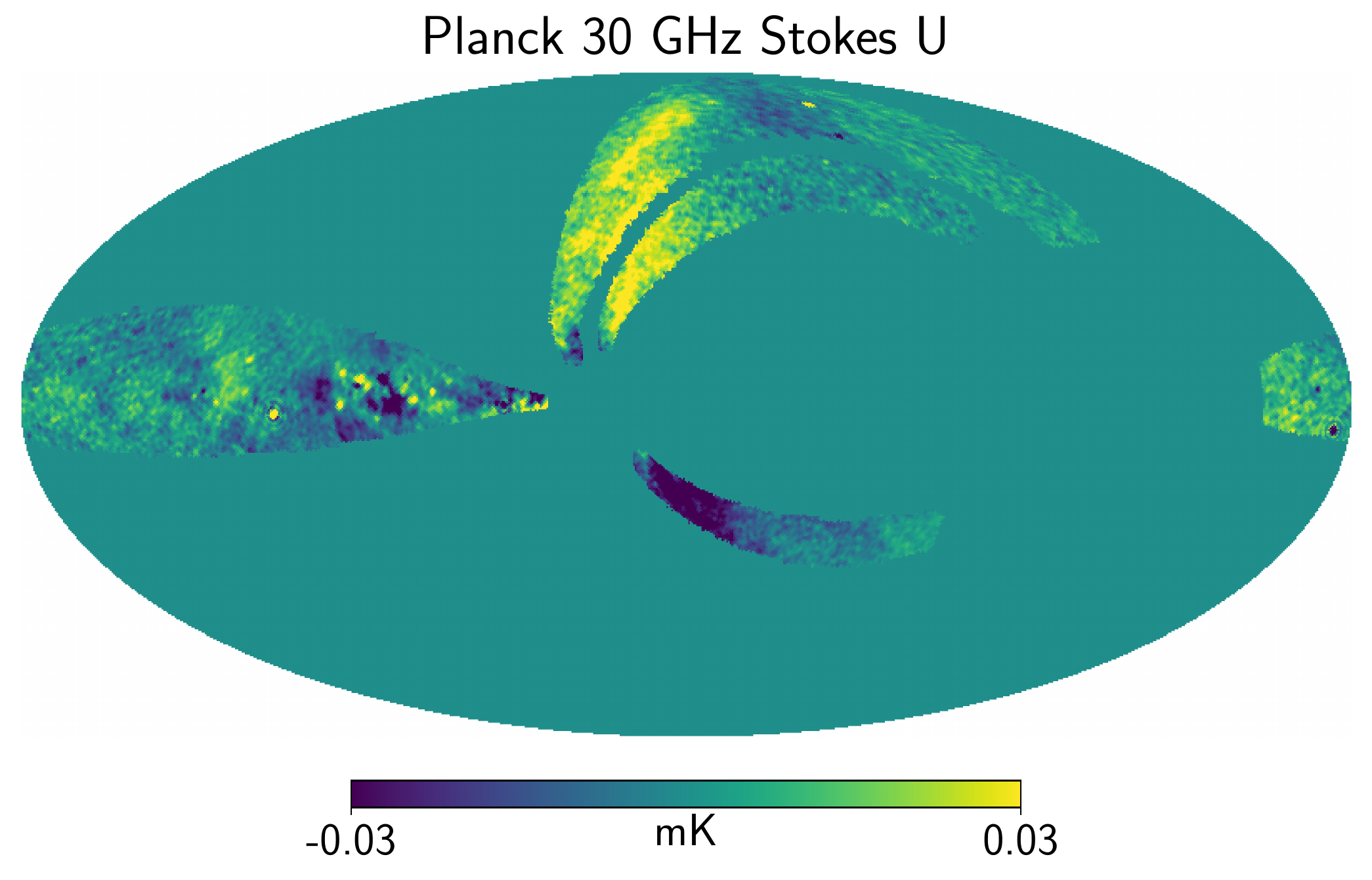}
\end{minipage}
\begin{minipage}{5.7cm}
\includegraphics[width=5.6cm]{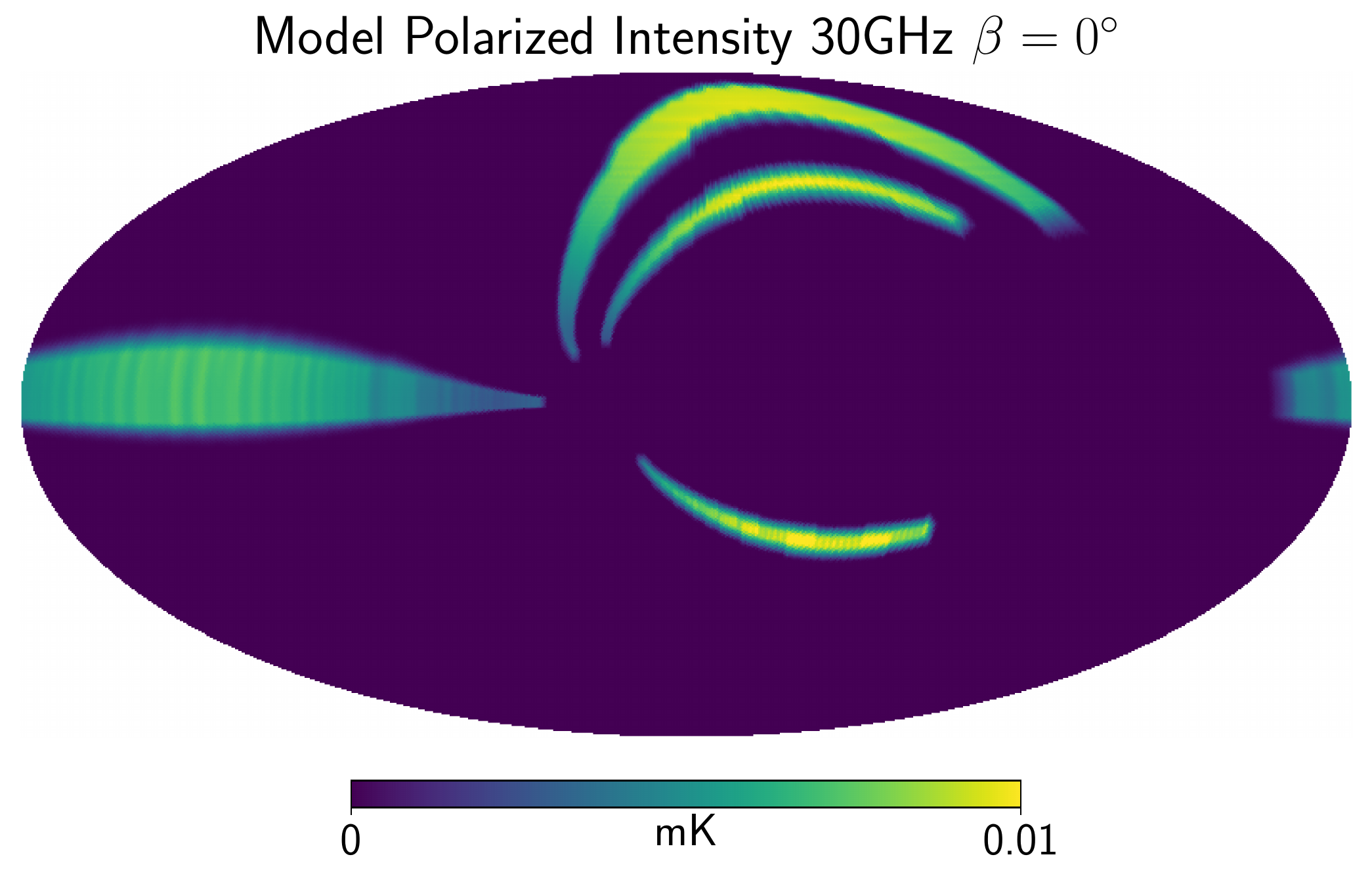}
\end{minipage}
\hfill
\begin{minipage}{5.7cm}
\includegraphics[width=5.6cm]{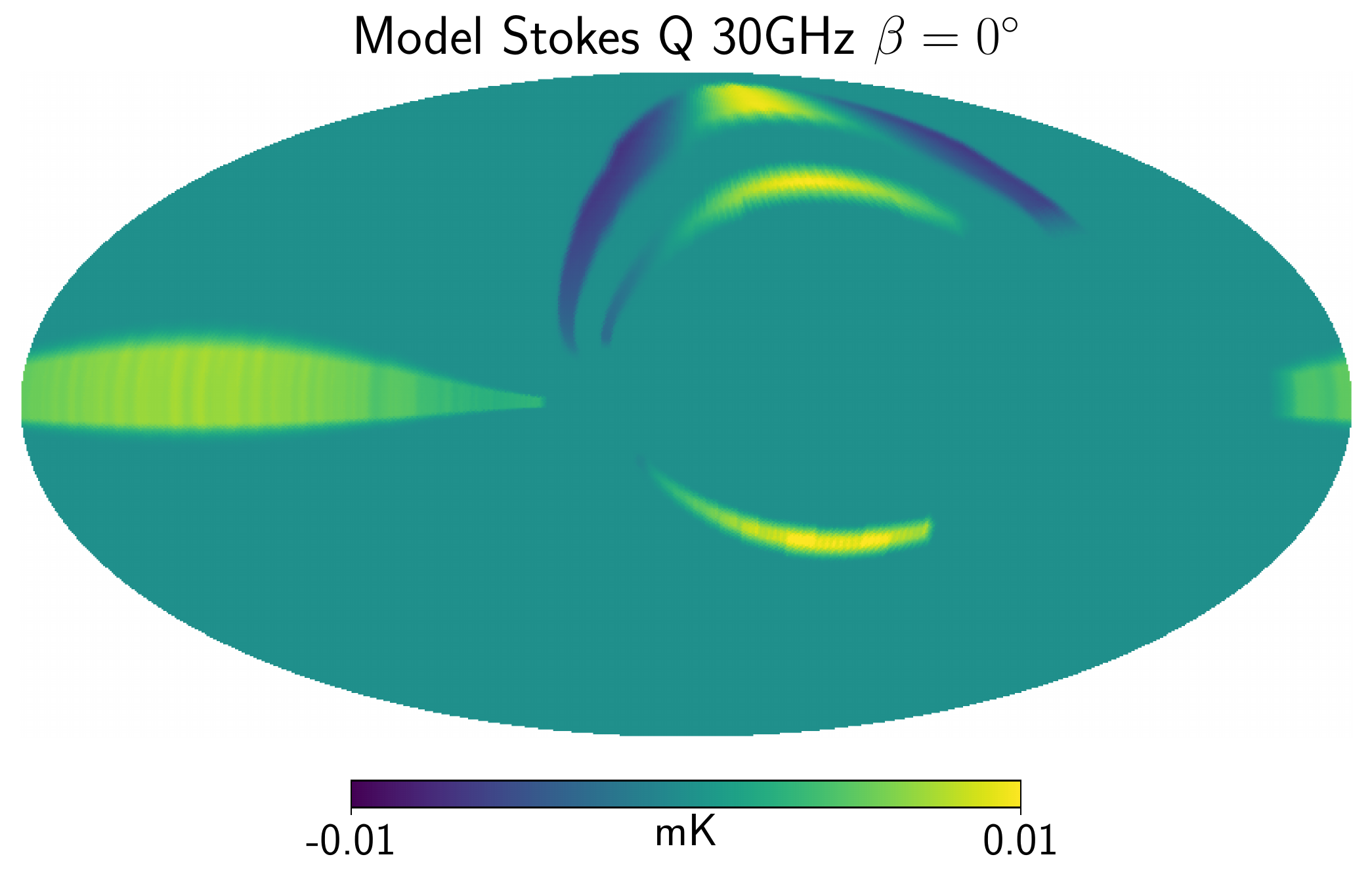}
\end{minipage}
\hfill
\begin{minipage}{5.7cm}
\includegraphics[width=5.6cm]{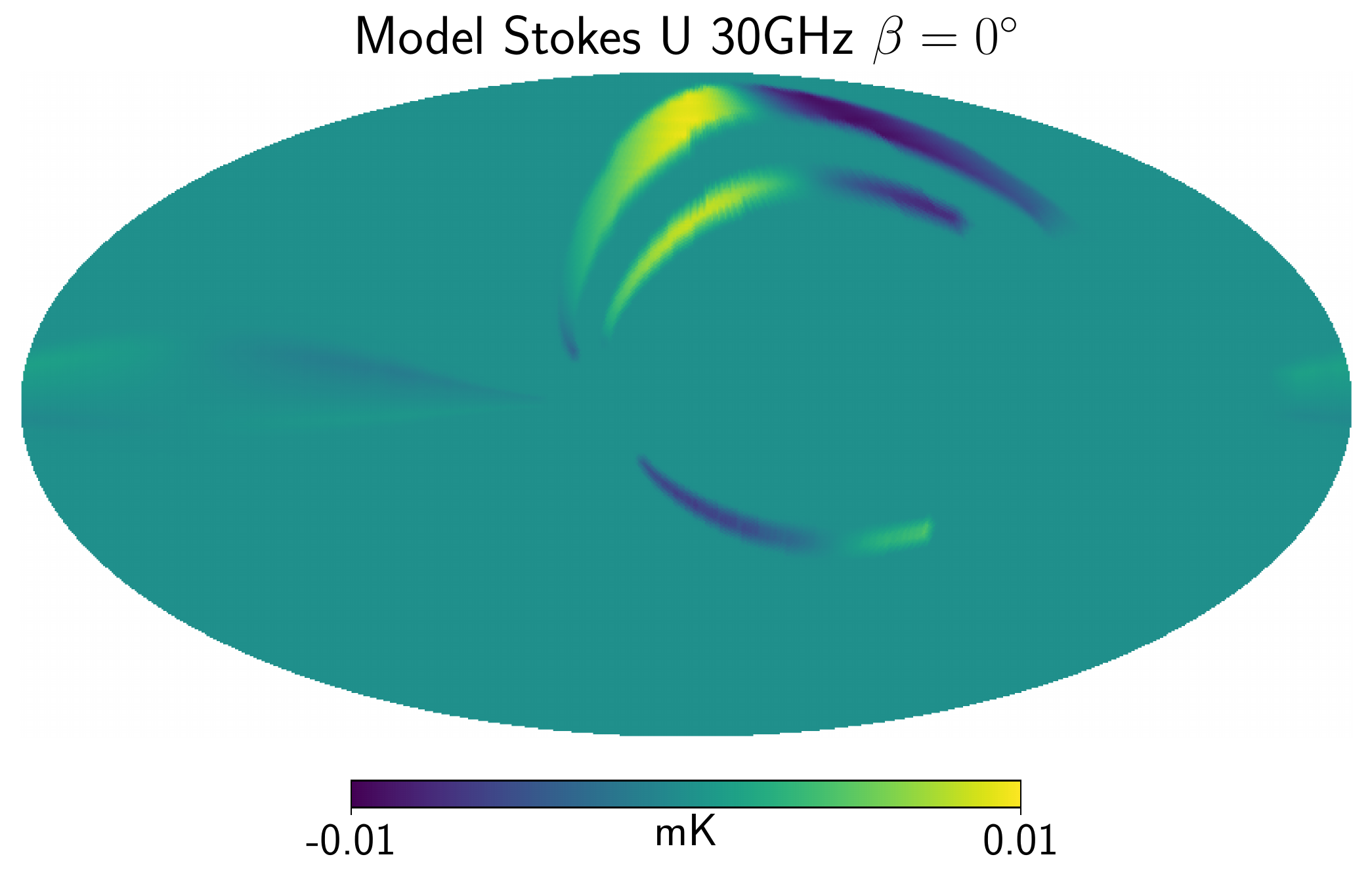}
\end{minipage}
\begin{minipage}{5.7cm}
\includegraphics[width=5.6cm]{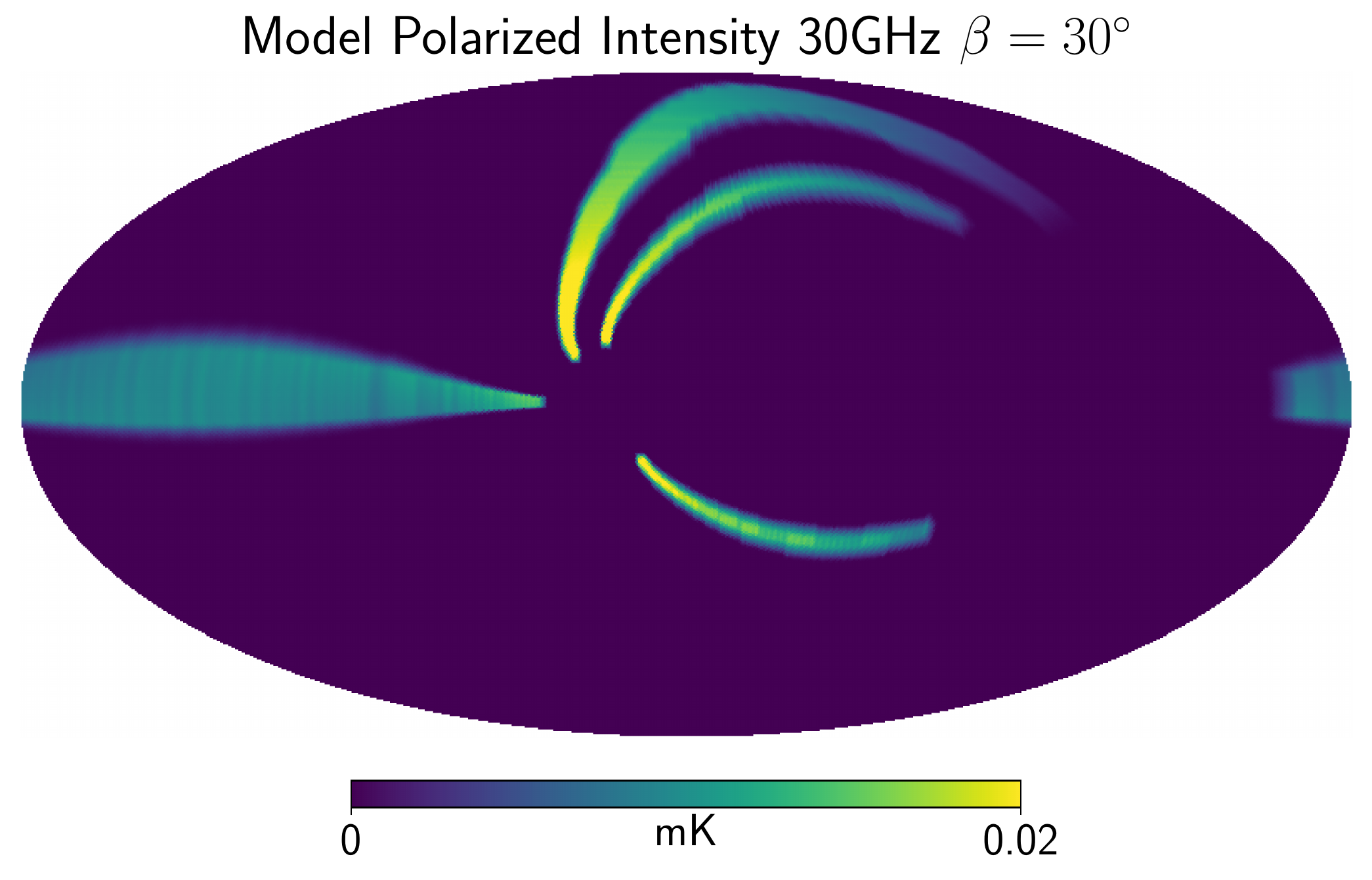}
\end{minipage}
\hfill
\begin{minipage}{5.7cm}
\includegraphics[width=5.6cm]{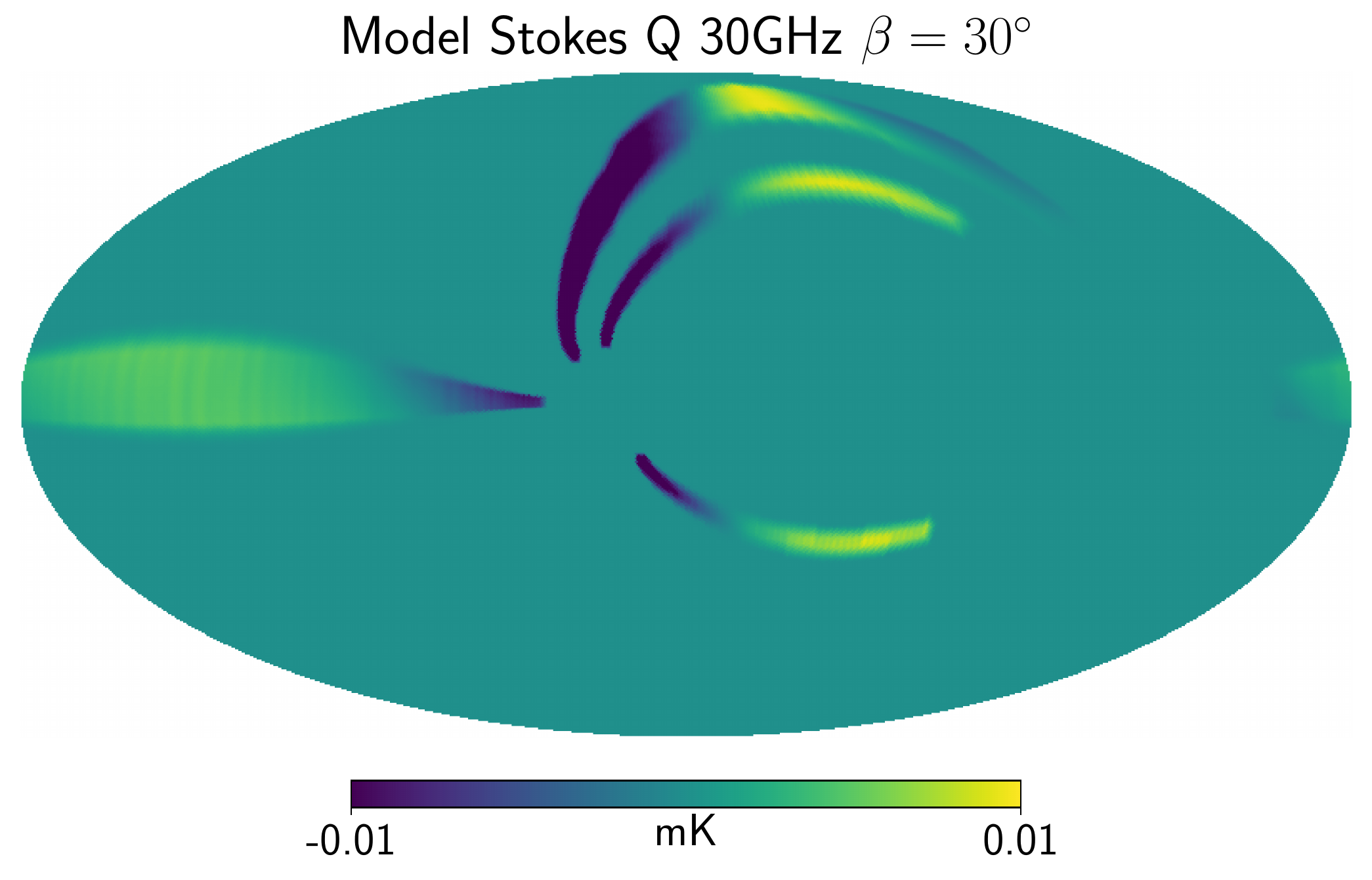}
\end{minipage}
\hfill
\begin{minipage}{5.7cm}
\includegraphics[width=5.6cm]{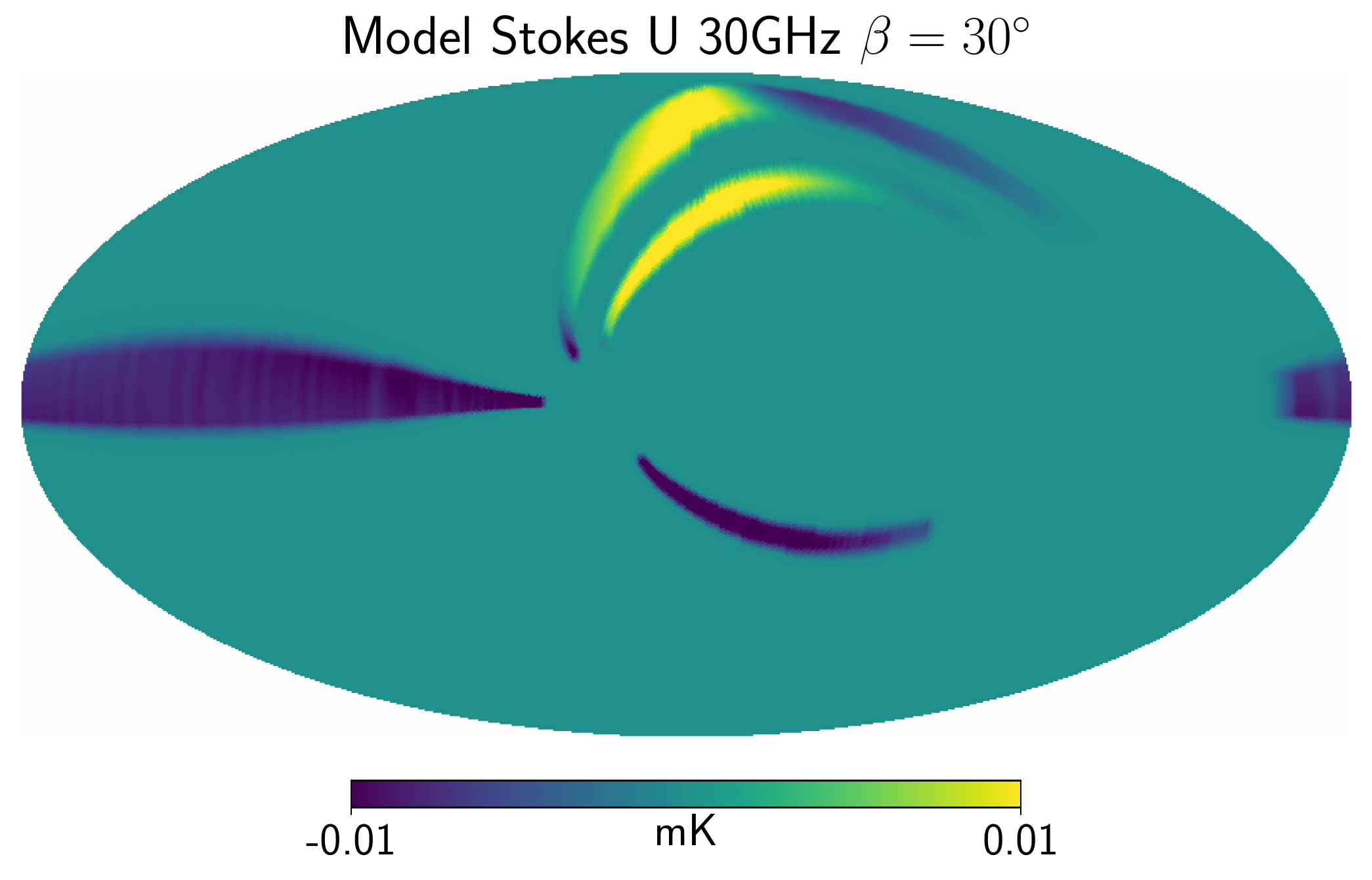}
\end{minipage}
\caption{ \label{fig:qu} Top row: Left: 30~GHz Planck PI. A mask has been applied to show only values in locations where the model has been computed. Centre: Masked Planck 30~GHz Stokes~$Q$. Right: Masked Planck 30~GHz Stokes~$U$. Middle row: Left: 30~GHz Model PI with $\beta=0^\circ$. Centre: 30~GHz Model Stokes~$Q$ with $\beta=0^\circ$. Right: 30~GHz Model Stokes~$Q$ with $\beta=0^\circ$. Bottom row: Same as middle row, but for  $\beta=30^\circ$. }
\end{figure*}

\subsection{\label{sec:straightlines}Straight-line Filaments}

For the first iteration of our model, we use regularly spaced straight filaments and model their synchrotron brightness. These straight-line filaments extend the entire distance through a volume that measures 2000~pc on a side, with the Sun at the centre.  In each case we model eight filaments that pass through the points (measured in pc): $(0,0,100)$ (directly above us), $(0,0,-100)$ (directly below us), $(100,0,0)$ (in the Galactic plane and towards the Galactic centre, on the right side of the plot as shown in Fig.~\ref{fig:uniformfields} (left panels)), $(-100,0,0)$ (in the Galactic plane and away from the Galactic centre, on the left side of the plot as shown in Fig.~\ref{fig:uniformfields}), $(100,0,100)$ (top-right), $(-100,0,100)$ (top-left), $(100,0,-100)$ (bottom-right), and $(-100,0,-100)$ (bottom-left).  

We orient the filaments at a range of angles, in $10^\circ$ increments from $l_\alpha=10^\circ$ to $l_\alpha=180^\circ$, where $l_\alpha$ is the Galactic longitude of the orientation of the filaments as viewed from the top-down. We choose a distance of 100~pc as this is a value often quoted in the literature as the distance to the NPS \citep[e.g.,][]{1968Obs....88..269S, 2012A&A...545A..21P, 2015ApJ...811...40S} (at least the distance to the nearest point, and in the non-Galactic centre scenario, see discussion in Sec.~\ref{sec:nps}). 

Looking at the full set of models (a subset is shown in Fig.~\ref{fig:appendixuniformfields}), we already see broad features that resemble the Fan Region and NPS and some of the other loops. We pay particular attention to the models where $l_\alpha = 70^\circ$, shown in the top row of Fig.~\ref{fig:uniformfields}, since this aligns with the Local Bubble magnetic field (see Sec.~\ref{sec:intro}). A significant observational feature of these models is that there are two points where the filaments appear to converge, which shifts with the orientation angle. By visual inspection of the data in Fig.~\ref{fig:planck-pi-overlay} (left panel), we can see this convergence point occurs around $l=30^\circ$. Looking at the models in Fig.~\ref{fig:appendixuniformfields}, we see that an orientation of $l_\alpha=30^\circ\pm10^\circ$ most closely matches this convergence point, which makes sense since in this model the observer is symmetrically surrounded by filaments going in that direction. Looking at the curvature of the NPS for $b>30^\circ$, we find that models with $l_\alpha=40^\circ\pm10^\circ$ are more consistent.

Using the morphology of the NPS as a guide, and by inspecting the 30~GHz polarized intensity map, as shown in Fig.~\ref{fig:planck-pi-overlay}, we see that, although the orientation of the model synchrotron filament is an approximate match to the position and orientation, it still differs from the observations. We adjust the model by altering the orientation of the filaments between $20^\circ<l<70^\circ$ in $5^\circ$ increments, and adjusting the $(x,y,z)$ position of each filament from 10~pc to 150~pc in increments of 10~pc. From these we select, by eye, four line segments that correspond to the approximate positions of the NPS and the Fan Region. We also choose to include two other prominent bright filaments on the sky, Loop IX and Loop Is (see Fig.~\ref{fig:planck-pi-overlay}), following the naming convention of \citet{2015MNRAS.452..656V}, as example cases to see whether this model can be applied more generally. We find that overall, an orientation towards $l_\alpha=45^\circ$ is a reasonably close match, and the chosen filaments pass through the following positions: NPS: $(20, 0, 60)$, Fan: $(-100, 0, 5)$, Loop IX: $(100, 0, 150)$, and Loop Is: $(140,0,-150)$. These are shown in the bottom panel of Fig.~\ref{fig:uniformfields}. 

\begin{table*}
\centering
\begin{tabular}{|c|c|c|c|c|c|} %c|c|}
\hline 
 & $dr$ (pc) & $dz$ (pc) & w (pc) & $\phi_1$ & $\phi_2$ \tabularnewline 
\hline 
\hline 
Fan & 50 & 5 & 20 & $310^\circ$ & $340^\circ$ \tabularnewline
\hline 
NPS & -20 & 100 & 20 & $310^\circ$ & $340^\circ$ \tabularnewline 
\hline 
Loop IX & -100 & 140 & 20 & $310^\circ$ & $340^\circ$ \tabularnewline 
\hline 
Loop Is & -170 & -120 & 20 & $310^\circ$ & $340^\circ$ \tabularnewline 
\hline 
\end{tabular}

\caption{\label{tab:modelparams}Parameters used for the filaments in the preferred curved filament model, with $\beta=0^\circ$. These filaments have \textbf{$l_\alpha=45^\circ$}, and $c=(0.72,-0.72)$. $dr$ is the distance from the Sun to the filament, and $dz$ is the distance above the Galactic mid-plane. The uncertainty on all positional values is approximately $\pm10$~pc}
\end{table*}

\begin{figure*}[!ht]
\centering \includegraphics[width=17cm]{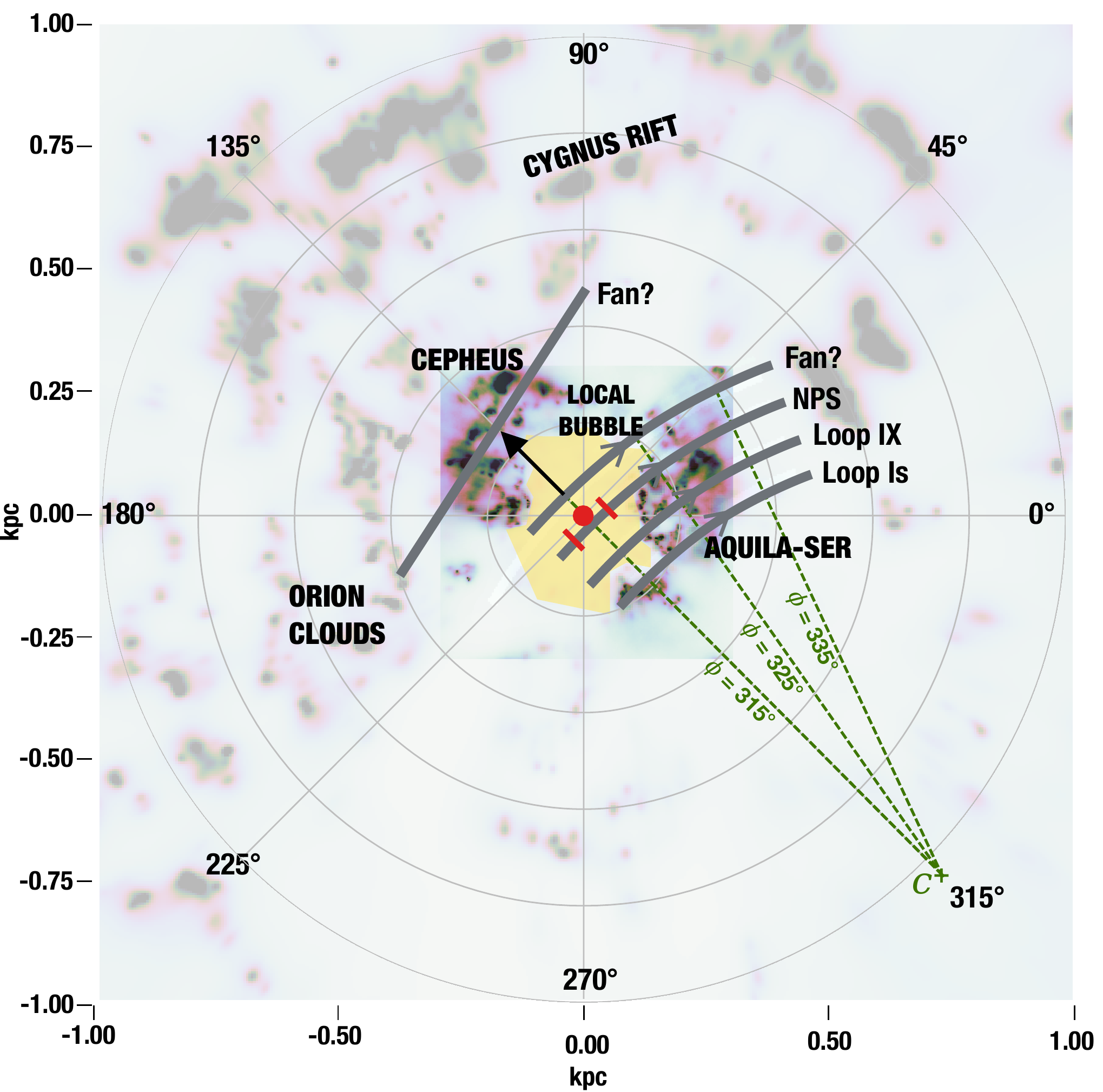}
\begin{scriptsize}\caption{\label{fig:local-galaxy} Filaments overlaid on the dust model from \citet{2018A&A...616A.132L} (\url{https://stilism.obspm.fr}), which has been integrated over the z-axis, and with labels from \citet{2014A&A...561A..91L}.
The inset central square region shows a higher resolution dust model derived from Gaia data \citep{2019A&A...631A..32L,2021A&A...647C...1L} for a smaller region, $\pm300$~pc from the Sun (marked with a red dot). The red ticks on the NPS filament indicate the segment seen at high Galactic latitudes ($b>50^\circ$). The arrows indicate the direction of the magnetic field. There are two filaments labeled for the Fan Region. The straight filament is at an orientation of $60^\circ$ and is coincident with the structure found by \citet{2020Natur.578..237A} (and discussed in Sec.~\ref{sec:origin}). The green cross, labeled $c$, marks the central axis of the cylinder that is used to defined the filaments. The angle $\phi$ is measured from $c$ in a clockwise direction, with $\phi=0^\circ$ towards the top of the plot (i.e., towards $l=90^\circ$). The green dashed lines show several values of $\phi$, with corresponding points shown in Fig.~\ref{fig:planck-pi-overlay}.
}
\end{scriptsize} 
\end{figure*}

\subsection{\label{sec:curvedlines}Curved Filaments}

In Sec.~\ref{sec:straightlines} we described an extremely simple model that can reproduce structures in locations broadly corresponding to the NPS, the Fan Region, and other prominent loops. However, there are some important differences in that, for $|b|<30^\circ$, the modelled filaments (excluding the Fan) do not appear to curve inwards (i.e., towards the Galactic centre) when projected on the sky, as seen in the data. Furthermore, we can hardly expect to find perfectly straight filaments nearly 3~kpc in length. The structure detected by \citet{2020Natur.578..237A} is seen to be 2.7~kpc long but it shows a distinct curvature in the $xy$-plane, and an even more extreme undulation through the disk of the Galaxy. 

We therefore refine our model by using four curved, but still parallel, filaments. At several hundred parsecs long, they are longer than the magnetized bubble filaments proposed by \citet{2015MNRAS.452..656V}, but shorter than our straight filaments. We place these parallel, curved line-segments on a set of concentric cylinders, as illustrated in Fig.~\ref{fig:schematic}. We created a grid of models by adjusting the following parameters: 

\begin{enumerate}
  
  \item $l_\alpha$, as defined previously, is the longitude of the orientation of the filaments as viewed from the top-down. In the case of curved filaments, it is the angle of the tangent to the curve at the point closest to the Sun. 

  \item $c$ $(c_x,c_y)$, is the centre point of the cylinder that is used to construct the parallel curved filaments, as labeled in Fig.~\ref{fig:local-galaxy}. We define only one component of this centre coordinate, $c_y$. $c_x$ is then calculated by $c_x=-{c_y}{\tan(\alpha)}$. The distance from the point $c$ to the Sun is $r=\sqrt{c_x^2+c_y^2}$. The filaments extend over a range of angles from $\phi_2$ to $\phi_1$, which are measured from $c$ in a clockwise direction, with $\phi=0^\circ$ towards the top of the plot (i.e., towards $l=90^\circ$) (see Fig.~\ref{fig:local-galaxy}).

  \item We define the relative distance to each filament, $dr$,  which is measured from the Sun to the filament. The distance, $c_r$, which is measured from the filament to point $c$ is given by $c_r=r+dr$. We also define the height of each filament above the Galactic plane, $dz$. The filaments have a width, $w$.
  
  \item We define an angle $\beta$, that measures the inclination of the magnetic field, with respect to the Galactic plane, where $B_z/B_{xy}=\tan(\beta)$. Here $B_z$ is the component of the magnetic field that is perpendicular to the plane, and $B_{xy}$ is the total magnetic field that is parallel to the plane. For this curved filament model, the filaments are entirely parallel to the plane and the magnetic field is aligned with the filament, thus $\beta=0^\circ$.

\end{enumerate}

Once again, we focus on the geometry of the NPS when searching this parameter space. We find that for an orientation of $l_\alpha=+45^\circ$, and curving the filaments along the arc of a circle having a centre at point, $c$, located at $r=1.02$~kpc from the Sun towards $l=315^\circ$, $c=(0.72, -0.72)$~kpc, (see Fig.~\ref{fig:local-galaxy}) an excellent correspondence to the position and curvature of the NPS is achieved (see Fig.~\ref{fig:planck-pi-overlay}). We also increase $w$ from 10 to 20~pc, since given the length, the angular size decreases along the length of the filament. These parameters are summarized in Table~\ref{tab:modelparams}.

The location of the centre point at $l=315^\circ$ is a geometrical construct that allows us to define the position of the filaments. We do not intend to imply that there is anything particularly special or physically meaningful about this location. However, we note that an analysis by \citet{2016Natur.532...73B} finds that one of the supernovae that likely contributed to the Local Bubble must have occurred in this general direction (towards $l=327^\circ$, $b=+11^\circ$). This is discussed further in Sec.~\ref{sec:origin}.
 
Although there is some degeneracy for the distance, radius, and width of these filaments, and we are able to achieve comparable matches to the data with different values of $c$, we choose to keep $dz_{NPS}=100$~pc, since there are several independent observational studies that agree on this value (see Sec.~\ref{sec:nps}). This fixes the particular value of $c$ that we use. 

We keep the other line-segments parallel to the NPS filament by placing them on the surface of a set of nested cylinders each having the same centre point defined above (see Fig.~\ref{fig:schematic}). Each segment is parallel to the Galactic plane, and located at varying distance, $dz$, above or below it. The segments also have varying distance, $dr$, measured from the Sun to the centre of the filament.

In Fig.~\ref{fig:qu}, we show the model PI, $Q$, and $U$ (middle row) computed at 30~GHz, in comparison to the Planck 30~GHz data (top row). For ease of comparison, we mask the Planck data to show only the regions where the model is defined. By comparing these maps, one can see that there is good general correspondence between the sign of Stokes $Q$ and $U$ for the model when compared to the data. However, when comparing the PI maps, there is a discrepancy where the Planck PI map has a considerable brightness gradient along the NPS that is not seen in the model (middle row, left column). This is discussed further in the next section. 

Fig.~\ref{fig:local-galaxy} shows the position of these filaments on the distribution of dust in the nearby Galaxy, a distribution based on reddening (from SDSS/APOGEE-DR14, Gaia, and 2MASS) and parallax observations of nearby stars from Gaia.

\subsection{\label{sec:tiltedlines}Tilted magnetic field}

\begin{figure*}[!ht]
\centering 
\begin{minipage}{8.5cm}
\includegraphics[width=8.4cm]{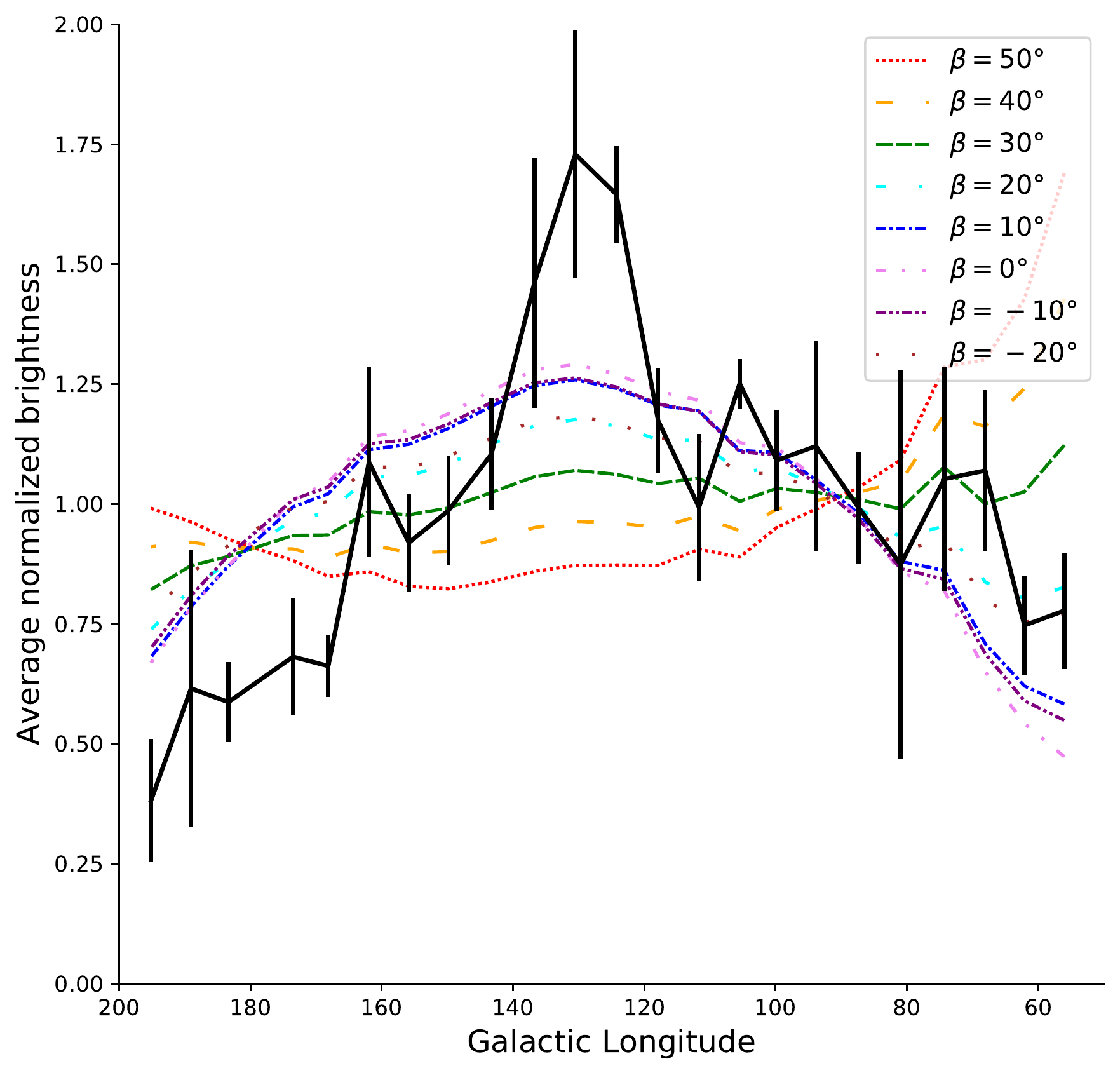}
\end{minipage}
\hfill
\begin{minipage}{8.5cm}
\includegraphics[width=8.4cm]{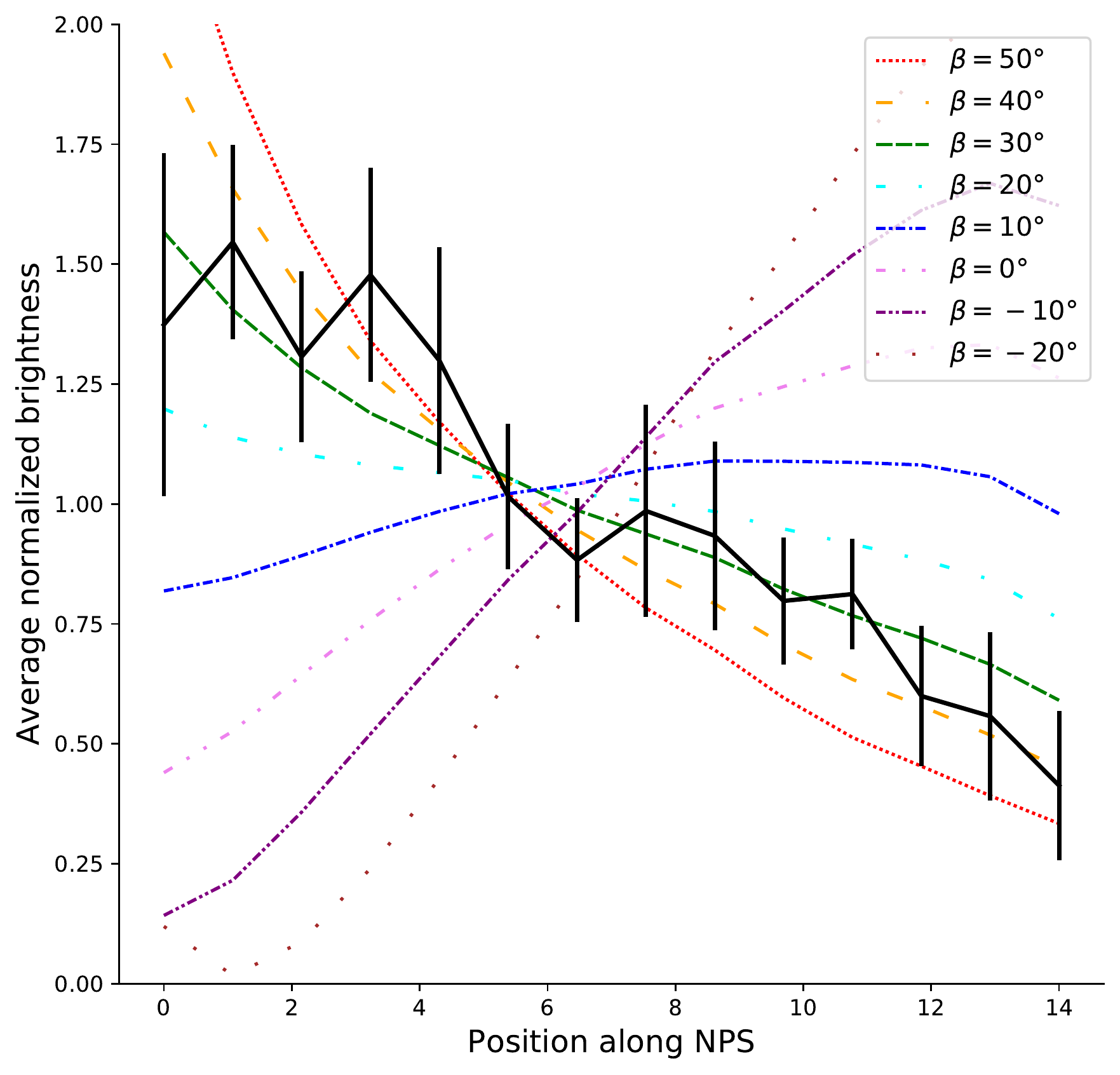}
%figures/m1-wT.pdf
\end{minipage}
\hfill
\begin{scriptsize}\caption{ \label{fig:brightness} Polarized intensity averaged over 2-degree diameter circles at points along the length of the Fan Region for $b=1^\circ$ (left) and the NPS (right) as shown in Fig.~\ref{fig:planck-pi-overlay}. The solid black line is the 30~GHz data and the other lines are the models having varying values of $B_z$ such that $B_z/B_{xy}=\tan(\beta)$. In the case of the NPS, the points are equally distributed along the spur starting nearest to the Galactic plane. For the Fan Region, the best of these models is for a small, or no, $B_z$ component ($-10^\circ<\beta<10^\circ$), while for the NPS, the best of these models is for $\beta\approx30^\circ$. The error bars show the standard deviation within each 2-degree circle. The bright peak in the Fan Region plot (left) around $120^\circ<l<140^\circ$ is due to bright polarized features around the W3/W4/W5 superbubble region, which has been excluded from the definition of the Fan Region used in this work (see Fig.~\ref{fig:fig1} and Sec.~\ref{sec:fan}). }
\end{scriptsize} 
\end{figure*}

One prominent observational feature is a distinct brightness gradient along the length of the NPS and related loops, IX and Is, as one can see in Fig.~\ref{fig:planck-pi-overlay}. This is not well reproduced with filaments that are strictly parallel to the Galactic plane with an aligned magnetic field. The brightness gradients may arise from density variations along the filaments, but they could also result from variations in synchrotron brightness due to variations in magnetic field orientation. If we change the orientation of the magnetic field, by including a vertical field component, $B_z$, we can reasonably reproduce the observed brightness gradients of the NPS (also IX and Is). In Fig.~\ref{fig:brightness}, we show the brightness gradients for a range of values of magnetic field orientation, and we show that using $B_z/B_{xy}\approxeq0.6$ (i.e., inclined at $\sim30^\circ$ to the Galactic plane) for the NPS, gives a reasonable match to the data.  In the bottom row of Fig.~\ref{fig:qu}, we show the 30~GHz model PI, $Q$, and $U$ for $\beta=30^\circ$. Here we can see visually that the brightness gradient along the NPS for this model has a better correspondence between model and data. 

In this case, the magnetic field is then not entirely parallel to the filament. This could be produced if the filaments consist of a bundle of finer filaments with a twisted or helical structure \citep[see Fig.~3 of][]{2019A&A...621A..97B}; a $B_z$ component is a
natural result of a twisted filament. The best values of $B_z$ depend on the $z$ position of the filament, i.e., $B_z>0$ for a filament above the plane, $B_z<0$ for a filament below the plane and $B_z\approx0$ for a filament in the plane, such as the Fan Region. 

Alternatively, the filament may also be tilted with respect to the Galactic plane. This is a reasonable scenario to assume if the filaments are curved in 3D around a very large and extended bubble.  

Even though the case where $\beta=30^\circ$ shows a better correspondence for the brightness along the NPS, we choose to discuss the $\beta=0^\circ$ case for the remainder of this paper. As discussed above, the gradient may arise from several different effects, such as a change in filament orientation, or possibly from an increase in density along this line of sight. We note that Fig.~\ref{fig:qu} shows that the pattern of sign changes in the Stokes $Q$ and $U$ is quite similar in both cases. 

\subsection{\label{sec:faradayrotation}Faraday Rotation}

In addition to the synchrotron intensity, we add a constant thermal electron density filling the volume of our model, and calculate Faraday rotation. This has no impact on the synchrotron brightness, but only on the FD. We can compute models at a variety of frequencies, and we include the contribution of Faraday rotation that is internally generated within the filament. 

The volume averaged thermal electron density at the Galactic midplane is $n_{e}\approx0.015\pm0.001~\text{cm}^{-3}$ \citep{2020ApJ...897..124O}. This value is higher in certain locations, such as star-forming regions, with H$\alpha$ being a good tracer of thermal electron content. We examine the all-sky H$\alpha$ map of \citet{2003ApJS..146..407F}, and find no obvious visible counterpart to either the NPS or the Fan Region. There is some bright H$\alpha$ from the Perseus Arm, but it is offset to lower Galactic longitude. Thus, there is no evidence that the NPS or the Fan Region has a higher thermal electron content than the Galactic average, and we assume a value of $n_{e,i}=0.015~\text{cm}^{-3}$.

As discussed in the previous section, we assume a magnetic field strength of $24~\mu$G, but note that the bulk of this field is in the plane-of-sky, rather than along the line-of-sight. These values give FDs of a few rad~m$^2$, which is consistent with observations \citep[][]{1984A&A...135..238S,2015ApJ...811...40S,2021AJ....162...35W}.

\begin{figure}[!ht]
\centering 
\begin{minipage}{8.5cm}
\includegraphics[width=8.5cm]{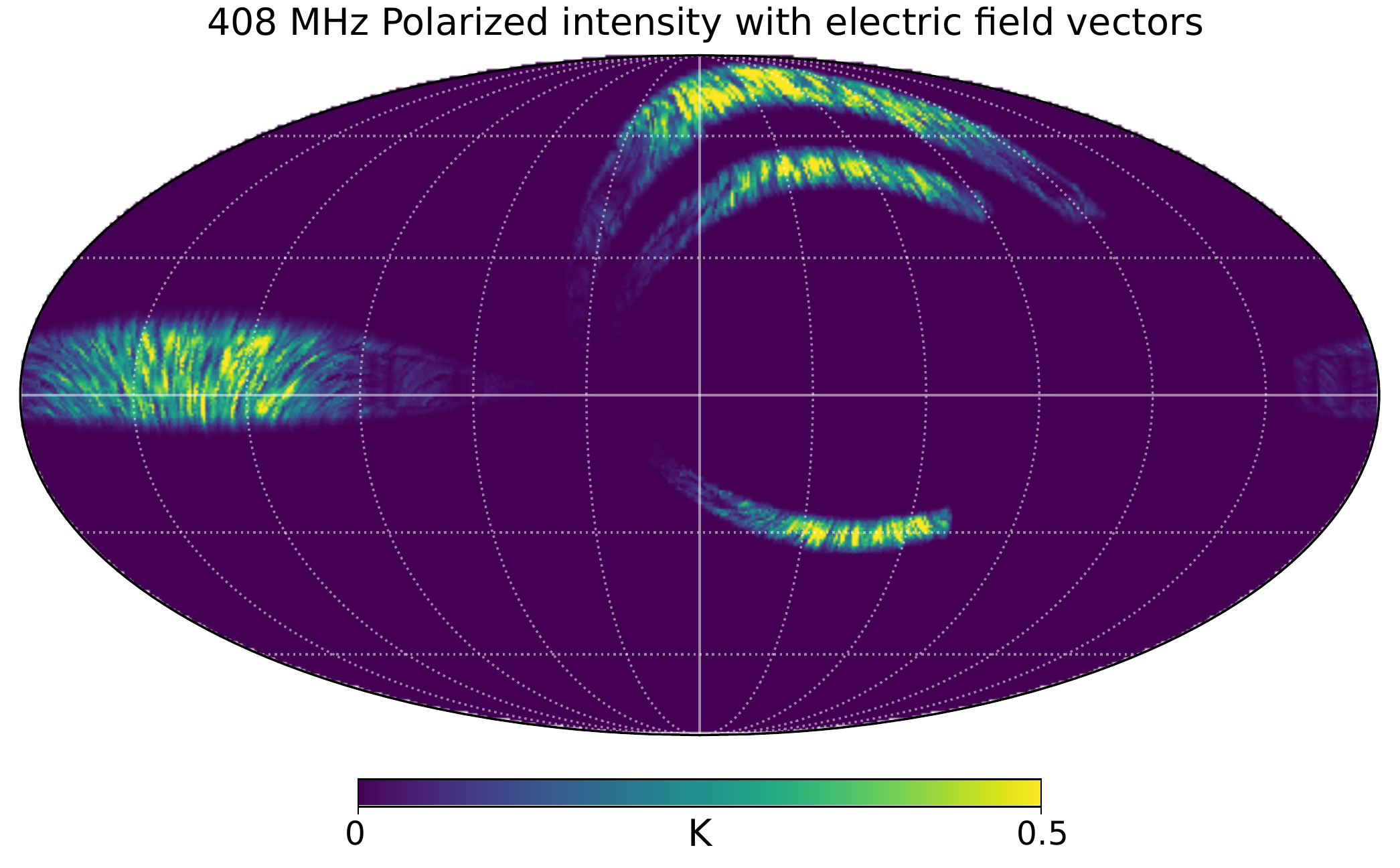}
\end{minipage}

\begin{scriptsize}\caption{\label{fig:vectors}Line integral convolution map of the electric field vectors from the 408~MHz model illustrating the characteristic fan shape of the Fan Region. }
\end{scriptsize} 
\end{figure}

In Fig.~\ref{fig:vectors}, we show the Faraday rotated electric field vectors for the 408~MHz model, using the case where $\beta=0^\circ$. Here we see the characteristic fan shape, which gives the Fan Region its name, and which is consistent with the observations of \citet{1976A&AS...26..129B} (see their Fig.~6a). 

Fig.~6a of \citet{1976A&AS...26..129B} also includes a map that shows the NPS, but it is considerably more difficult to identify and compare the NPS using that map, so we use other data for this comparison. We use the model Stokes~$Q$ and $U$ to compute maps of the model polarization angles, $\chi$, and also difference maps between the models (M) and the data (D), $\Delta\chi$ similar to that described in Sec.~\ref{sec:data}:

\begin{equation}
\begin{aligned}
\Delta\chi_{1.4~\text{GHz}} = 90^\circ - | 90^\circ - |\chi_{\text{M}_\text{1.4~GHz} } - \chi_{\text{D}_\text{1.4~GHz} }||
\\
\Delta\chi_{30~\text{GHz}} = 90^\circ - | 90^\circ - |\chi_{\text{M}_\text{30~GHz} } - \chi_{\text{D}_\text{30~GHz} }||.
\end{aligned}
\end{equation}

\begin{figure*}[!ht]
\centering 
\begin{minipage}{8.5cm}
\includegraphics[width=8.4cm]{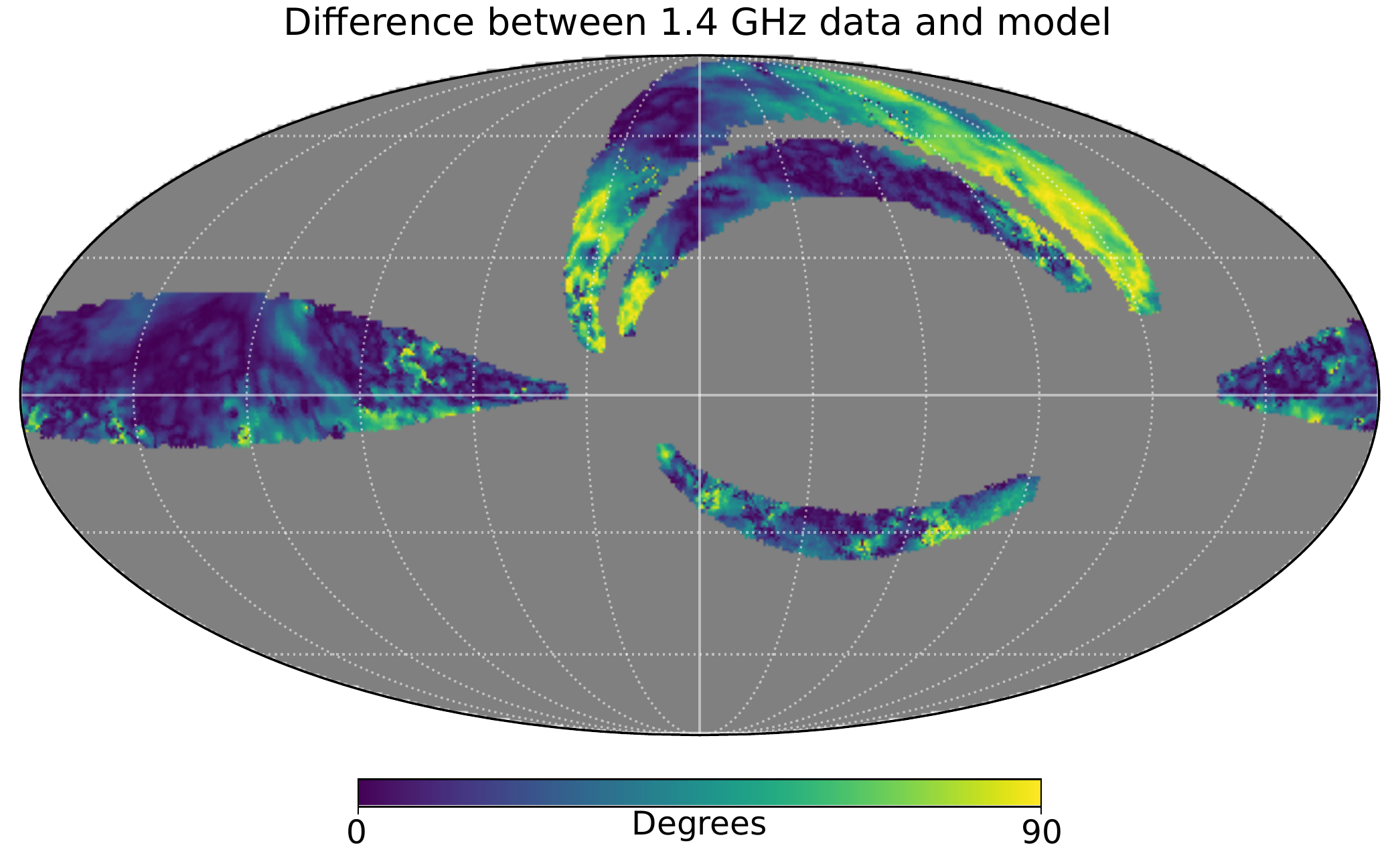}
\end{minipage}
\hfill
\begin{minipage}{8.5cm}
\includegraphics[width=8.4cm]{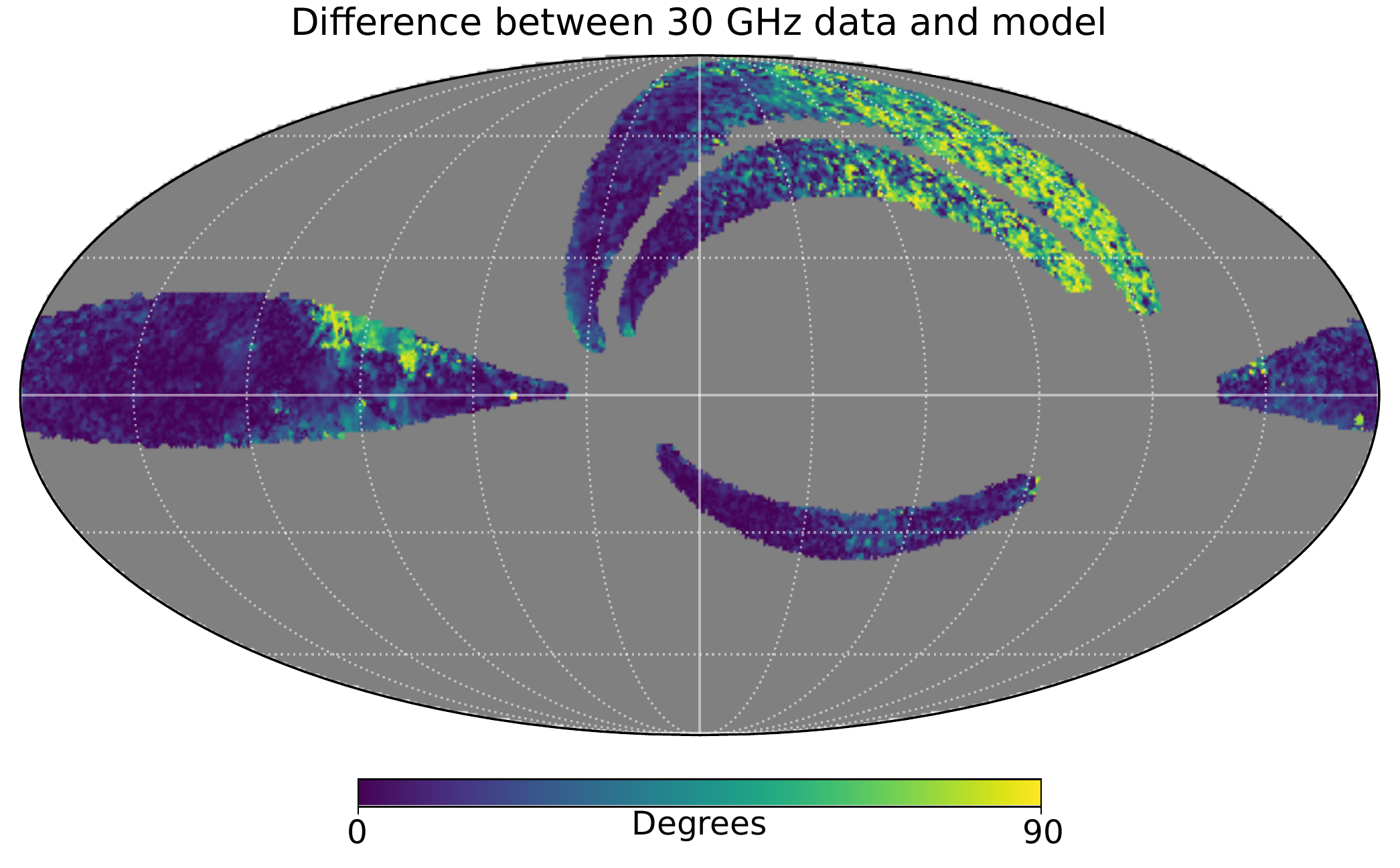}
\end{minipage}
\caption{ \label{fig:datamodeldiff} Left: Absolute difference between the polarization angles in the data and the model at 1.4~GHz. Right: Absolute difference between the polarization angles in the data and the model at 30~GHz. }
\end{figure*}

These difference maps are shown in Fig.~\ref{fig:datamodeldiff}. These maps have been masked to show just the regions where the model filaments are located. These show that there is good agreement in the orientation of the angles between the model and the data over some portions of the NPS, but there are other sections where the agreement is poor. This is discussed further in the next section. 

\begin{figure}[!ht]
\centering 
\includegraphics[width=8.4cm]{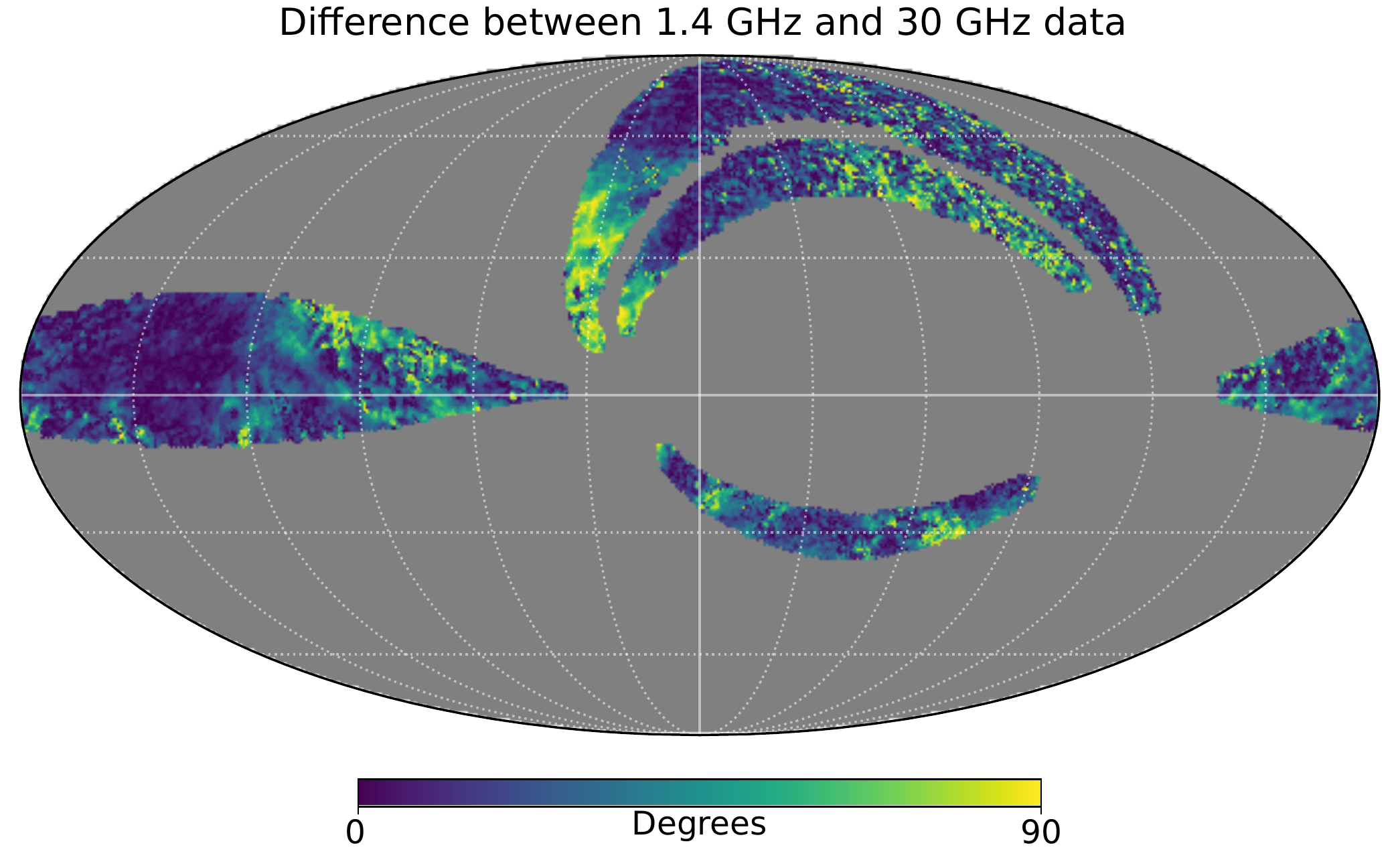}
\caption{ \label{fig:datadiffmasked} Difference in polarization angles between 1.4~GHz and 30~GHz data.  }
\end{figure}

\section{\label{sec:discussion}Discussion}

We present a model that is able to achieve broad agreement with a wide variety of observational properties including the overall morphology, polarization structure (i.e., magnetic field orientation), and brightness gradients. The agreement is particularly remarkable given the simplicity of the model.  However there are some notable discrepancies.

In Fig.~\ref{fig:datamodeldiff}, there are several yellow regions, which indicate large disagreement between the model and the data. These regions are located at the ends of the filaments, particularly in the case of the comparison of the 1.4~GHz model with the data, while the centre portions seem to have good agreement. The 30~GHz data and the 30~GHz model agree remarkably well, except for the right-hand side of the NPS and IX filaments. 

The differences in agreement we see between the low and high frequency maps are consistent with a foreground depolarization structure. In our model, this part of the NPS is also behind the Aquila-Ser dust cloud \citep[see Fig.~6, and also][]{2014A&A...561A..91L}, which could be the source of electrons causing the depolarization, and explains why the model does not agree with the data here. 

To highlight this further, Fig.~\ref{fig:datadiffmasked} shows the same difference map between the 1.4~GHz and 30~GHz data as in Fig.~\ref{fig:datadiff} but with the same projection and masking as in Fig.~\ref{fig:datamodeldiff}. Comparing Figs.~\ref{fig:datamodeldiff} and \ref{fig:datadiffmasked} we see similar regions of disagreement. This suggests that the model, which only includes internal Faraday rotation along the filament, is lacking the contribution of this foreground depolarization structure, which could be a region of increased thermal electrons, an increased line-of-sight magnetic field component towards this direction, or both. 

Still, the difference between the 30~GHz data and the 30~GHz model on the right-hand side of the NPS and IX filaments most likely represents some intrinsic difference between reality and the simple field of the model that is tangential to the filament. It should be noted that despite the apparently large extent on the sky of this region of disagreement, it is actually a very small fraction of the filament itself, which is only the small tip located entirely to the left of the left hand red hash mark shown in Fig.~\ref{fig:local-galaxy} (i.e., $\phi<\sim312^\circ$).

\subsection{\label{sec:distance}Distance}

Our model is able to resolve the apparent contradictions in the distance to the NPS outlined in Sec.~\ref{sec:nps}, which suggest the high latitude portion is nearby, while lower latitude portions are quite distant. In our model, the brightest end (i.e., the part observed nearest to the plane) of a long filament can be quite distant ($>300$~pc). Therefore the end that suffers the depolarization and X-ray absorption can still be located behind the Aquila Rift while the centre portion is still nearby ($\sim100$~pc) (see Fig.~\ref{fig:local-galaxy}). 

The distance to the Fan Region is more difficult to determine. \citet{2017MNRAS.467.4631H} conclude that at least 30\% of the brightest emission originates in or beyond the W4 superbubble, which is $\approx 2$~kpc away. Existing models of the mean Galactic field \citep[e.g.,][]{Collaboration:2016eh} show that the field towards $l\sim135^\circ$ is largely oriented perpendicular to the line-of-sight, all the way through to the edge of the Galaxy, which is the orientation that gives maximum synchrotron intensity. Yet, these models fall short of reproducing the total emission in this direction. Since the Fan Region is near the plane, we certainly must be observing a superposition of emission over a range of distances.  Moreover, the 40\% polarization fraction in the Fan Region and the 70\% maximal polarization fraction of synchroton emission \citep{1965ARA&A...3..297G} suggest that the polarized emission originates in $\gtrsim 4/7$ of the volume that produces the total intensity emission in this direction \citep{1967MNRAS.136..347B,2017MNRAS.467.4631H}. Therefore, the source of the polarized emission must be reasonably mixed along the line of sight. We suggest that this proposed nearby filament contributes a significant fraction of this total emission. The enhancement relative to total intensity is largely due to the field orientation perpendicular to the line of sight. The small FD values of a few rad/m$^2$ observed by \citet{1984A&A...135..238S} and also \citet[][]{2015ApJ...811...40S, 2021AJ....162...35W} suggest that it is nearby.

\subsection{\label{sec:origin}Origin}

These filaments must have either a compressed magnetic field and/or a localized enhancement of CRE density (or both) that can produce the observed synchrotron radiation. \citet{1998LNP...506..229H} has suggested that density enhancements from expanding HI shells are able to trap and accelerate relativistic electrons. HI observations show that the NPS has an associated HI shell \citep{1970Natur.225..364B}, which could thus be a source of the CREs necessary for the observed X-rays and Gamma rays. These may be relics of supernovae (SNe) in the Local Bubble and/or aligned, on average, with an arm of the Galaxy. 

The mean magnetic field of spiral galaxies, including the Milky Way, follows the large-scale spiral structure \citep{2015A&ARv..24....4B}. SNR shells compress the mean field, and are aligned with it \citep{1998ApJ...493..781G, 2016A&A...587A.148W}. As remnants age and dissipate back into the interstellar medium, relics of these shells will become elongated along field lines, and may leave filaments aligned with the mean field. Strong filamentary structure of a very old (and nearby) SNR could possibly remain in evidence as long as $\sim10^6$~years \citep[e.g.,][]{2002ApJ...576L..41M}, though this is up to a factor of 10 longer than generally accepted \citep{2012SSRv..166..231R}. 

The Local Bubble is filled with a soft X-ray emitting gas and surrounded by cold neutral gas and dust \citep{2020A&A...636A..17P}. Through an analysis of the total B-type stellar population in the solar neighborhood, \citet{2006MNRAS.373..993F} estimate that approximately 14–20 SNe have exploded to create the Local Bubble and conclude that the energy has been sufficient to create this cavity. We speculate that we are seeing old fragments of these SNRs, and that multiple SNe explosions could contribute to the longevity of filaments by repeatedly shocking an already dense region. Filaments of a possibly similar nature are observed elsewhere in the Galaxy (e.g., West et al. in prep). There are also examples of simulations showing filamentary structures that are created in the context of Galactic fountains and superbubbles formed from OB associations with many ($\sim100$) SNe explosions  \citep[e.g.,][]{2008MNRAS.388..573M} and simulations showing that multiple SNe explosions can create superbubbles \citep{2017MNRAS.465.1720Y}. 

Further, \citet{2016Natur.532...73B} find the most recent of these SNe occurred in a direction towards $l=327^\circ$, $b=+11^\circ$. This is close to the centre of the concentric cylinders that we use, which is towards $l=315^\circ$. Based on the straight-line filaments, we constrained the tangent of the arc of our filaments to be at $45^\circ$. The case having the tangent perpendicular to $l=327^\circ$ would have the filaments oriented at $57^\circ$.

In Fig.~\ref{fig:appendixuniformfields} we can see there is a range of filament orientations spanning  $30^\circ<l_\alpha<80^\circ$, which would be reasonably consistent with the appearance of the Fan Region. Thus, the orientation of the Fan Region filament could also be consistent with the orientation of the structure found by \citet{2020Natur.578..237A}, which is $l_\alpha\approx60^\circ$. In Fig.~\ref{fig:local-galaxy} we show a filament with this orientation. We suggest that the Fan Region could be the remains of a shock at the wall of the Local Bubble, which plowed into the interstellar medium and triggered star formation along the boundary.

\subsection{\label{sec:chminey}Association with a Galactic chimney or outflow}

\citet{2009Ap&SS.323....1W} present a model of the Local Bubble as a local cavity with a relatively open top, i.e., a Galactic chimney. There are many other examples of superbubbles in the Galaxy \citep[e.g.,][]{1979ApJ...229..533H}. We also observe outflows and examples of superbubbles in other galaxies such as those presented by \citet{2020A&A...639A.111S}. 

\citet{2007ApJ...656..914W} present an example of what is thought to be a fragmenting superbubble. In this case there are HI and also synchrotron emitting filaments that, at least in projection, are stretching through the centre of the bubble. These are located both near the Galactic plane, and also at a height of $\sim$100~pc. It seems plausible that the filamentary structure we propose for the Fan Region, NPS, and other loops are local examples of such fragments in a structure that could be analogous to the W4 chimney.

\subsection{\label{sec:multiwavelength}Multi-wavelength picture}

This study has focused on explaining the radio observations, and the properties those data reveal. However, a complete model must explain the full spectrum of observations. One mystery that remains is that the NPS has associated X-ray emission, where there is no obvious X-ray counterpart to the Fan Region. One possibility to explain this difference is that in the open-top, chimney picture of the Local Bubble, the Fan and NPS exist in very different environments. The Fan Region environment is much more dense, being in the Galactic mid-plane, and located adjacent to dense molecular clouds, whereas the NPS, which is located across the top of the chimney, would be a much less dense environment. In this picture, the NPS would cool more slowly, since the cooling rate is enhanced in higher density environments. Thus it is possible that the Fan Region has cooled to the point that there is no longer X-ray emission, whereas the NPS is still hot enough to emit.

\section{\label{sec:conclusions}Conclusions}

We present a simple model that, for the first time, includes both the Fan Region and the North Polar Spur (NPS). This very simple model suggests that the NPS and the Fan Region, along with other loops, are long magnetized filamentary structures that surround the top, bottom, and left side (as viewed from the North Galactic pole) of the Local Bubble (or local cavity). We find a particular set of parallel curved filaments with a tangent point oriented at $45^\circ$ that can reproduce the orientation and morphology of the NPS, in addition to the Fan Region, and two other prominent loops, IX and Is. The observed brightness gradient in the NPS can be reproduced reasonably well when the filament is tilted by $30^\circ$. The brightness distribution in the Fan Region does not have such a gradient and is reasonably reproduced when the filament is parallel to the plane (or tilted by a small amount,  $\pm10^\circ$).  We show that this model is consistent with a large number of observational studies on these regions.

This model has implications for developing a holistic model of magnetic fields in galaxies. We still do not fully understand the origin and evolution of regular magnetic fields in galaxies and how this field is maintained. In this picture, where long-lived filaments are elongated along the field lines, they could be one source where electrons are trapped, and contribute to maintaining such a regular field \citep{1998LNP...506..229H}.

In addition, understanding the full nature of this foreground structure provides us with context to understand similar structures that are increasingly being revealed with new observations, e.g., large filaments, superbubbles, and other large bubble-like structures, that are observed in more distant parts of the Milky Way \citep[e.g.,][]{1996Natur.380..687N} and other galaxies \citep[e.g.,][]{2020A&A...639A.111S,2012ApJ...754L..35H, 2008A&A...490..555B}. We assume that the filaments of the NPS in the 30~GHz data are about ~20 pc wide. That same filament in the Perseus arm, for example, would have an apparent thickness of $\sim30'$ and would be located within a few degrees of the Galactic plane. Therefore, it would be superimposed on all of the other Galactic plane emission and much more difficult to discern. Thus, it is likely that we do not currently have the resolution and sensitivity to see this level of structure in many locations except the local environment, and possibly in the Perseus arm. We note that, on careful examination of the data, it is a simplification to say that these features have a constant width of $6^\circ$. We predict that higher resolution observations of this region will confirm a much more complex filamentary structure. We also predict that more sensitive observations in our and other galaxies will increasingly reveal synchrotron filaments, particularly associated with HI filamentary structure and associated with other HI shells and bubbles.

Since we observe the NPS, Fan Region, and other filaments superimposed on the rest of the Milky Way's emission, these components should be modelled together. We plan to do a follow-up study to model these filaments in conjunction with a Galactic field model using a much more robust parameter optimization using the Interstellar MAGnetic field INference Engine (IMAGINE) \citep{2019Galax...7...17H}. 

\section*{Acknowledgments}
The Dunlap Institute is funded through an endowment established by the David Dunlap family and the University of Toronto. J.L.W. and B.M.G. acknowledge the support of the Natural Sciences and Engineering Research Council of Canada (NSERC) through grant RGPIN-2015-05948, and of the Canada Research Chairs program. We thank L. Rudnick for providing data, and K. Ferri\`{e}re, C. Heiles, J. Hunt, I. Grenier, and T. Vernstrom for many useful discussions. We thank the anonymous referee for their valuable comments that resulted in a much improved manuscript. This research has made use of the NASA Astrophysics Data System (ADS). This research made use of Astropy,\footnote{http://www.astropy.org} a community-developed core Python package for Astronomy. Some of the results in this paper have been derived using the HEALPix package.
We acknowledge the use of data provided by the Centre d'Analyse de Données Etendues (CADE), a service of IRAP-UPS/CNRS (http://cade.irap.omp.eu, Paradis et al., 2012, A\&A, 543, 103).

\bibliography{references}{}

\begin{thebibliography}{}
\expandafter\ifx\csname natexlab\endcsname\relax\def\natexlab#1{#1}\fi
\providecommand{\url}[1]{\href{#1}{#1}}
\providecommand{\dodoi}[1]{doi:~\href{http://doi.org/#1}{\nolinkurl{#1}}}
\providecommand{\doeprint}[1]{\href{http://ascl.net/#1}{\nolinkurl{http://ascl.net/#1}}}
\providecommand{\doarXiv}[1]{\href{https://arxiv.org/abs/#1}{\nolinkurl{https://arxiv.org/abs/#1}}}

\bibitem[{{Alves} {et~al.}(2020){Alves}, {Zucker}, {Goodman}, {Speagle},
  {Meingast}, {Robitaille}, {Finkbeiner}, {Schlafly}, \&
  {Green}}]{2020Natur.578..237A}
{Alves}, J., {Zucker}, C., {Goodman}, A.~A., {et~al.} 2020, \nat, 578, 237,
  \dodoi{10.1038/s41586-019-1874-z}

\bibitem[{{Alves} {et~al.}(2018){Alves}, {Boulanger}, {Ferri{\`e}re}, \&
  {Montier}}]{2018A&A...611L...5A}
{Alves}, M.~I.~R., {Boulanger}, F., {Ferri{\`e}re}, K., \& {Montier}, L. 2018,
  \aap, 611, L5, \dodoi{10.1051/0004-6361/201832637}

\bibitem[{{Beck}(2015)}]{2015A&ARv..24....4B}
{Beck}, R. 2015, \aapr, 24, 4, \dodoi{10.1007/s00159-015-0084-4}

\bibitem[{{Bennett}(1962)}]{1962MmRAS..68..163B}
{Bennett}, A.~S. 1962, \memras, 68, 163

\bibitem[{{Berdyugin} {et~al.}(2014){Berdyugin}, {Piirola}, \&
  {Teerikorpi}}]{2014A&A...561A..24B}
{Berdyugin}, A., {Piirola}, V., \& {Teerikorpi}, P. 2014, \aap, 561, A24,
  \dodoi{10.1051/0004-6361/201322604}

\bibitem[{{Berkhuijsen} {et~al.}(1964){Berkhuijsen}, {Brouw}, {Muller}, \&
  {Tinbergen}}]{1964BAN....17..465B}
{Berkhuijsen}, E.~M., {Brouw}, W.~N., {Muller}, C.~A., \& {Tinbergen}, J. 1964,
  \bain, 17, 465

\bibitem[{{Berkhuijsen} {et~al.}(1970){Berkhuijsen}, {Haslam}, \&
  {Salter}}]{1970Natur.225..364B}
{Berkhuijsen}, E.~M., {Haslam}, C.~G.~T., \& {Salter}, C.~J. 1970, \nat, 225,
  364, \dodoi{10.1038/225364a0}

\bibitem[{{Berkhuijsen} {et~al.}(1971){Berkhuijsen}, {Haslam}, \&
  {Salter}}]{1971A&A....14..252B}
---. 1971, \aap, 14, 252

\bibitem[{{Bingham} \& {Shakeshaft}(1967)}]{1967MNRAS.136..347B}
{Bingham}, R.~G., \& {Shakeshaft}, J.~R. 1967, \mnras, 136, 347,
  \dodoi{10.1093/mnras/136.4.347}

\bibitem[{{Bland-Hawthorn} \& {Cohen}(2003)}]{2003ApJ...582..246B}
{Bland-Hawthorn}, J., \& {Cohen}, M. 2003, \apj, 582, 246,
  \dodoi{10.1086/344573}

\bibitem[{{Boomsma} {et~al.}(2008){Boomsma}, {Oosterloo}, {Fraternali}, {van
  der Hulst}, \& {Sancisi}}]{2008A&A...490..555B}
{Boomsma}, R., {Oosterloo}, T.~A., {Fraternali}, F., {van der Hulst}, J.~M., \&
  {Sancisi}, R. 2008, \aap, 490, 555, \dodoi{10.1051/0004-6361:200810120}

\bibitem[{{Bracco} {et~al.}(2019){Bracco}, {Candelaresi}, {Del Sordo}, \&
  {Brandenburg}}]{2019A&A...621A..97B}
{Bracco}, A., {Candelaresi}, S., {Del Sordo}, F., \& {Brandenburg}, A. 2019,
  \aap, 621, A97, \dodoi{10.1051/0004-6361/201833961}

\bibitem[{{Breitschwerdt} {et~al.}(2016){Breitschwerdt}, {Feige}, {Schulreich},
  {Avillez}, {Dettbarn}, \& {Fuchs}}]{2016Natur.532...73B}
{Breitschwerdt}, D., {Feige}, J., {Schulreich}, M.~M., {et~al.} 2016, \nat,
  532, 73, \dodoi{10.1038/nature17424}

\bibitem[{{Brouw} \& {Spoelstra}(1976)}]{1976A&AS...26..129B}
{Brouw}, W.~N., \& {Spoelstra}, T.~A.~T. 1976, \aaps, 26, 129

\bibitem[{{Bunner} {et~al.}(1972){Bunner}, {Coleman}, {Kraushaar}, \&
  {McCammon}}]{1972ApJ...172L..67B}
{Bunner}, A.~N., {Coleman}, P.~L., {Kraushaar}, W.~L., \& {McCammon}, D. 1972,
  \apjl, 172, L67, \dodoi{10.1086/180893}

\bibitem[{{Casandjian} {et~al.}(2009){Casandjian}, {Grenier}, \& {the Fermi
  Large Area Telescope Collaboration}}]{2009arXiv0912.3478C}
{Casandjian}, J.-M., {Grenier}, I., \& {the Fermi Large Area Telescope
  Collaboration}. 2009, arXiv e-prints.
\newblock \doarXiv{0912.3478}

\bibitem[{{Clark}(2018)}]{2018ApJ...857L..10C}
{Clark}, S.~E. 2018, \apjl, 857, L10, \dodoi{10.3847/2041-8213/aabb54}

\bibitem[{{Clark} \& {Hensley}(2019)}]{2019ApJ...887..136C}
{Clark}, S.~E., \& {Hensley}, B.~S. 2019, \apj, 887, 136,
  \dodoi{10.3847/1538-4357/ab5803}

\bibitem[{{Das} {et~al.}(2020){Das}, {Zucker}, {Speagle}, {Goodman}, {Green},
  \& {Alves}}]{2020MNRAS.498.5863D}
{Das}, K.~K., {Zucker}, C., {Speagle}, J.~S., {et~al.} 2020, \mnras, 498, 5863,
  \dodoi{10.1093/mnras/staa2702}

\bibitem[{{Dickinson} {et~al.}(2019){Dickinson}, {Barr}, {Chiang}, {Copley},
  {Grumitt}, {Harper}, {Heilgendorff}, {Jew}, {Jonas}, {Jones}, {Leahy},
  {Leech}, {Leitch}, {Muchovej}, {Pearson}, {Peel}, {Readhead}, {Sievers},
  {Stevenson}, \& {Taylor}}]{2019MNRAS.485.2844D}
{Dickinson}, C., {Barr}, A., {Chiang}, H.~C., {et~al.} 2019, \mnras, 485, 2844,
  \dodoi{10.1093/mnras/stz522}

\bibitem[{{Egger} \& {Aschenbach}(1995)}]{1995A&A...294L..25E}
{Egger}, R.~J., \& {Aschenbach}, B. 1995, \aap, 294, L25

\bibitem[{{Fesen} {et~al.}(2015){Fesen}, {Neustadt}, {Black}, \&
  {Koeppel}}]{2015ApJ...812...37F}
{Fesen}, R.~A., {Neustadt}, J.~M.~M., {Black}, C.~S., \& {Koeppel}, A.~H.~D.
  2015, \apj, 812, 37, \dodoi{10.1088/0004-637X/812/1/37}

\bibitem[{{Finkbeiner}(2003)}]{2003ApJS..146..407F}
{Finkbeiner}, D.~P. 2003, \apjs, 146, 407, \dodoi{10.1086/374411}

\bibitem[{{Fuchs} {et~al.}(2006){Fuchs}, {Breitschwerdt}, {de Avillez},
  {Dettbarn}, \& {Flynn}}]{2006MNRAS.373..993F}
{Fuchs}, B., {Breitschwerdt}, D., {de Avillez}, M.~A., {Dettbarn}, C., \&
  {Flynn}, C. 2006, \mnras, 373, 993, \dodoi{10.1111/j.1365-2966.2006.11044.x}

\bibitem[{Gaensler(1998)}]{1998ApJ...493..781G}
Gaensler, B.~M. 1998, \apj, 493, 1

\bibitem[{{Ginzburg} \& {Syrovatskii}(1965)}]{1965ARA&A...3..297G}
{Ginzburg}, V.~L., \& {Syrovatskii}, S.~I. 1965, \araa, 3, 297,
  \dodoi{10.1146/annurev.aa.03.090165.001501}

\bibitem[{{Goodman} {et~al.}(2014){Goodman}, {Alves}, {Beaumont}, {Benjamin},
  {Borkin}, {Burkert}, {Dame}, {Jackson}, {Kauffmann}, {Robitaille}, \&
  {Smith}}]{2014ApJ...797...53G}
{Goodman}, A.~A., {Alves}, J., {Beaumont}, C.~N., {et~al.} 2014, \apj, 797, 53,
  \dodoi{10.1088/0004-637X/797/1/53}

\bibitem[{{G{\'o}rski} {et~al.}(2019){G{\'o}rski}, {Banday}, {Lawrence},
  {Gaier}, {Jaffe}, \& {Rocha}}]{2019BAAS...51g.188G}
{G{\'o}rski}, K., {Banday}, A.~J., {Lawrence}, C.~R., {et~al.} 2019, in \baas,
  Vol.~51, 188

\bibitem[{{Green}(2019)}]{2019JApA...40...36G}
{Green}, D.~A. 2019, JApA, 40, 36, \dodoi{10.1007/s12036-019-9601-6}

\bibitem[{{Hanbury Brown} {et~al.}(1960){Hanbury Brown}, {Davies}, \&
  {Hazard}}]{1960Obs....80..191H}
{Hanbury Brown}, R., {Davies}, R.~D., \& {Hazard}, C. 1960, Obs, 80, 191

\bibitem[{{Haslam} {et~al.}(1971){Haslam}, {Kahn}, \&
  {Meaburn}}]{1971A&A....12..388H}
{Haslam}, C.~G.~T., {Kahn}, F.~D., \& {Meaburn}, J. 1971, \aap, 12, 388

\bibitem[{{Haslam} {et~al.}(1982){Haslam}, {Salter}, {Stoffel}, \&
  {Wilson}}]{1982A&AS...47....1H}
{Haslam}, C.~G.~T., {Salter}, C.~J., {Stoffel}, H., \& {Wilson}, W.~E. 1982,
  \aaps, 47, 1

\bibitem[{{Haverkorn} {et~al.}(2019){Haverkorn}, {Boulanger}, {En{\ss}lin},
  {H{\"o}randel}, {Jaffe}, {Jasche}, {Rachen}, \&
  {Shukurov}}]{2019Galax...7...17H}
{Haverkorn}, M., {Boulanger}, F., {En{\ss}lin}, T., {et~al.} 2019, Galax, 7,
  17, \dodoi{10.3390/galaxies7010017}

\bibitem[{{Heald}(2012)}]{2012ApJ...754L..35H}
{Heald}, G.~H. 2012, \apjl, 754, L35, \dodoi{10.1088/2041-8205/754/2/L35}

\bibitem[{{Heiles}(1979)}]{1979ApJ...229..533H}
{Heiles}, C. 1979, \apj, 229, 533, \dodoi{10.1086/156986}

\bibitem[{{Heiles}(1996)}]{1996ApJ...462..316H}
---. 1996, \apj, 462, 316, \dodoi{10.1086/177153}

\bibitem[{{Heiles}(1998)}]{1998LNP...506..229H}
{Heiles}, C. 1998, in Lecture Notes in Physics, Berlin Springer Verlag, Vol.
  506, IAU Colloq. 166: The Local Bubble and Beyond, ed. D.~{Breitschwerdt},
  M.~J. {Freyberg}, \& J.~{Truemper}, 229--238

\bibitem[{{Hill} {et~al.}(2017){Hill}, {Landecker}, {Carretti}, {Douglas},
  {Sun}, {Gaensler}, {Mao}, {McClure-Griffiths}, {Reich}, {Wolleben}, {Dickey},
  {Gray}, {Haverkorn}, {Leahy}, \& {Schnitzeler}}]{2017MNRAS.467.4631H}
{Hill}, A.~S., {Landecker}, T.~L., {Carretti}, E., {et~al.} 2017, \mnras, 467,
  4631, \dodoi{10.1093/mnras/stx389}

\bibitem[{{Hutschenreuter} \& {En{\ss}lin}(2020)}]{2020A&A...633A.150H}
{Hutschenreuter}, S., \& {En{\ss}lin}, T.~A. 2020, \aap, 633, A150,
  \dodoi{10.1051/0004-6361/201935479}

\bibitem[{{Hutschenreuter} {et~al.}(2021){Hutschenreuter}, {Anderson}, {Betti},
  {Bower}, {Brown}, {Br{\"u}ggen}, {Carretti}, {Clarke}, {Clegg}, {Costa},
  {Croft}, {Van Eck}, {Gaensler}, {de Gasperin}, {Haverkorn}, {Heald}, {Hull},
  {Inoue}, {Johnston-Hollitt}, {Kaczmarek}, {Law}, {Ma}, {MacMahon}, {Mao},
  {Riseley}, {Roy}, {Shanahan}, {Shimwell}, {Stil}, {Sobey}, {O'Sullivan},
  {Tasse}, {Vacca}, {Vernstrom}, {Williams}, {Wright}, \&
  {En{\ss}lin}}]{2021arXiv210201709H}
{Hutschenreuter}, S., {Anderson}, C.~S., {Betti}, S., {et~al.} 2021, arXiv
  e-prints, arXiv:2102.01709.
\newblock \doarXiv{2102.01709}

\bibitem[{{Iacobelli} {et~al.}(2013){Iacobelli}, {Haverkorn}, \&
  {Katgert}}]{2013A&A...549A..56I}
{Iacobelli}, M., {Haverkorn}, M., \& {Katgert}, P. 2013, \aap, 549, A56,
  \dodoi{10.1051/0004-6361/201220175}

\bibitem[{{Jaffe} {et~al.}(2010){Jaffe}, {Leahy}, {Banday}, {Leach}, {Lowe}, \&
  {Wilkinson}}]{2010MNRAS.401.1013J}
{Jaffe}, T.~R., {Leahy}, J.~P., {Banday}, A.~J., {et~al.} 2010, \mnras, 401,
  1013, \dodoi{10.1111/j.1365-2966.2009.15745.x}

\bibitem[{{Jeli{\'c}} {et~al.}(2015){Jeli{\'c}}, {de Bruyn}, {Pandey},
  {Mevius}, {Haverkorn}, {Brentjens}, {Koopmans}, {Zaroubi}, {Abdalla}, {Asad},
  {Bus}, {Chapman}, {Ciardi}, {Fernandez}, {Ghosh}, {Harker}, {Iliev},
  {Jensen}, {Kazemi}, {Mellema}, {Offringa}, {Patil}, {Vedantham}, \&
  {Yatawatta}}]{2015A&A...583A.137J}
{Jeli{\'c}}, V., {de Bruyn}, A.~G., {Pandey}, V.~N., {et~al.} 2015, \aap, 583,
  A137, \dodoi{10.1051/0004-6361/201526638}

\bibitem[{{Kataoka} {et~al.}(2018){Kataoka}, {Sofue}, {Inoue}, {Akita},
  {Nakashima}, \& {Totani}}]{2018Galax...6...27K}
{Kataoka}, J., {Sofue}, Y., {Inoue}, Y., {et~al.} 2018, Galax, 6, 27,
  \dodoi{10.3390/galaxies6010027}

\bibitem[{{Lallement} {et~al.}(2016){Lallement}, {Snowden}, {Kuntz}, {Dame},
  {Koutroumpa}, {Grenier}, \& {Casandjian}}]{2016A&A...595A.131L}
{Lallement}, R., {Snowden}, S., {Kuntz}, K.~D., {et~al.} 2016, \aap, 595, A131,
  \dodoi{10.1051/0004-6361/201629453}

\bibitem[{{Lallement} {et~al.}(2014){Lallement}, {Vergely}, {Valette},
  {Puspitarini}, {Eyer}, \& {Casagrande}}]{2014A&A...561A..91L}
{Lallement}, R., {Vergely}, J.-L., {Valette}, B., {et~al.} 2014, \aap, 561,
  A91, \dodoi{10.1051/0004-6361/201322032}

\bibitem[{{Lallement} {et~al.}(2018){Lallement}, {Capitanio}, {Ruiz-Dern},
  {Danielski}, {Babusiaux}, {Vergely}, {Elyajouri}, {Arenou}, \&
  {Leclerc}}]{2018A&A...616A.132L}
{Lallement}, R., {Capitanio}, L., {Ruiz-Dern}, L., {et~al.} 2018, \aap, 616,
  A132, \dodoi{10.1051/0004-6361/201832832}

\bibitem[{{LaRocca} {et~al.}(2020){LaRocca}, {Kaaret}, {Kuntz}, {Hodges-Kluck},
  {Zajczyk}, {Bluem}, {Ringuette}, \& {Jahoda}}]{2020ApJ...904...54L}
{LaRocca}, D.~M., {Kaaret}, P., {Kuntz}, K.~D., {et~al.} 2020, \apj, 904, 54,
  \dodoi{10.3847/1538-4357/abbdfd}

\bibitem[{{Leike} \& {En{\ss}lin}(2019)}]{2019A&A...631A..32L}
{Leike}, R.~H., \& {En{\ss}lin}, T.~A. 2019, \aap, 631, A32,
  \dodoi{10.1051/0004-6361/201935093}

\bibitem[{{Leike} \& {En{\ss}lin}(2021)}]{2021A&A...647C...1L}
---. 2021, \aap, 647, C1, \dodoi{10.1051/0004-6361/201935093e}

\bibitem[{{Mathewson} \& {Milne}(1965)}]{1965AuJPh..18..635M}
{Mathewson}, D.~S., \& {Milne}, D.~K. 1965, \aujph, 18, 635,
  \dodoi{10.1071/PH650635}

\bibitem[{{McCullough} \& {Benjamin}(2001)}]{2001AJ....122.1500M}
{McCullough}, P.~R., \& {Benjamin}, R.~A. 2001, \aj, 122, 1500,
  \dodoi{10.1086/322097}

\bibitem[{{McCullough} {et~al.}(2002){McCullough}, {Fields}, \&
  {Pavlidou}}]{2002ApJ...576L..41M}
{McCullough}, P.~R., {Fields}, B.~D., \& {Pavlidou}, V. 2002, \apjl, 576, L41,
  \dodoi{10.1086/343100}

\bibitem[{{MeerKAT Collaboration}(2018)}]{2018MNSSA..77..102.}
{MeerKAT Collaboration}, . 2018, \mnssa, 77, 102

\bibitem[{{Melioli} {et~al.}(2008){Melioli}, {Brighenti}, {D'Ercole}, \& {de
  Gouveia Dal Pino}}]{2008MNRAS.388..573M}
{Melioli}, C., {Brighenti}, F., {D'Ercole}, A., \& {de Gouveia Dal Pino}, E.~M.
  2008, \mnras, 388, 573, \dodoi{10.1111/j.1365-2966.2008.13446.x}

\bibitem[{{Normandeau} {et~al.}(1996){Normandeau}, {Taylor}, \&
  {Dewdney}}]{1996Natur.380..687N}
{Normandeau}, M., {Taylor}, A.~R., \& {Dewdney}, P.~E. 1996, \nat, 380, 687,
  \dodoi{10.1038/380687a0}

\bibitem[{{Ocker} {et~al.}(2020){Ocker}, {Cordes}, \&
  {Chatterjee}}]{2020ApJ...897..124O}
{Ocker}, S.~K., {Cordes}, J.~M., \& {Chatterjee}, S. 2020, \apj, 897, 124,
  \dodoi{10.3847/1538-4357/ab98f9}

\bibitem[{{Oppermann} {et~al.}(2015){Oppermann}, {Junklewitz}, {Greiner},
  {En{\ss}lin}, {Akahori}, {Carretti}, {Gaensler}, {Goobar}, {Harvey-Smith},
  {Johnston-Hollitt}, {Pratley}, {Schnitzeler}, {Stil}, \&
  {Vacca}}]{2015A&A...575A.118O}
{Oppermann}, N., {Junklewitz}, H., {Greiner}, M., {et~al.} 2015, \aap, 575,
  A118, \dodoi{10.1051/0004-6361/201423995}

\bibitem[{{Panopoulou} {et~al.}(2021){Panopoulou}, {Dickinson}, {Readhead},
  {Pearson}, \& {Peel}}]{2021arXiv210614267P}
{Panopoulou}, G.~V., {Dickinson}, C., {Readhead}, A.~C.~S., {Pearson}, T.~J.,
  \& {Peel}, M.~W. 2021, arXiv e-prints, arXiv:2106.14267.
\newblock \doarXiv{2106.14267}

\bibitem[{{Paradis} {et~al.}(2012){Paradis}, {Dobashi}, {Shimoikura},
  {Kawamura}, {Onishi}, {Fukui}, \& {Bernard}}]{2012A&A...543A.103P}
{Paradis}, D., {Dobashi}, K., {Shimoikura}, T., {et~al.} 2012, \aap, 543, A103,
  \dodoi{10.1051/0004-6361/201118740}

\bibitem[{{Pelgrims} {et~al.}(2020){Pelgrims}, {Ferri{\`e}re}, {Boulanger},
  {Lallement}, \& {Montier}}]{2020A&A...636A..17P}
{Pelgrims}, V., {Ferri{\`e}re}, K., {Boulanger}, F., {Lallement}, R., \&
  {Montier}, L. 2020, \aap, 636, A17, \dodoi{10.1051/0004-6361/201937157}

\bibitem[{{Planck Collaboration} {et~al.}(2016{\natexlab{a}}){Planck
  Collaboration}, {Aghanim}, {Alves}, \& {et. al.}}]{planck-et-al-dust}
{Planck Collaboration}, {Aghanim}, N., {Alves}, M.~I.~R., \& {et. al.}
  2016{\natexlab{a}}, \aap, 596, A105, \dodoi{10.1051/0004-6361/201628636}

\bibitem[{{Planck Collaboration} {et~al.}(2015){Planck Collaboration}, {Ade},
  {Aghanim}, {Alina}, {Alves}, {Armitage-Caplan}, {Arnaud}, {Arzoumanian},
  {Ashdown}, {Atrio-Barandela}, \& et~al.}]{2015A&A...576A.104P}
{Planck Collaboration}, {Ade}, P.~A.~R., {Aghanim}, N., {et~al.} 2015, \aap,
  576, A104, \dodoi{10.1051/0004-6361/201424082}

\bibitem[{{Planck Collaboration} {et~al.}(2016{\natexlab{b}}){Planck
  Collaboration}, {Adam}, {Ade}, {Aghanim}, {Akrami}, {Alves}, {Arg{\"u}eso},
  {Arnaud}, {Arroja}, {Ashdown}, {Aumont}, {Baccigalupi}, {Ballardini},
  {Banday}, {Barreiro}, {Bartlett}, {Bartolo}, {Basak}, {Battaglia},
  {Battaner}, {Battye}, {Benabed}, {Beno{\^\i}t}, {Benoit-L{\'e}vy}, {Bernard},
  {Bersanelli}, {Bertincourt}, {Bielewicz}, {Bikmaev}, {Bock}, {B{\"o}hringer},
  {Bonaldi}, {Bonavera}, {Bond}, {Borrill}, {Bouchet}, {Boulanger}, {Bucher},
  {Burenin}, {Burigana}, {Butler}, {Calabrese}, {Cardoso}, {Carvalho},
  {Casaponsa}, {Castex}, {Catalano}, {Challinor}, {Chamballu}, {Chary},
  {Chiang}, {Chluba}, {Chon}, {Christensen}, {Church}, {Clemens}, {Clements},
  {Colombi}, {Colombo}, {Combet}, {Comis}, {Contreras}, {Couchot}, {Coulais},
  {Crill}, {Cruz}, {Curto}, {Cuttaia}, {Danese}, {Davies}, {Davis}, {de
  Bernardis}, {de Rosa}, {de Zotti}, {Delabrouille}, {Delouis}, {D{\'e}sert},
  {Di Valentino}, {Dickinson}, {Diego}, {Dolag}, {Dole}, {Donzelli},
  {Dor{\'e}}, {Douspis}, {Ducout}, {Dunkley}, {Dupac}, {Efstathiou},
  {Eisenhardt}, {Elsner}, {En{\ss}lin}, {Eriksen}, {Falgarone}, {Fantaye},
  {Farhang}, {Feeney}, {Fergusson}, {Fernandez-Cobos}, {Feroz}, {Finelli},
  {Florido}, {Forni}, {Frailis}, {Fraisse}, {Franceschet}, {Franceschi},
  {Frejsel}, {Frolov}, {Galeotta}, {Galli}, {Ganga}, {Gauthier},
  {G{\'e}nova-Santos}, {Gerbino}, {Ghosh}, {Giard}, {Giraud-H{\'e}raud},
  {Giusarma}, {Gjerl{\o}w}, {Gonz{\'a}lez-Nuevo}, {G{\'o}rski}, {Grainge},
  {Gratton}, {Gregorio}, {Gruppuso}, {Gudmundsson}, {Hamann}, {Handley},
  {Hansen}, {Hanson}, {Harrison}, {Heavens}, {Helou}, {Henrot-Versill{\'e}},
  {Hern{\'a}ndez-Monteagudo}, {Herranz}, {Hildebrandt}, {Hivon}, {Hobson},
  {Holmes}, {Hornstrup}, {Hovest}, {Huang}, {Huffenberger}, {Hurier},
  {Ili{\'c}}, {Jaffe}, {Jaffe}, {Jin}, {Jones}, {Juvela}, {Karakci},
  {Keih{\"a}nen}, {Keskitalo}, {Khamitov}, {Kiiveri}, {Kim}, {Kisner},
  {Kneissl}, {Knoche}, {Knox}, {Krachmalnicoff}, {Kunz}, {Kurki-Suonio},
  {Lacasa}, {Lagache}, {L{\"a}hteenm{\"a}ki}, {Lamarre}, {Langer}, {Lasenby},
  {Lattanzi}, {Lawrence}, {Le Jeune}, {Leahy}, {Lellouch}, {Leonardi},
  {Le{\'o}n-Tavares}, {Lesgourgues}, {Levrier}, {Lewis}, {Liguori}, {Lilje},
  {Lilley}, {Linden-V{\o}rnle}, {Lindholm}, {Liu}, {L{\'o}pez-Caniego},
  {Lubin}, {Ma}, {Mac{\'\i}as-P{\'e}rez}, {Maggio}, {Maino}, {Mak},
  {Mandolesi}, {Mangilli}, {Marchini}, {Marcos-Caballero}, {Marinucci},
  {Maris}, {Marshall}, {Martin}, {Martinelli}, {Mart{\'\i}nez-Gonz{\'a}lez},
  {Masi}, {Matarrese}, {Mazzotta}, {McEwen}, {McGehee}, {Mei}, {Meinhold},
  {Melchiorri}, {Melin}, {Mendes}, {Mennella}, {Migliaccio}, {Mikkelsen},
  {Millea}, {Mitra}, {Miville-Desch{\^e}nes}, {Molinari}, {Moneti}, {Montier},
  {Moreno}, {Morgante}, {Mortlock}, {Moss}, {Mottet}, {M{\"u}nchmeyer},
  {Munshi}, {Murphy}, {Narimani}, {Naselsky}, {Nastasi}, {Nati}, {Natoli},
  {Negrello}, {Netterfield}, {N{\o}rgaard-Nielsen}, {Noviello}, {Novikov},
  {Novikov}, {Olamaie}, {Oppermann}, {Orlando}, {Oxborrow}, {Paci}, {Pagano},
  {Pajot}, {Paladini}, {Pandolfi}, {Paoletti}, {Partridge}, {Pasian},
  {Patanchon}, {Pearson}, {Peel}, {Peiris}, {Pelkonen}, {Perdereau}, {Perotto},
  {Perrott}, {Perrotta}, {Pettorino}, {Piacentini}, {Piat}, {Pierpaoli},
  {Pietrobon}, {Plaszczynski}, {Pogosyan}, {Pointecouteau}, {Polenta}, {Popa},
  {Pratt}, {Pr{\'e}zeau}, {Prunet}, {Puget}, {Rachen}, {Racine}, {Reach},
  {Rebolo}, {Reinecke}, {Remazeilles}, {Renault}, {Renzi}, {Ristorcelli},
  {Rocha}, {Roman}, {Romelli}, {Rosset}, {Rossetti}, {Rotti}, {Roudier},
  {Rouill{\'e} d'Orfeuil}, {Rowan-Robinson}, {Rubi{\~n}o-Mart{\'\i}n},
  {Ruiz-Granados}, {Rumsey}, {Rusholme}, {Said}, {Salvatelli}, {Salvati},
  {Sandri}, {Sanghera}, {Santos}, {Saunders}, {Sauv{\'e}}, {Savelainen},
  {Savini}, {Schaefer}, {Schammel}, {Scott}, {Seiffert}, {Serra}, {Shellard},
  {Shimwell}, {Shiraishi}, {Smith}, {Souradeep}, {Spencer}, {Spinelli},
  {Stanford}, {Stern}, {Stolyarov}, {Stompor}, {Strong}, {Sudiwala}, {Sunyaev},
  {Sutter}, {Sutton}, {Suur-Uski}, {Sygnet}, {Tauber}, {Tavagnacco}, {Terenzi},
  {Texier}, {Toffolatti}, {Tomasi}, {Tornikoski}, {Tramonte}, {Tristram},
  {Troja}, {Trombetti}, {Tucci}, {Tuovinen}, {T{\"u}rler}, {Umana},
  {Valenziano}, {Valiviita}, {Van Tent}, {Vassallo}, {Vibert}, {Vidal}, {Viel},
  {Vielva}, {Villa}, {Wade}, {Walter}, {Wandelt}, {Watson}, {Wehus},
  {Welikala}, {Weller}, {White}, {White}, {Wilkinson}, {Yvon}, {Zacchei},
  {Zibin}, \& {Zonca}}]{2016A&A...594A...1P}
{Planck Collaboration}, {Adam}, R., {Ade}, P.~A.~R., {et~al.}
  2016{\natexlab{b}}, \aap, 594, A1, \dodoi{10.1051/0004-6361/201527101}

\bibitem[{{Planck Collaboration} {et~al.}(2016{\natexlab{c}}){Planck
  Collaboration}, {Adam}, {Ade}, {Aghanim}, {Alves}, {Arnaud}, {Ashdown},
  {Aumont}, {Baccigalupi}, {Banday}, {Barreiro}, {Bartlett}, {Bartolo},
  {Battaner}, {Benabed}, {Beno{\^\i}t}, {Benoit-L{\'e}vy}, {Bernard},
  {Bersanelli}, {Bielewicz}, {Bock}, {Bonaldi}, {Bonavera}, {Bond}, {Borrill},
  {Bouchet}, {Boulanger}, {Bucher}, {Burigana}, {Butler}, {Calabrese},
  {Cardoso}, {Catalano}, {Challinor}, {Chamballu}, {Chary}, {Chiang},
  {Christensen}, {Clements}, {Colombi}, {Colombo}, {Combet}, {Couchot},
  {Coulais}, {Crill}, {Curto}, {Cuttaia}, {Danese}, {Davies}, {Davis}, {de
  Bernardis}, {de Rosa}, {de Zotti}, {Delabrouille}, {D{\'e}sert}, {Dickinson},
  {Diego}, {Dole}, {Donzelli}, {Dor{\'e}}, {Douspis}, {Ducout}, {Dupac},
  {Efstathiou}, {Elsner}, {En{\ss}lin}, {Eriksen}, {Falgarone}, {Fergusson},
  {Finelli}, {Forni}, {Frailis}, {Fraisse}, {Franceschi}, {Frejsel},
  {Galeotta}, {Galli}, {Ganga}, {Ghosh}, {Giard}, {Giraud-H{\'e}raud},
  {Gjerl{\o}w}, {Gonz{\'a}lez-Nuevo}, {G{\'o}rski}, {Gratton}, {Gregorio},
  {Gruppuso}, {Gudmundsson}, {Hansen}, {Hanson}, {Harrison}, {Helou},
  {Henrot-Versill{\'e}}, {Hern{\'a}ndez-Monteagudo}, {Herranz}, {Hildebrandt},
  {Hivon}, {Hobson}, {Holmes}, {Hornstrup}, {Hovest}, {Huffenberger}, {Hurier},
  {Jaffe}, {Jaffe}, {Jones}, {Juvela}, {Keih{\"a}nen}, {Keskitalo}, {Kisner},
  {Kneissl}, {Knoche}, {Kunz}, {Kurki-Suonio}, {Lagache},
  {L{\"a}hteenm{\"a}ki}, {Lamarre}, {Lasenby}, {Lattanzi}, {Lawrence}, {Le
  Jeune}, {Leahy}, {Leonardi}, {Lesgourgues}, {Levrier}, {Liguori}, {Lilje},
  {Linden-V{\o}rnle}, {L{\'o}pez-Caniego}, {Lubin}, {Mac{\'\i}as-P{\'e}rez},
  {Maggio}, {Maino}, {Mandolesi}, {Mangilli}, {Maris}, {Marshall}, {Martin},
  {Mart{\'\i}nez-Gonz{\'a}lez}, {Masi}, {Matarrese}, {McGehee}, {Meinhold},
  {Melchiorri}, {Mendes}, {Mennella}, {Migliaccio}, {Mitra},
  {Miville-Desch{\^e}nes}, {Moneti}, {Montier}, {Morgante}, {Mortlock}, {Moss},
  {Munshi}, {Murphy}, {Naselsky}, {Nati}, {Natoli}, {Netterfield},
  {N{\o}rgaard-Nielsen}, {Noviello}, {Novikov}, {Novikov}, {Orlando},
  {Oxborrow}, {Paci}, {Pagano}, {Pajot}, {Paladini}, {Paoletti}, {Partridge},
  {Pasian}, {Patanchon}, {Pearson}, {Perdereau}, {Perotto}, {Perrotta},
  {Pettorino}, {Piacentini}, {Piat}, {Pierpaoli}, {Pietrobon}, {Plaszczynski},
  {Pointecouteau}, {Polenta}, {Pratt}, {Pr{\'e}zeau}, {Prunet}, {Puget},
  {Rachen}, {Reach}, {Rebolo}, {Reinecke}, {Remazeilles}, {Renault}, {Renzi},
  {Ristorcelli}, {Rocha}, {Rosset}, {Rossetti}, {Roudier},
  {Rubi{\~n}o-Mart{\'\i}n}, {Rusholme}, {Sandri}, {Santos}, {Savelainen},
  {Savini}, {Scott}, {Seiffert}, {Shellard}, {Spencer}, {Stolyarov}, {Stompor},
  {Strong}, {Sudiwala}, {Sunyaev}, {Sutton}, {Suur-Uski}, {Sygnet}, {Tauber},
  {Terenzi}, {Toffolatti}, {Tomasi}, {Tristram}, {Tucci}, {Tuovinen}, {Umana},
  {Valenziano}, {Valiviita}, {Van Tent}, {Vielva}, {Villa}, {Wade}, {Wandelt},
  {Wehus}, {Wilkinson}, {Yvon}, {Zacchei}, \& {Zonca}}]{2016A&A...594A..10P}
---. 2016{\natexlab{c}}, \aap, 594, A10, \dodoi{10.1051/0004-6361/201525967}

\bibitem[{{Planck Collaboration} {et~al.}(2016{\natexlab{d}}){Planck
  Collaboration}, Adam, Ade, Alves, Ashdown, Aumont, Baccigalupi, Banday,
  Barreiro, Bartolo, Battaner, Benabed, Benoit-L{\'e}vy, Bernard, Bersanelli,
  Bielewicz, Bonavera, Bond, Borrill, Bouchet, Boulanger, Bucher, Burigana,
  Butler, Calabrese, Cardoso, Catalano, Chiang, Christensen, {Colombo, L. P.
  L.}, Combet, Couchot, Crill, Curto, Cuttaia, Danese, Davis, de~Bernardis,
  de~Rosa, de~Zotti, Delabrouille, Dickinson, Diego, Dolag, Dore, Ducout,
  Dupac, Elsner, En{\ss}lin, Eriksen, Ferriere, Finelli, Forni, Frailis,
  Fraisse, Franceschi, Galeotta, Ganga, Ghosh, Giard, Gjerl{\o}w,
  Gonz{\'a}lez-Nuevo, G{\'o}rski, Gregorio, Gruppuso, Gudmundsson, Hansen,
  Harrison, Hern{\'a}ndez-Monteagudo, Herranz, Hildebrandt, Hobson, Hornstrup,
  Hurier, Jaffe, Jaffe, Jones, Juvela, Keih{\"a}nen, Keskitalo, Kisner, Knoche,
  Kunz, Kurki-Suonio, Lamarre, Lasenby, Lattanzi, Lawrence, Leahy, Leonardi,
  Levrier, Liguori, Lilje, Linden-V{\o}rnle, L{\'o}pez-Caniego, Lubin,
  Mac{\'\i}as-P{\'e}rez, Maggio, Maino, Mandolesi, Mangilli, Maris, Martin,
  Mart{\'\i}nez-Gonz{\'a}lez, Masi, Matarrese, Melchiorri, Mennella,
  Migliaccio, Miville-Desch{\^e}nes, Moneti, Montier, Morgante, Munshi, Murphy,
  Naselsky, Nati, Natoli, N{\o}rgaard-Nielsen, Oppermann, Orlando, Pagano,
  Pajot, Paladini, Paoletti, Pasian, Perotto, Pettorino, Piacentini, Piat,
  Pierpaoli, Plaszczynski, Pointecouteau, Polenta, Ponthieu, Pratt, Prunet,
  Puget, Rachen, Reinecke, Remazeilles, Renault, Renzi, Ristorcelli, Rocha,
  Rossetti, Roudier, Rubi{\~n}o-Mart{\'\i}n, Rusholme, Sandri, Santos,
  Savelainen, Scott, Spencer, Stolyarov, Stompor, Strong, Sudiwala, Sunyaev,
  Suur-Uski, Sygnet, Tauber, Terenzi, Toffolatti, Tomasi, Tristram, Tucci,
  Valenziano, Valiviita, Van~Tent, Vielva, Villa, Wade, Wandelt, Wehus, Yvon,
  Zacchei, \& Zonca}]{Collaboration:2016eh}
{Planck Collaboration}, Adam, R., Ade, P. A.~R., {et~al.} 2016{\natexlab{d}},
  \aap, 596, A103

\bibitem[{{Planck Collaboration} {et~al.}(2016{\natexlab{e}}){Planck
  Collaboration}, {Ade}, {Aghanim}, {Alves}, {Arnaud}, {Ashdown}, {Aumont},
  {Baccigalupi}, {Banday}, {Barreiro}, \& et~al.}]{2016A&A...594A..25P}
{Planck Collaboration}, {Ade}, P.~A.~R., {Aghanim}, N., {et~al.}
  2016{\natexlab{e}}, \aap, 594, A25, \dodoi{10.1051/0004-6361/201526803}

\bibitem[{{Predehl} {et~al.}(2020){Predehl}, {Sunyaev}, {Becker}, {Brunner},
  {Burenin}, {Bykov}, {Cherepashchuk}, {Chugai}, {Churazov}, {Doroshenko},
  {Eismont}, {Freyberg}, {Gilfanov}, {Haberl}, {Khabibullin}, {Krivonos},
  {Maitra}, {Medvedev}, {Merloni}, {Nandra}, {Nazarov}, {Pavlinsky}, {Ponti},
  {Sanders}, {Sasaki}, {Sazonov}, {Strong}, \& {Wilms}}]{2020Natur.588..227P}
{Predehl}, P., {Sunyaev}, R.~A., {Becker}, W., {et~al.} 2020, \nat, 588, 227,
  \dodoi{10.1038/s41586-020-2979-0}

\bibitem[{{Puspitarini} \& {Lallement}(2012)}]{2012A&A...545A..21P}
{Puspitarini}, L., \& {Lallement}, R. 2012, \aap, 545, A21,
  \dodoi{10.1051/0004-6361/201219284}

\bibitem[{{Puspitarini} {et~al.}(2014){Puspitarini}, {Lallement}, {Vergely}, \&
  {Snowden}}]{2014A&A...566A..13P}
{Puspitarini}, L., {Lallement}, R., {Vergely}, J.-L., \& {Snowden}, S.~L. 2014,
  \aap, 566, A13, \dodoi{10.1051/0004-6361/201322942}

\bibitem[{{Reich} \& {Reich}(1988)}]{1988A&A...196..211R}
{Reich}, P., \& {Reich}, W. 1988, \aap, 196, 211

\bibitem[{{Reynolds}(2017)}]{2017hsn..book.1981R}
{Reynolds}, S.~P. 2017, {Dynamical Evolution and Radiative Processes of
  Supernova Remnants}, ed. A.~W. {Alsabti} \& P.~{Murdin}, 1981,
  \dodoi{10.1007/978-3-319-21846-5\_89}

\bibitem[{{Reynolds} {et~al.}(2012){Reynolds}, {Gaensler}, \&
  {Bocchino}}]{2012SSRv..166..231R}
{Reynolds}, S.~P., {Gaensler}, B.~M., \& {Bocchino}, F. 2012, \ssr, 166, 231,
  \dodoi{10.1007/s11214-011-9775-y}

\bibitem[{{Rudnick} \& {Brown}(2009)}]{2009AJ....137..145R}
{Rudnick}, L., \& {Brown}, S. 2009, \aj, 137, 145,
  \dodoi{10.1088/0004-6256/137/1/145}

\bibitem[{{Santos} {et~al.}(2011){Santos}, {Corradi}, \&
  {Reis}}]{2011ApJ...728..104S}
{Santos}, F.~P., {Corradi}, W., \& {Reis}, W. 2011, \apj, 728, 104,
  \dodoi{10.1088/0004-637X/728/2/104}

\bibitem[{{Sarkar}(2019)}]{2019MNRAS.482.4813S}
{Sarkar}, K.~C. 2019, \mnras, 482, 4813, \dodoi{10.1093/mnras/sty2944}

\bibitem[{{Seaquist}(1968)}]{1968Obs....88..269S}
{Seaquist}, E.~R. 1968, \obs, 88, 269

\bibitem[{{Skalidis} \& {Pelgrims}(2019)}]{2019A&A...631L..11S}
{Skalidis}, R., \& {Pelgrims}, V. 2019, \aap, 631, L11,
  \dodoi{10.1051/0004-6361/201936547}

\bibitem[{{Sofue}(1977)}]{1977A&A....60..327S}
{Sofue}, Y. 1977, \aap, 60, 327

\bibitem[{{Sofue}(2015)}]{2015MNRAS.447.3824S}
---. 2015, \mnras, 447, 3824, \dodoi{10.1093/mnras/stu2661}

\bibitem[{{Sofue}(2020)}]{2020PASJ..tmp..161S}
---. 2020, \pasj, \dodoi{10.1093/pasj/psaa011}

\bibitem[{{Sofue} {et~al.}(2016){Sofue}, {Habe}, {Kataoka}, {Totani}, {Inoue},
  {Nakashima}, {Matsui}, \& {Akita}}]{2016MNRAS.459..108S}
{Sofue}, Y., {Habe}, A., {Kataoka}, J., {et~al.} 2016, \mnras, 459, 108,
  \dodoi{10.1093/mnras/stw623}

\bibitem[{{Spoelstra}(1972)}]{1972A&A....21...61S}
{Spoelstra}, T.~A.~T. 1972, \aap, 21, 61

\bibitem[{{Spoelstra}(1984)}]{1984A&A...135..238S}
---. 1984, \aap, 135, 238

\bibitem[{{Stein} {et~al.}(2020){Stein}, {Dettmar}, {Beck}, {Irwin}, {Wiegert},
  {Miskolczi}, {Wang}, {English}, {Henriksen}, {Radica}, \&
  {Li}}]{2020A&A...639A.111S}
{Stein}, Y., {Dettmar}, R.~J., {Beck}, R., {et~al.} 2020, \aap, 639, A111,
  \dodoi{10.1051/0004-6361/202037675}

\bibitem[{{Strong} {et~al.}(2011){Strong}, {Orlando}, \&
  {Jaffe}}]{2011A&A...534A..54S}
{Strong}, A.~W., {Orlando}, E., \& {Jaffe}, T.~R. 2011, \aap, 534, A54,
  \dodoi{10.1051/0004-6361/201116828}

\bibitem[{{Su} {et~al.}(2010){Su}, {Slatyer}, \&
  {Finkbeiner}}]{2010ApJ...724.1044S}
{Su}, M., {Slatyer}, T.~R., \& {Finkbeiner}, D.~P. 2010, \apj, 724, 1044,
  \dodoi{10.1088/0004-637X/724/2/1044}

\bibitem[{{Sun} {et~al.}(2015){Sun}, {Landecker}, {Gaensler}, {Carretti},
  {Reich}, {Leahy}, {McClure-Griffiths}, {Crocker}, {Wolleben}, {Haverkorn},
  {Douglas}, \& {Gray}}]{2015ApJ...811...40S}
{Sun}, X.~H., {Landecker}, T.~L., {Gaensler}, B.~M., {et~al.} 2015, \apj, 811,
  40, \dodoi{10.1088/0004-637X/811/1/40}

\bibitem[{{Testori} {et~al.}(2008){Testori}, {Reich}, \&
  {Reich}}]{2008A&A...484..733T}
{Testori}, J.~C., {Reich}, P., \& {Reich}, W. 2008, \aap, 484, 733,
  \dodoi{10.1051/0004-6361:20078842}

\bibitem[{{Tritsis} {et~al.}(2018){Tritsis}, {Federrath}, {Schneider}, \&
  {Tassis}}]{2018MNRAS.481.5275T}
{Tritsis}, A., {Federrath}, C., {Schneider}, N., \& {Tassis}, K. 2018, \mnras,
  481, 5275, \dodoi{10.1093/mnras/sty2677}

\bibitem[{{Tunmer}(1958)}]{1958PMag....3..370T}
{Tunmer}, H. 1958, \pmag, 3, 370, \dodoi{10.1080/14786435808236824}

\bibitem[{{Verschuur}(1970)}]{1970ApL.....6..215V}
{Verschuur}, G.~L. 1970, \aplett, 6, 215

\bibitem[{{Verschuur} \& {Magnani}(1994)}]{1994AJ....107..287V}
{Verschuur}, G.~L., \& {Magnani}, L. 1994, \aj, 107, 287,
  \dodoi{10.1086/116853}

\bibitem[{{Vidal} {et~al.}(2015){Vidal}, {Dickinson}, {Davies}, \&
  {Leahy}}]{2015MNRAS.452..656V}
{Vidal}, M., {Dickinson}, C., {Davies}, R.~D., \& {Leahy}, J.~P. 2015, \mnras,
  452, 656, \dodoi{10.1093/mnras/stv1328}

\bibitem[{Waelkens {et~al.}(2009)Waelkens, Jaffe, Reinecke, Kitaura, \&
  En{\ss}lin}]{Waelkens:2009bn}
Waelkens, A., Jaffe, T., Reinecke, M., Kitaura, F.~S., \& En{\ss}lin, T.~A.
  2009, A\&A, 495, 697

\bibitem[{{Welsh} \& {Shelton}(2009)}]{2009Ap&SS.323....1W}
{Welsh}, B.~Y., \& {Shelton}, R.~L. 2009, \apss, 323, 1,
  \dodoi{10.1007/s10509-009-0053-3}

\bibitem[{{West} {et~al.}(2007){West}, {English}, {Normandeau}, \&
  {Landecker}}]{2007ApJ...656..914W}
{West}, J.~L., {English}, J., {Normandeau}, M., \& {Landecker}, T.~L. 2007,
  \apj, 656, 914, \dodoi{10.1086/510609}

\bibitem[{{West} {et~al.}(2016){West}, {Safi-Harb}, {Jaffe}, {Kothes},
  {Landecker}, \& {Foster}}]{2016A&A...587A.148W}
{West}, J.~L., {Safi-Harb}, S., {Jaffe}, T., {et~al.} 2016, \aap, 587, A148,
  \dodoi{10.1051/0004-6361/201527001}

\bibitem[{{Wilkinson} \& {Smith}(1974)}]{1974MNRAS.167..593W}
{Wilkinson}, A., \& {Smith}, F.~G. 1974, \mnras, 167, 593,
  \dodoi{10.1093/mnras/167.3.593}

\bibitem[{{Wolleben}(2007)}]{2007ApJ...664..349W}
{Wolleben}, M. 2007, \apj, 664, 349, \dodoi{10.1086/518711}

\bibitem[{{Wolleben} {et~al.}(2006){Wolleben}, {Landecker}, {Reich}, \&
  {Wielebinski}}]{2006A&A...448..411W}
{Wolleben}, M., {Landecker}, T.~L., {Reich}, W., \& {Wielebinski}, R. 2006,
  \aap, 448, 411, \dodoi{10.1051/0004-6361:20053851}

\bibitem[{{Wolleben} {et~al.}(2021){Wolleben}, {Landecker}, {Douglas}, {Gray},
  {Ordog}, {Dickey}, {Hill}, {Carretti}, {Brown}, {Gaensler}, {Han},
  {Haverkorn}, {Kothes}, {Leahy}, {McClure-Griffiths}, {McConnell}, {Reich},
  {Taylor}, {Thomson}, \& {West}}]{2021AJ....162...35W}
{Wolleben}, M., {Landecker}, T.~L., {Douglas}, K.~A., {et~al.} 2021, \aj, 162,
  35, \dodoi{10.3847/1538-3881/abf7c1}

\bibitem[{{Yadav} {et~al.}(2017){Yadav}, {Mukherjee}, {Sharma}, \&
  {Nath}}]{2017MNRAS.465.1720Y}
{Yadav}, N., {Mukherjee}, D., {Sharma}, P., \& {Nath}, B.~B. 2017, \mnras, 465,
  1720, \dodoi{10.1093/mnras/stw2522}

\bibitem[{{Yusef-Zadeh} {et~al.}(1984){Yusef-Zadeh}, {Morris}, \&
  {Chance}}]{1984Natur.310..557Y}
{Yusef-Zadeh}, F., {Morris}, M., \& {Chance}, D. 1984, \nat, 310, 557,
  \dodoi{10.1038/310557a0}

\bibitem[{{Zaroubi} {et~al.}(2015){Zaroubi}, {Jelic}, {de Bruyn}, {Boulanger},
  {Bracco}, {Kooistra}, {Alves}, {Brentjens}, {Ferriere}, {Ghosh}, {Koopmans},
  {Levrier}, {Miville-Deschenes}, {Montier}, {Pandey}, \&
  {Soler}}]{2015MNRAS.454L..46Z}
{Zaroubi}, S., {Jelic}, V., {de Bruyn}, A.~G., {et~al.} 2015, \mnras, 454, L46,
  \dodoi{10.1093/mnrasl/slv123}

\bibitem[{Zonca {et~al.}(2019)Zonca, Singer, Lenz, Reinecke, Rosset, Hivon, \&
  G{\'o}rski}]{Zonca2019}
Zonca, A., Singer, L., Lenz, D., {et~al.} 2019, JOSS, 4, 1298,
  \dodoi{10.21105/joss.01298}

\bibitem[{{Zucker} {et~al.}(2015){Zucker}, {Battersby}, \&
  {Goodman}}]{2015ApJ...815...23Z}
{Zucker}, C., {Battersby}, C., \& {Goodman}, A. 2015, \apj, 815, 23,
  \dodoi{10.1088/0004-637X/815/1/23}

\end{thebibliography}
\bibliographystyle{aasjournal}

\newpage
\newpage
\clearpage

\appendix
\renewcommand{\thefigure}{A\arabic{figure}}
\renewcommand{\thetable}{A\arabic{table}}
\setcounter{figure}{0}

\section{\label{sec:method}Models}

Fig.~A1 shows a representative sample of straight-line filament models described in Sec.~\ref{sec:straightlines}.

\begin{figure*}[!ht]
\centering 

\begin{minipage}{2.5cm}
\includegraphics[width=2.5cm]{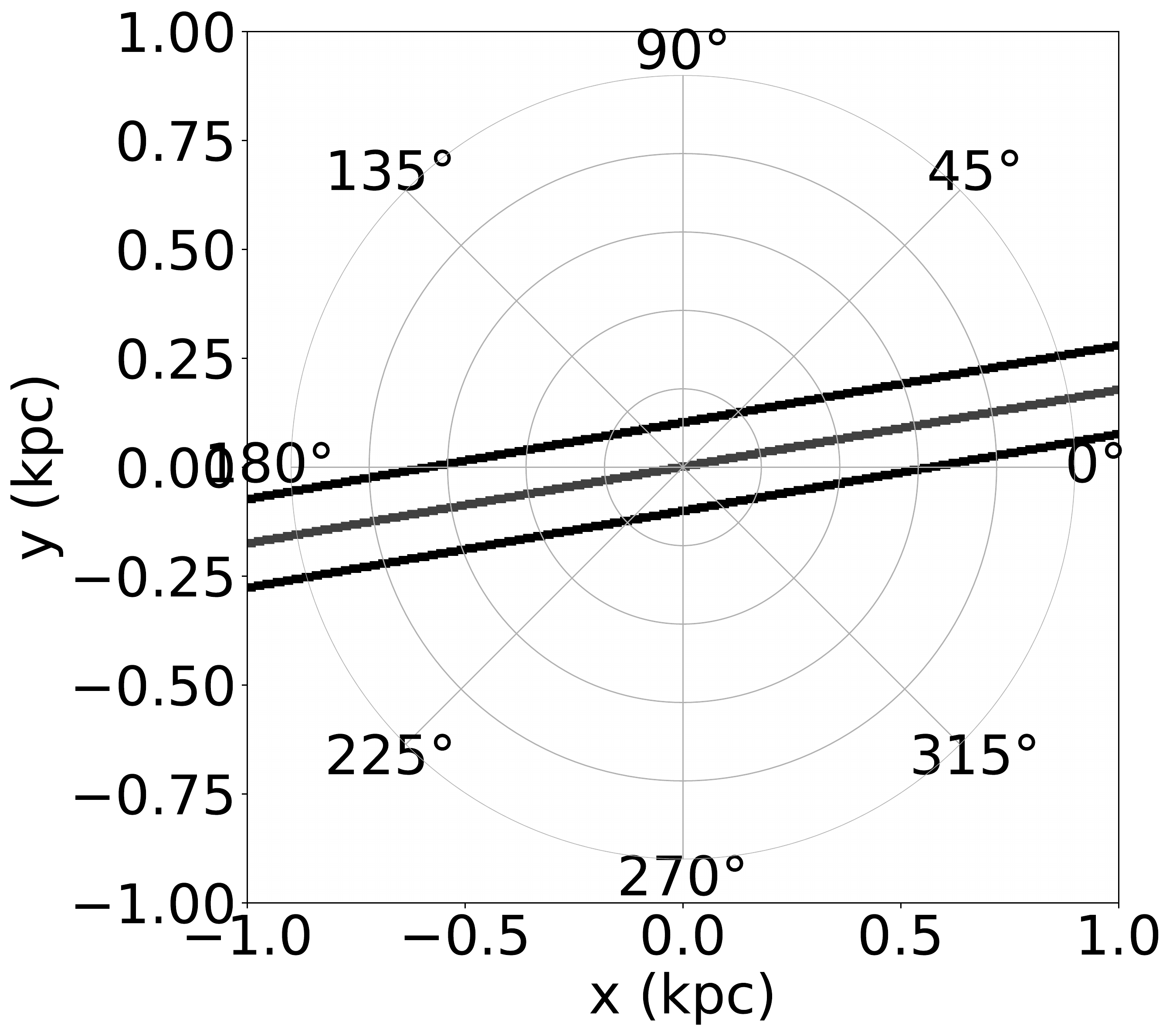}
%figures/m1-wT.pdf
\end{minipage}
%\hfill
\begin{minipage}{5cm}
\includegraphics[width=4.4cm]{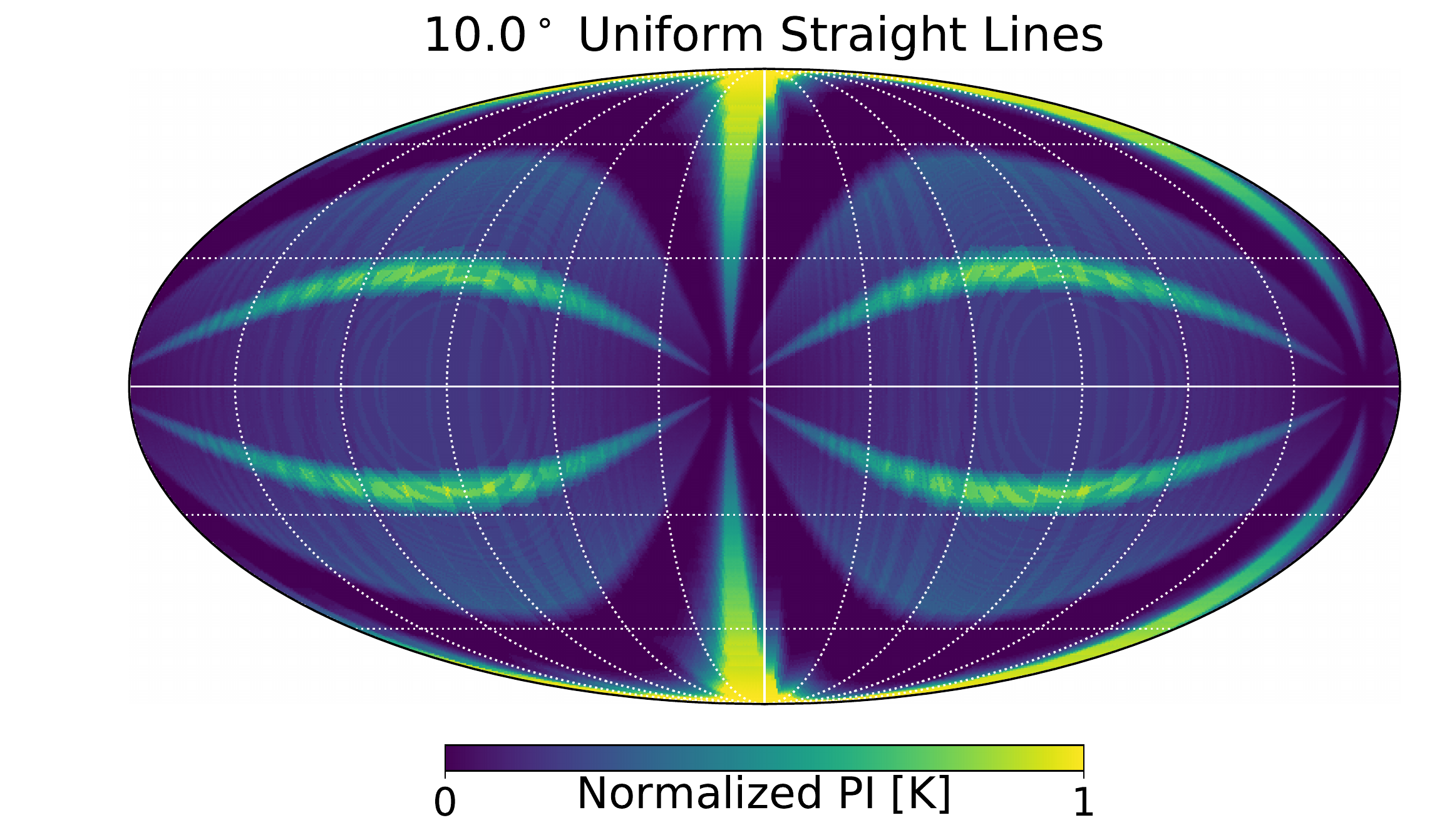}
%figures/m1-wT.pdf
\hfill
\end{minipage}
\begin{minipage}{2.5cm}
\includegraphics[width=2.5cm]{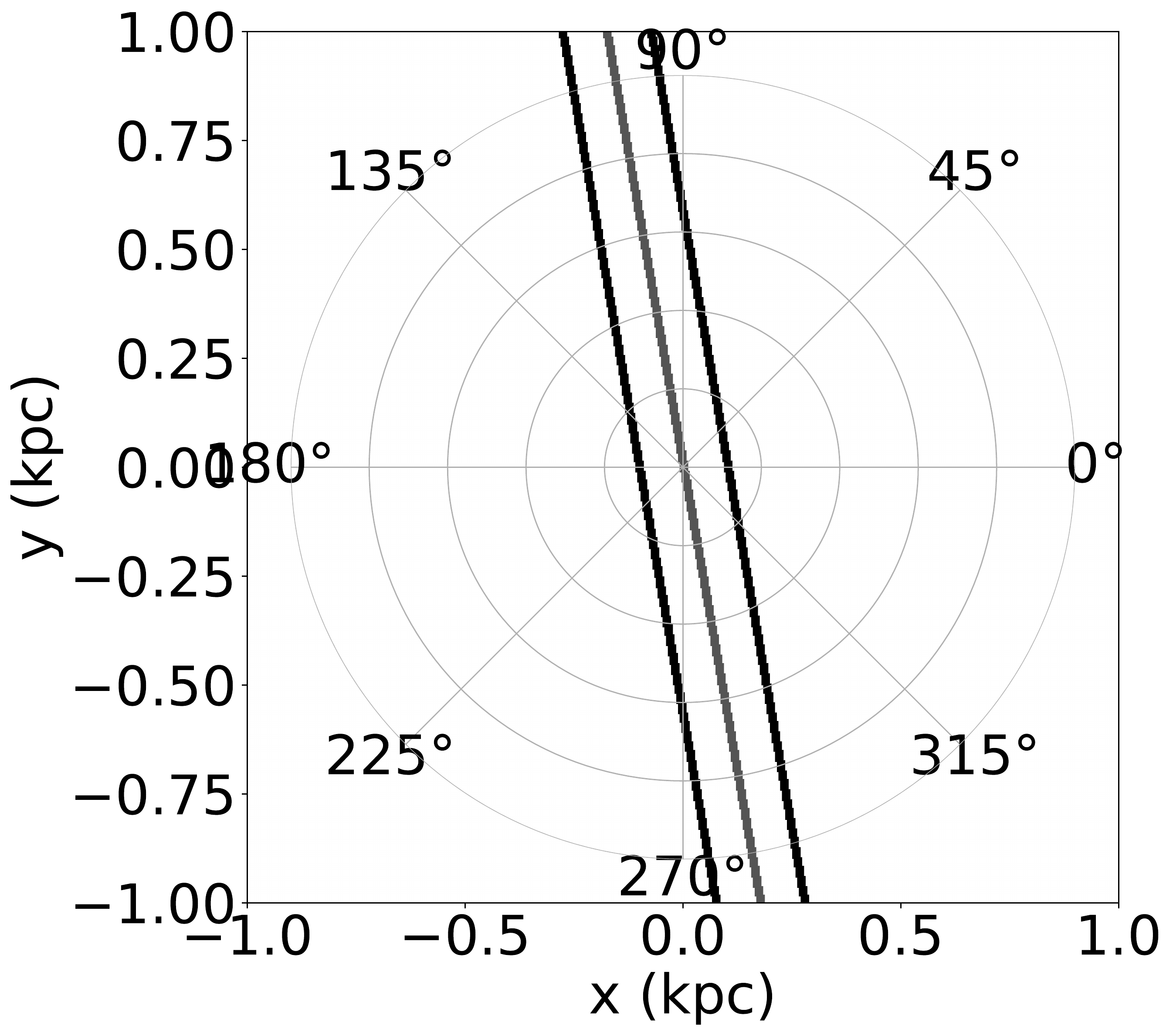}
\end{minipage}
%\hfill
\begin{minipage}{5cm}
\includegraphics[width=4.4cm]{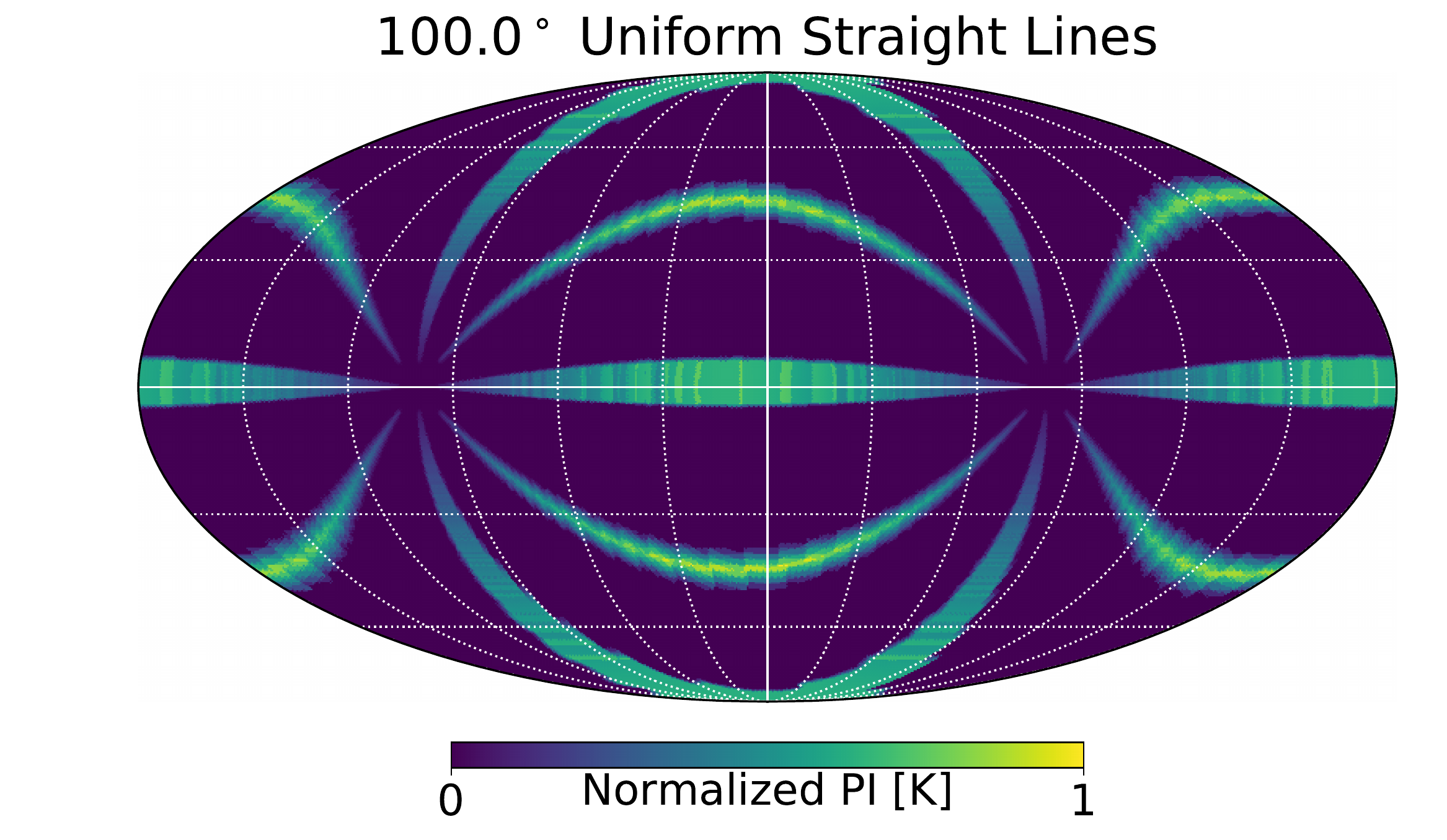}
\end{minipage}

\begin{minipage}{2.5cm}
\includegraphics[width=2.5cm]{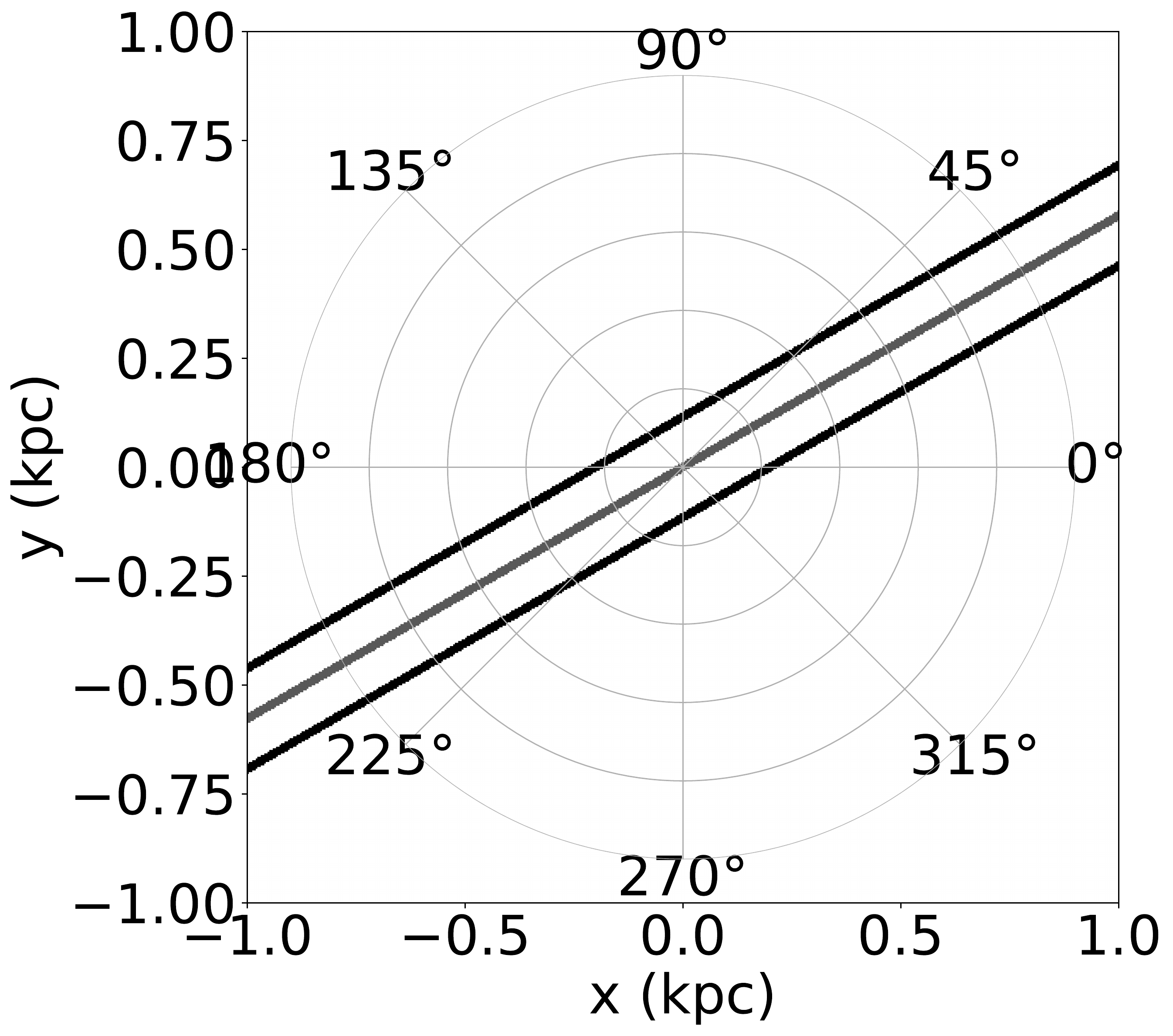}
\end{minipage}
%\hfill
\begin{minipage}{5cm}
\includegraphics[width=4.4cm]{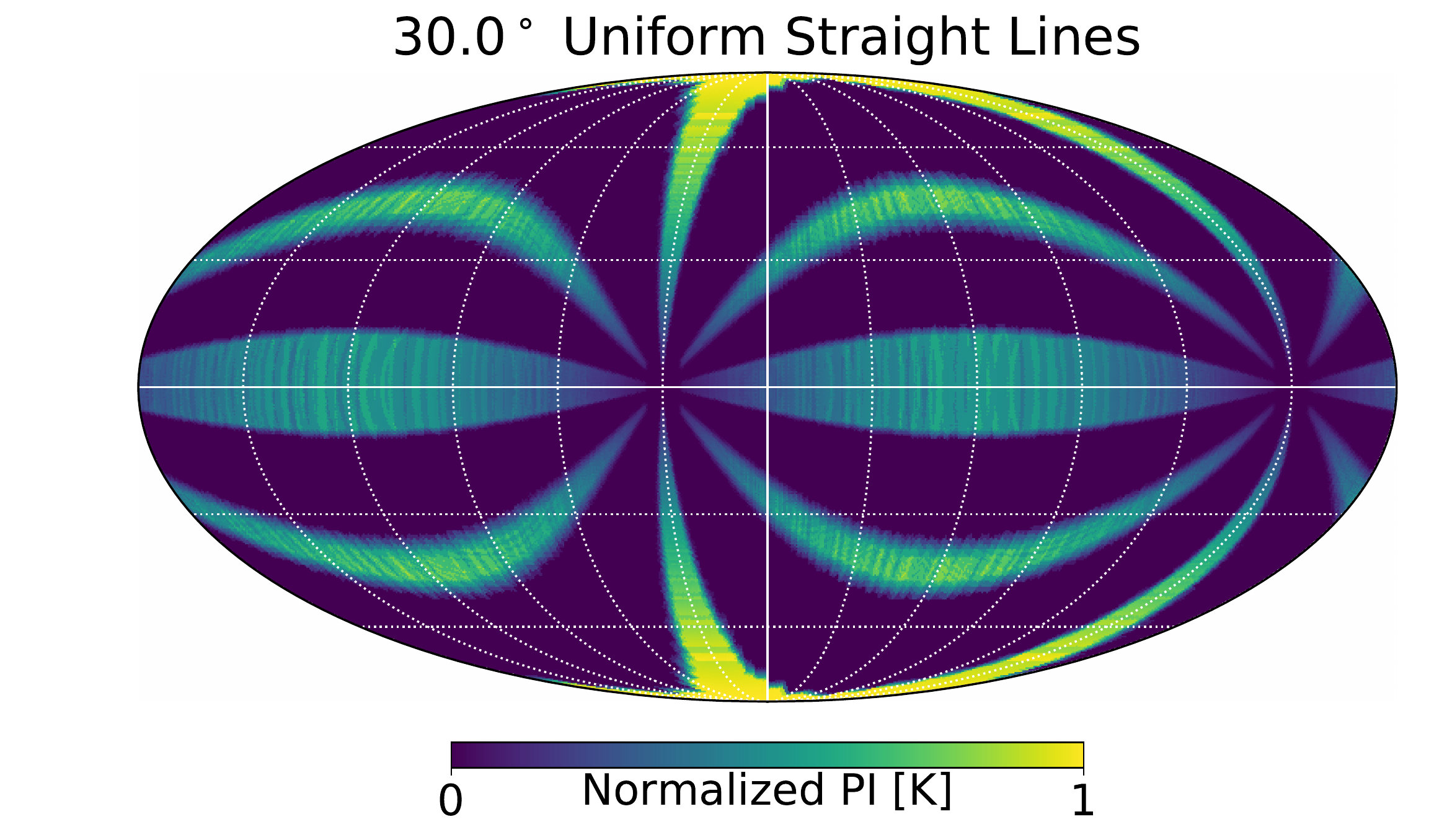}
\end{minipage}
\begin{minipage}{2.5cm}
\includegraphics[width=2.5cm]{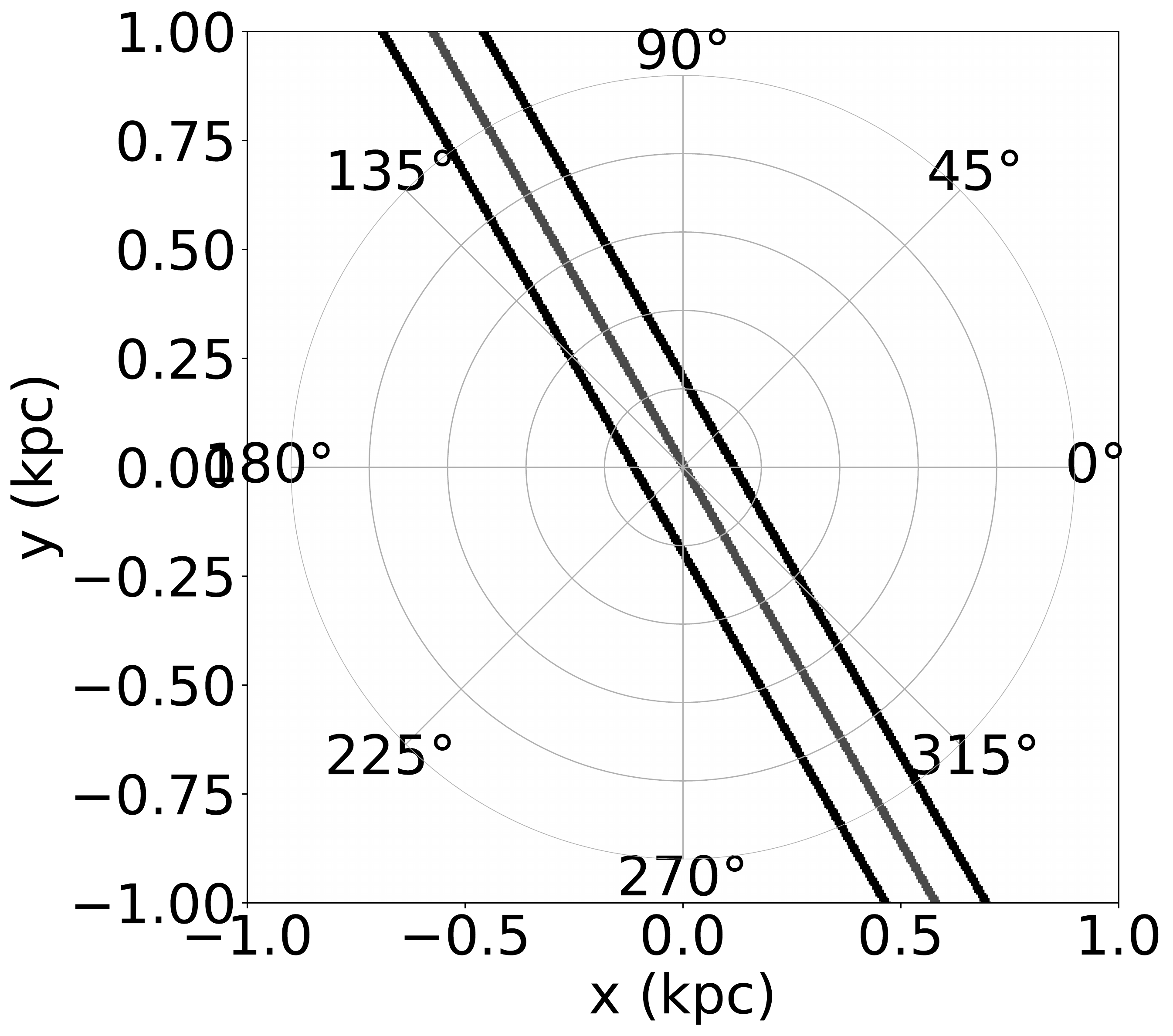}
\end{minipage}
%\hfill
\begin{minipage}{5cm}
\includegraphics[width=4.4cm]{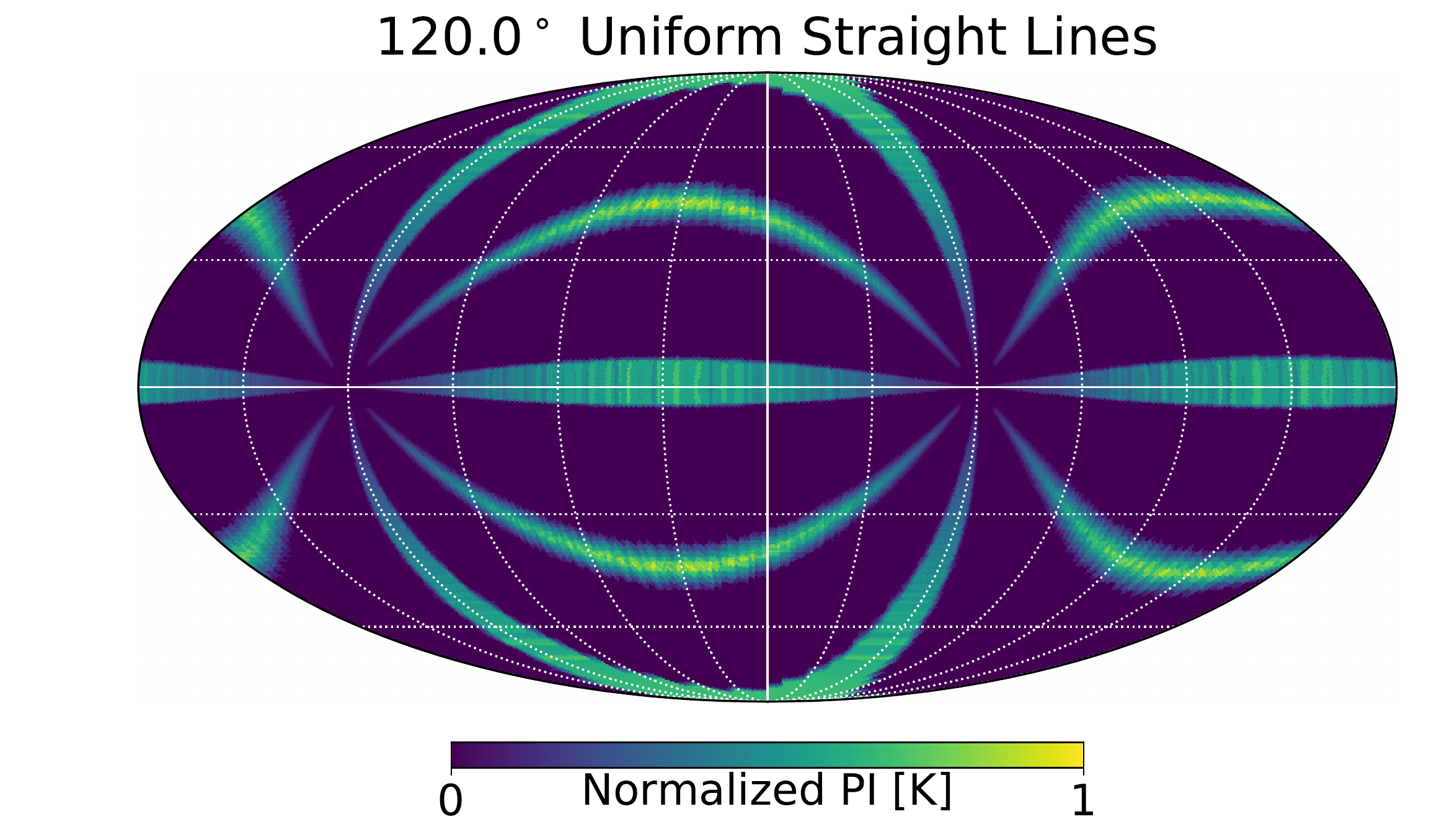}
\end{minipage}

\begin{minipage}{2.5cm}
\includegraphics[width=2.5cm]{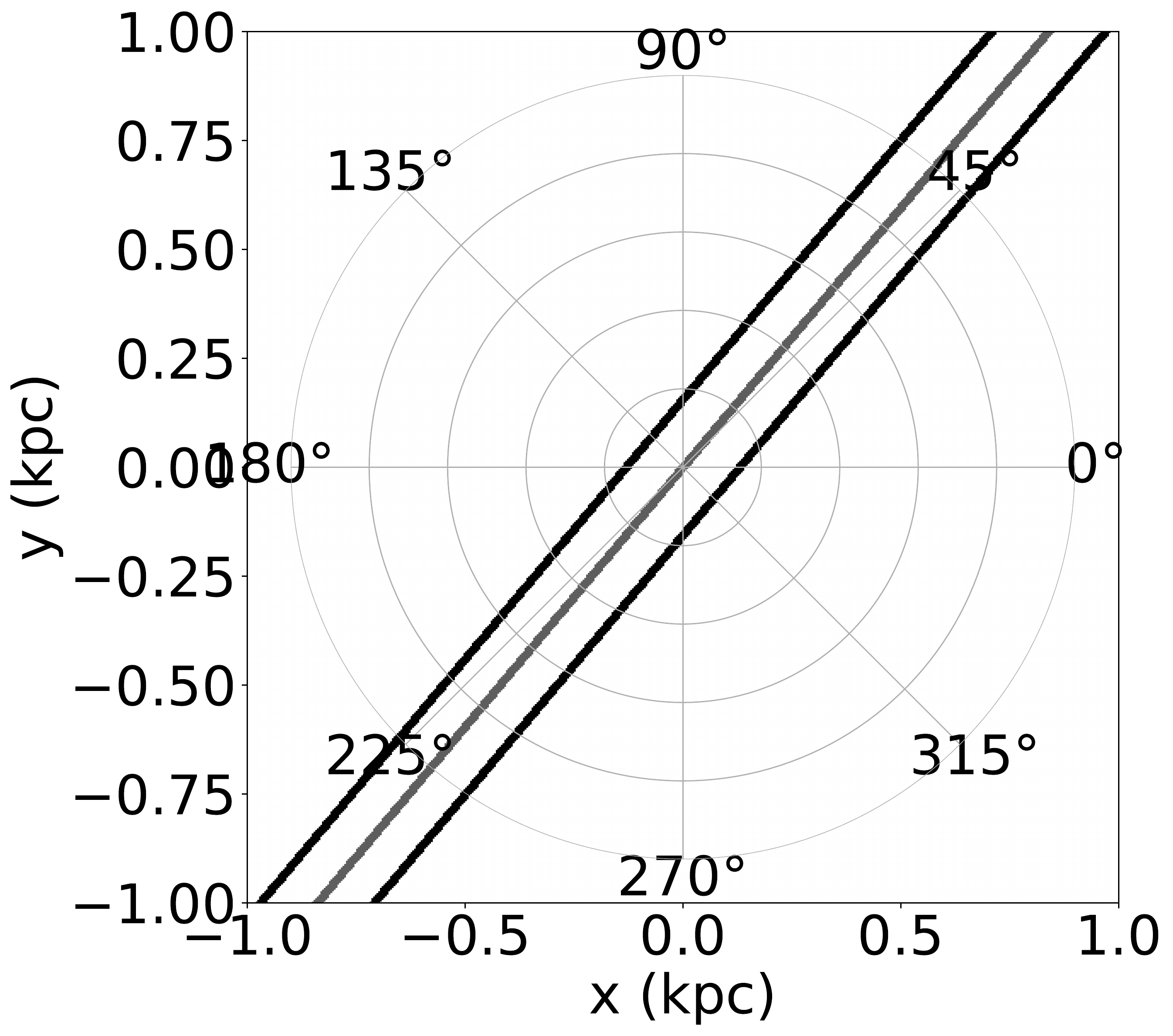}
\end{minipage}
%\hfill
\begin{minipage}{5cm}
\includegraphics[width=4.4cm]{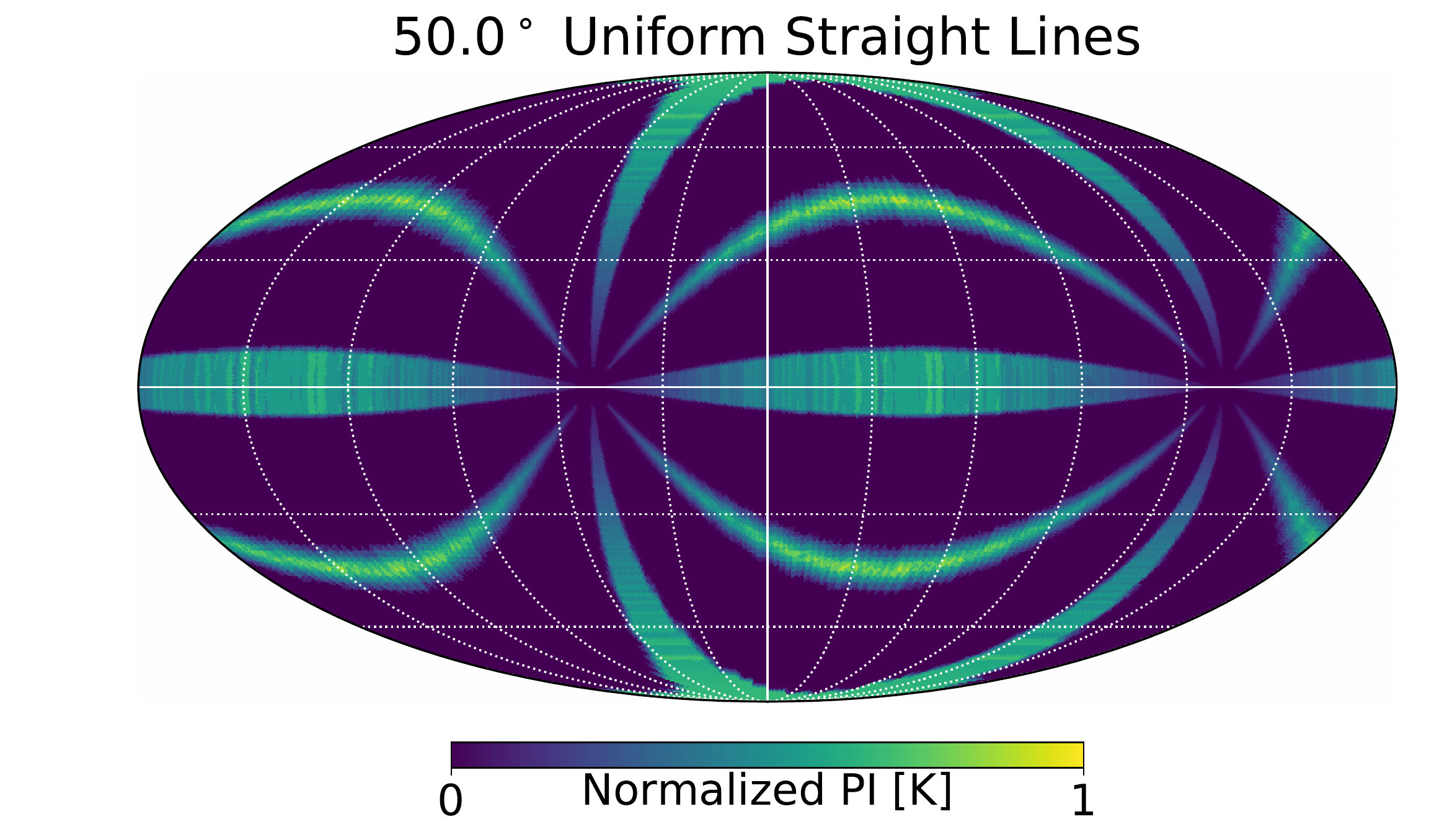}
\end{minipage}
\begin{minipage}{2.5cm}
\includegraphics[width=2.5cm]{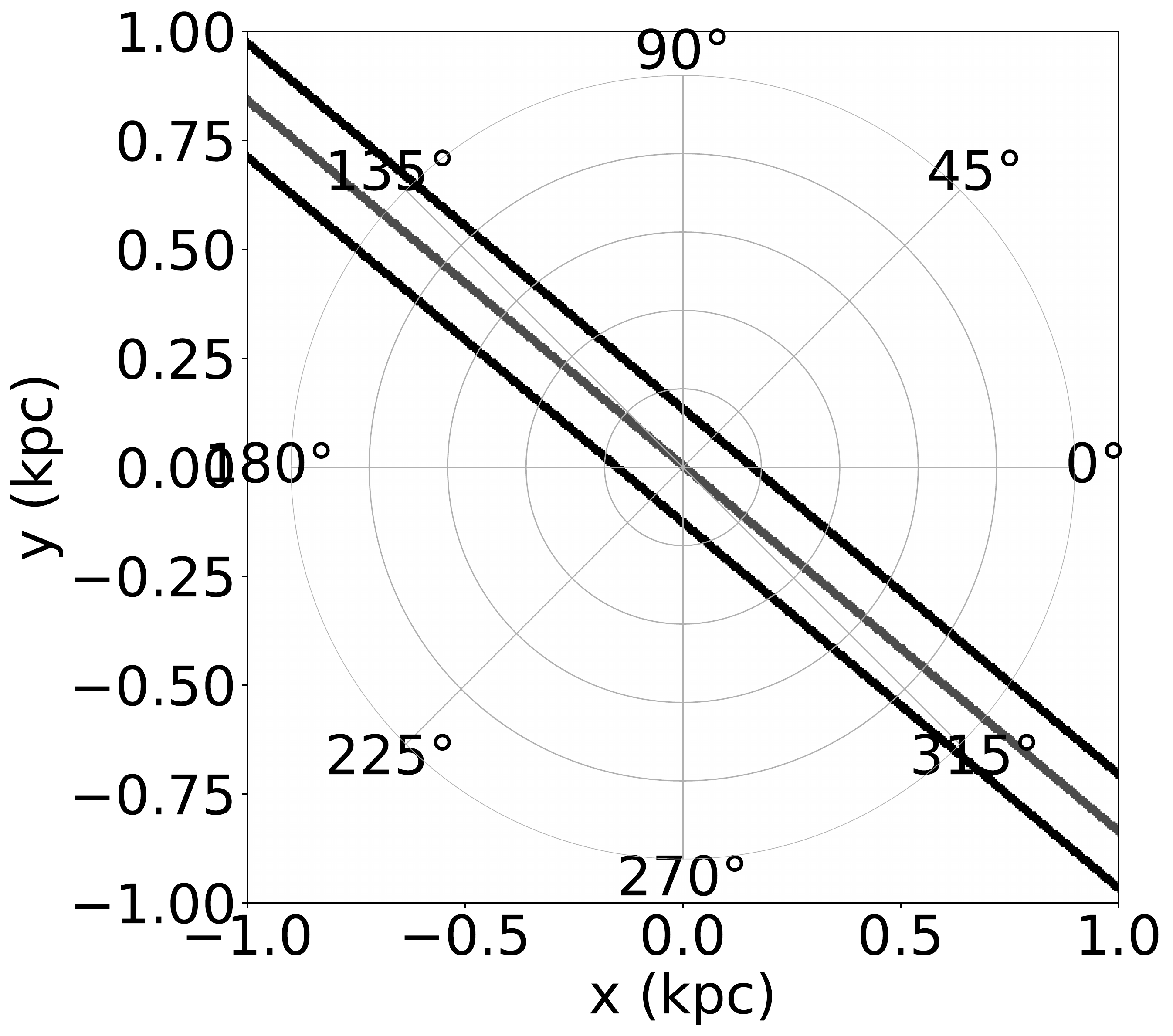}
\end{minipage}
%\hfill
\begin{minipage}{5cm}
\includegraphics[width=4.4cm]{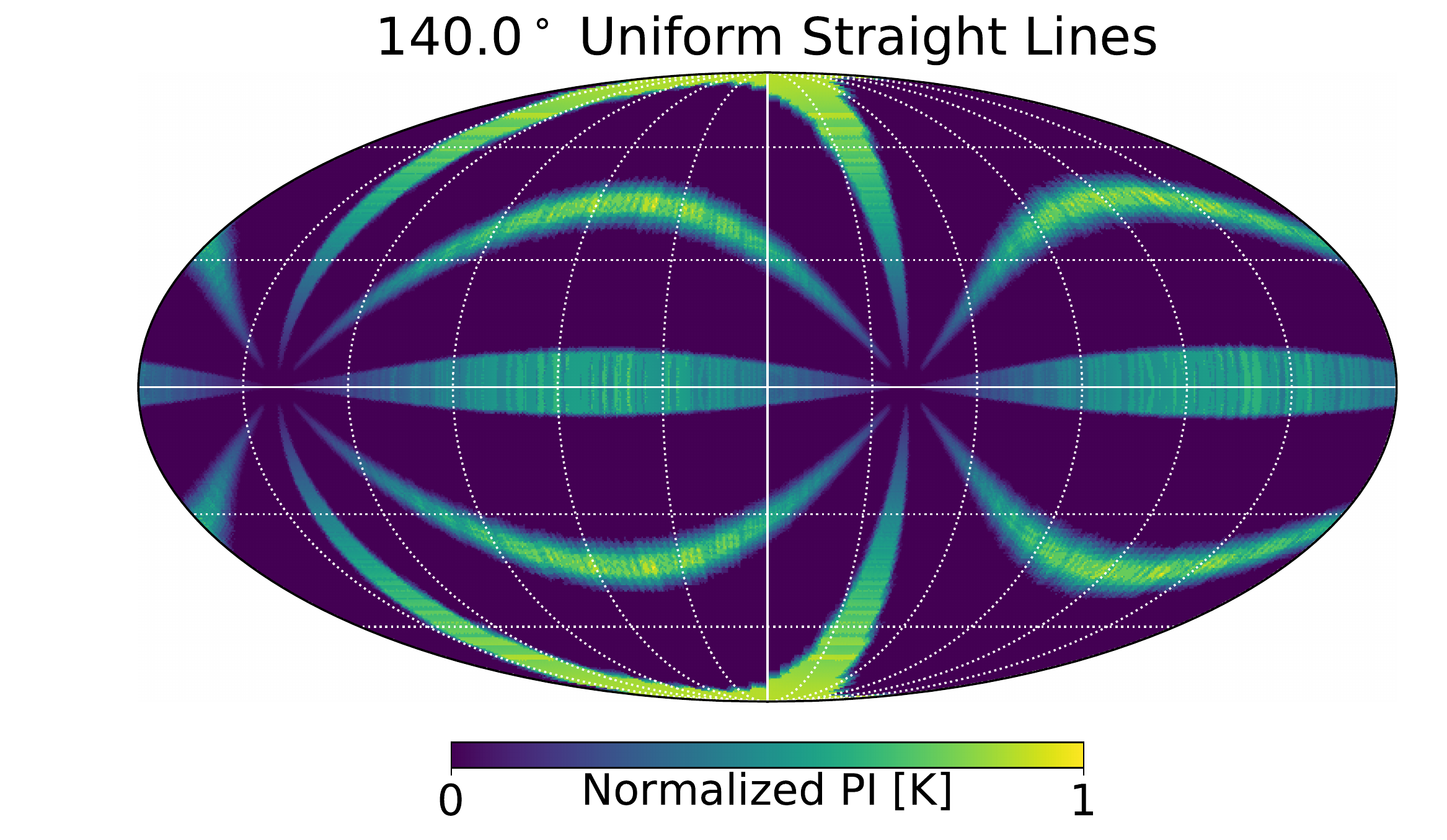}
\end{minipage}

\begin{minipage}{2.5cm}
\includegraphics[width=2.5cm]{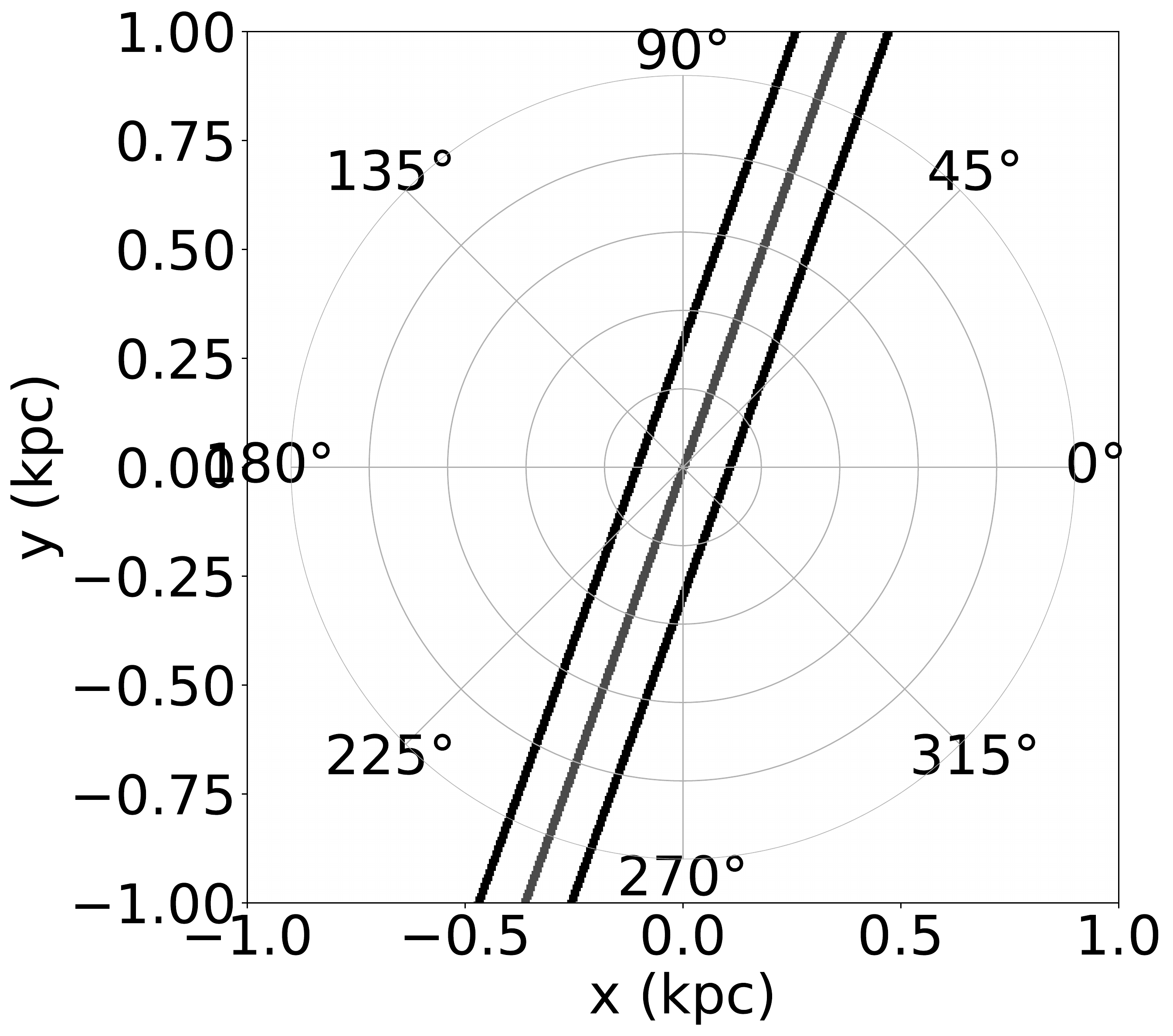}
\end{minipage}
%\hfill
\begin{minipage}{5cm}
\includegraphics[width=4.4cm]{straight-lines-uniform_70_0_100_0.pdf}
\end{minipage}
\begin{minipage}{2.5cm}
\includegraphics[width=2.5cm]{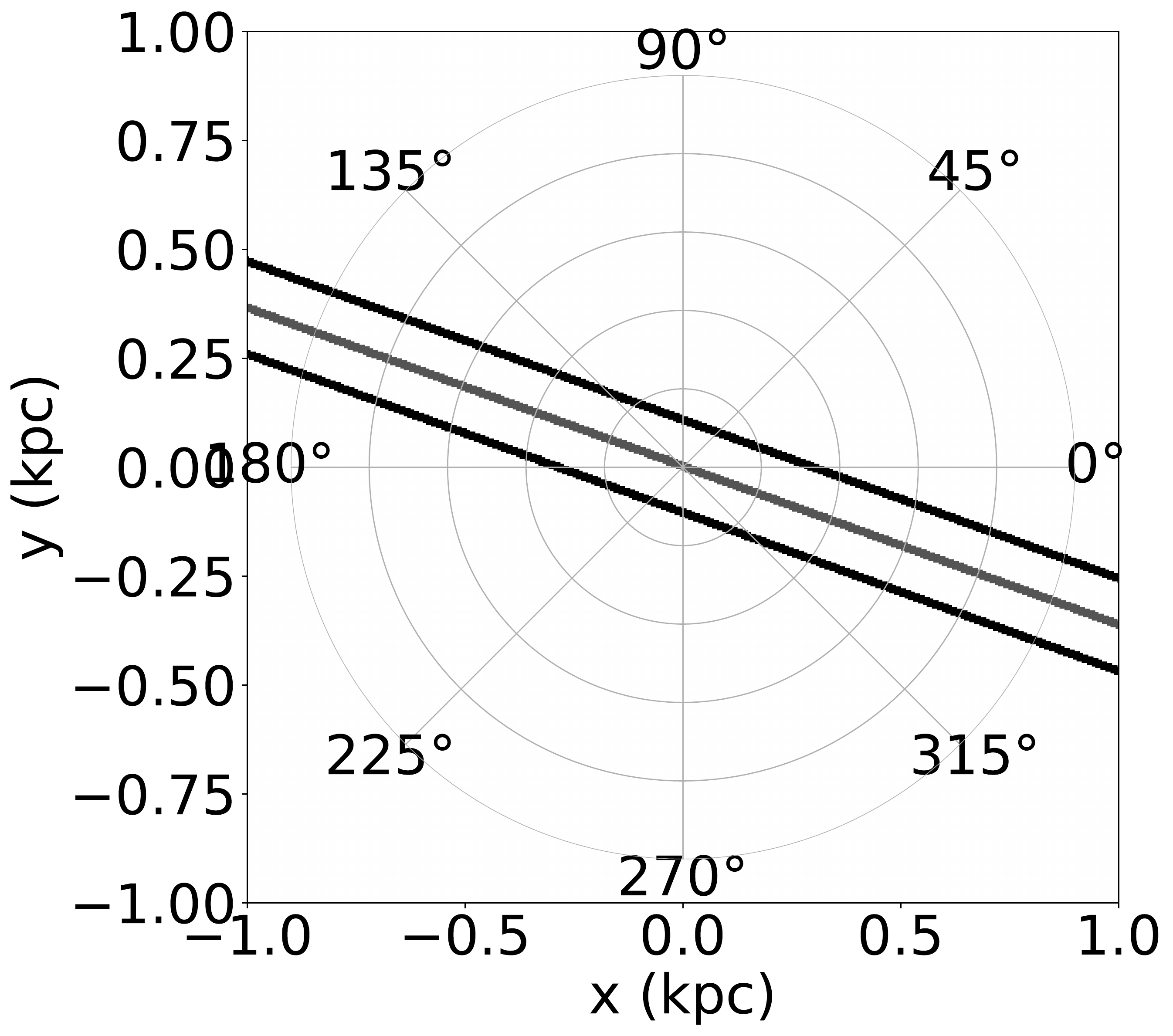}
\end{minipage}
%\hfill
\begin{minipage}{5cm}
\includegraphics[width=4.4cm]{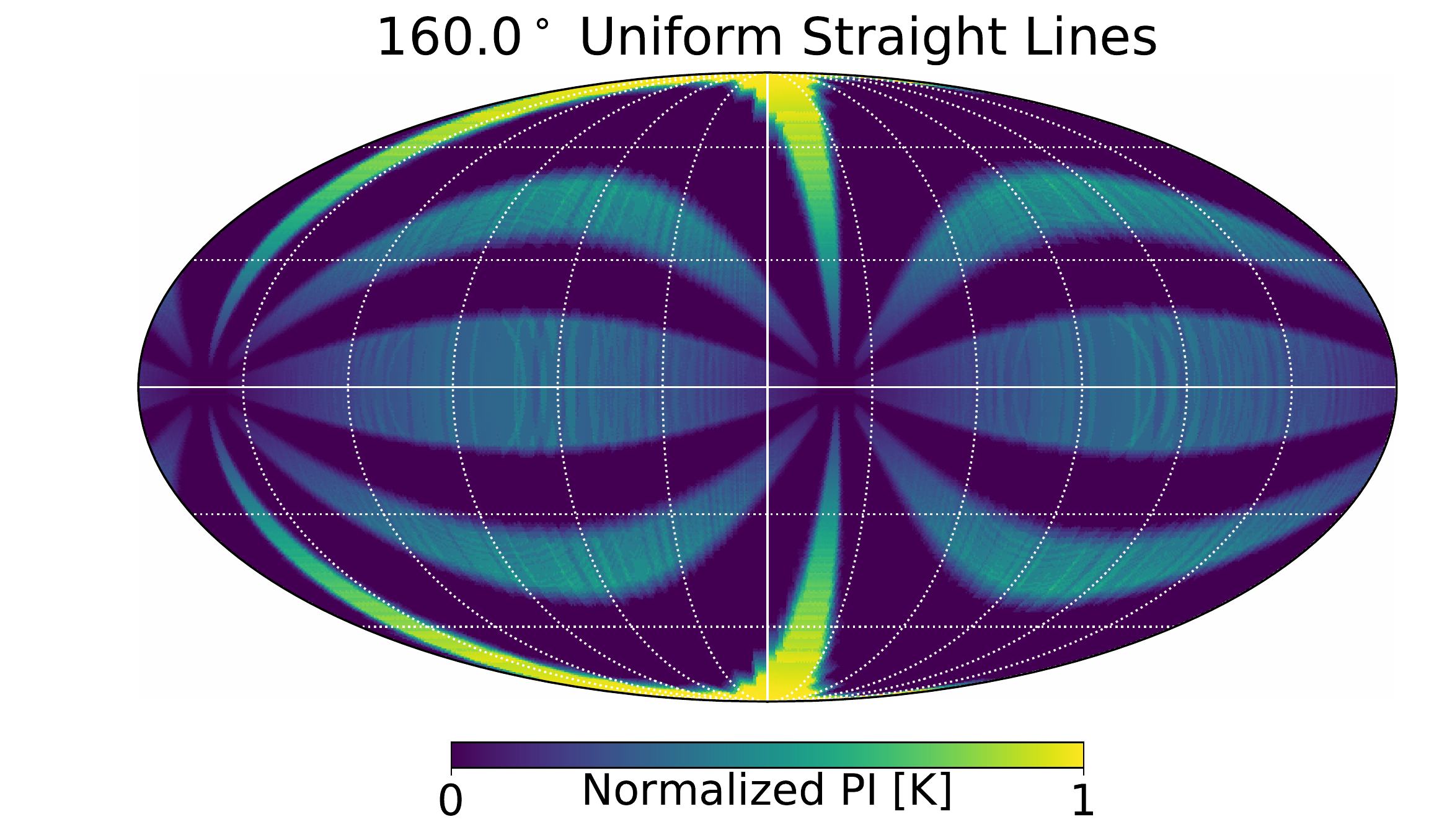}
\end{minipage}

\begin{minipage}{2.5cm}
\includegraphics[width=2.5cm]{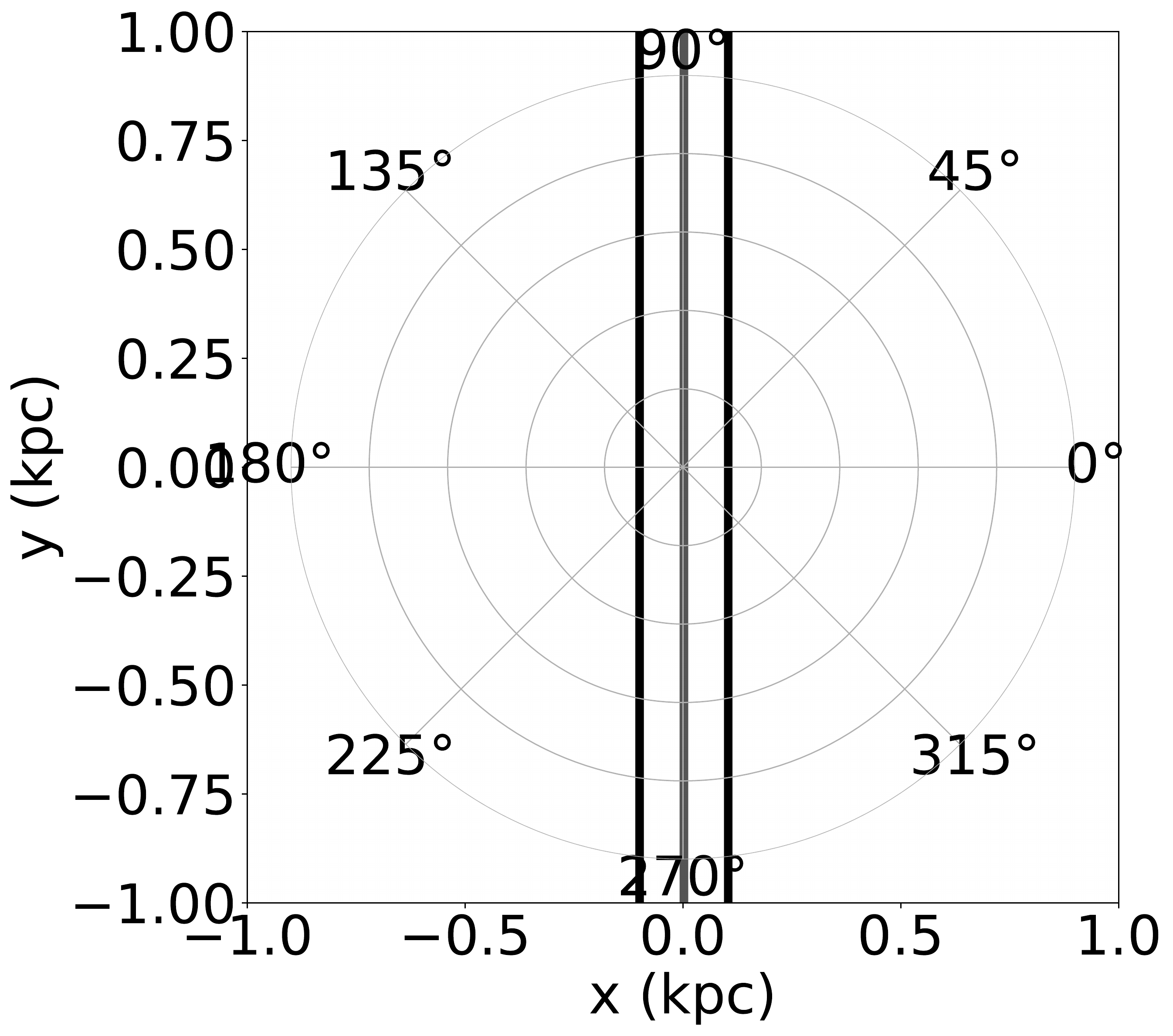}
\end{minipage}
%\hfill
\begin{minipage}{5cm}
\includegraphics[width=4.4cm]{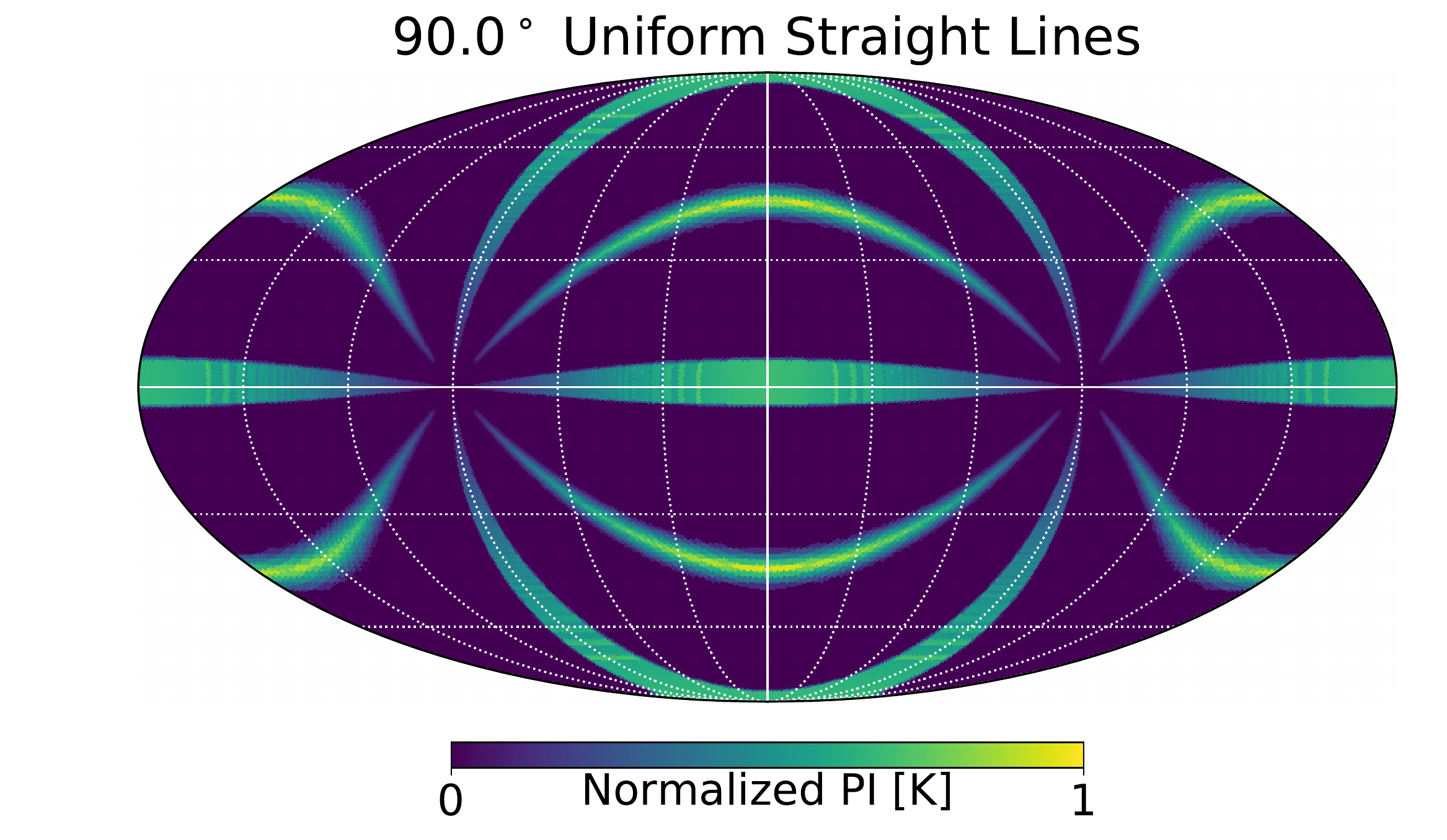}
\end{minipage}
\begin{minipage}{2.5cm}
\includegraphics[width=2.5cm]{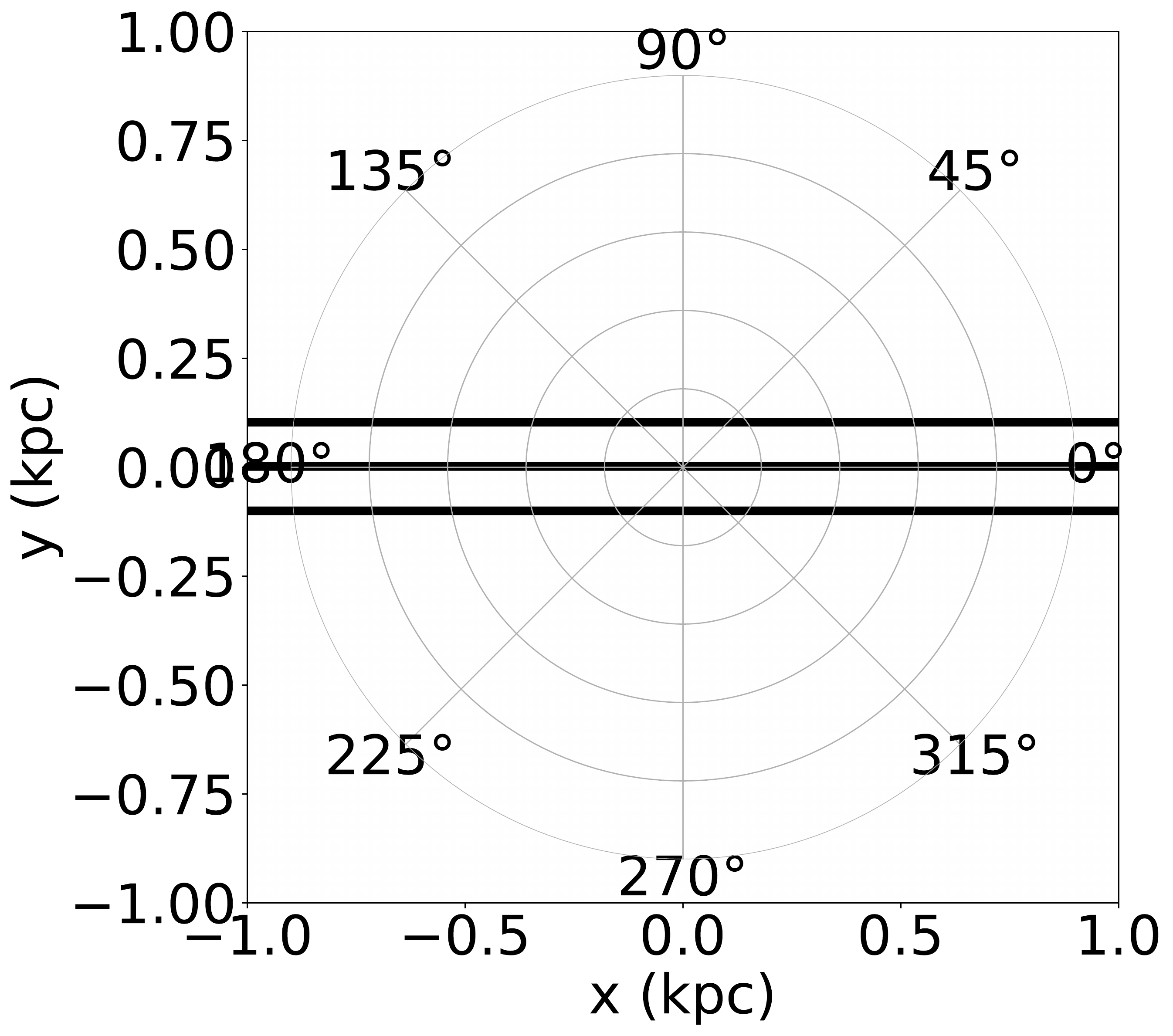}
\end{minipage}
%\hfill
\begin{minipage}{5cm}
\includegraphics[width=4.4cm]{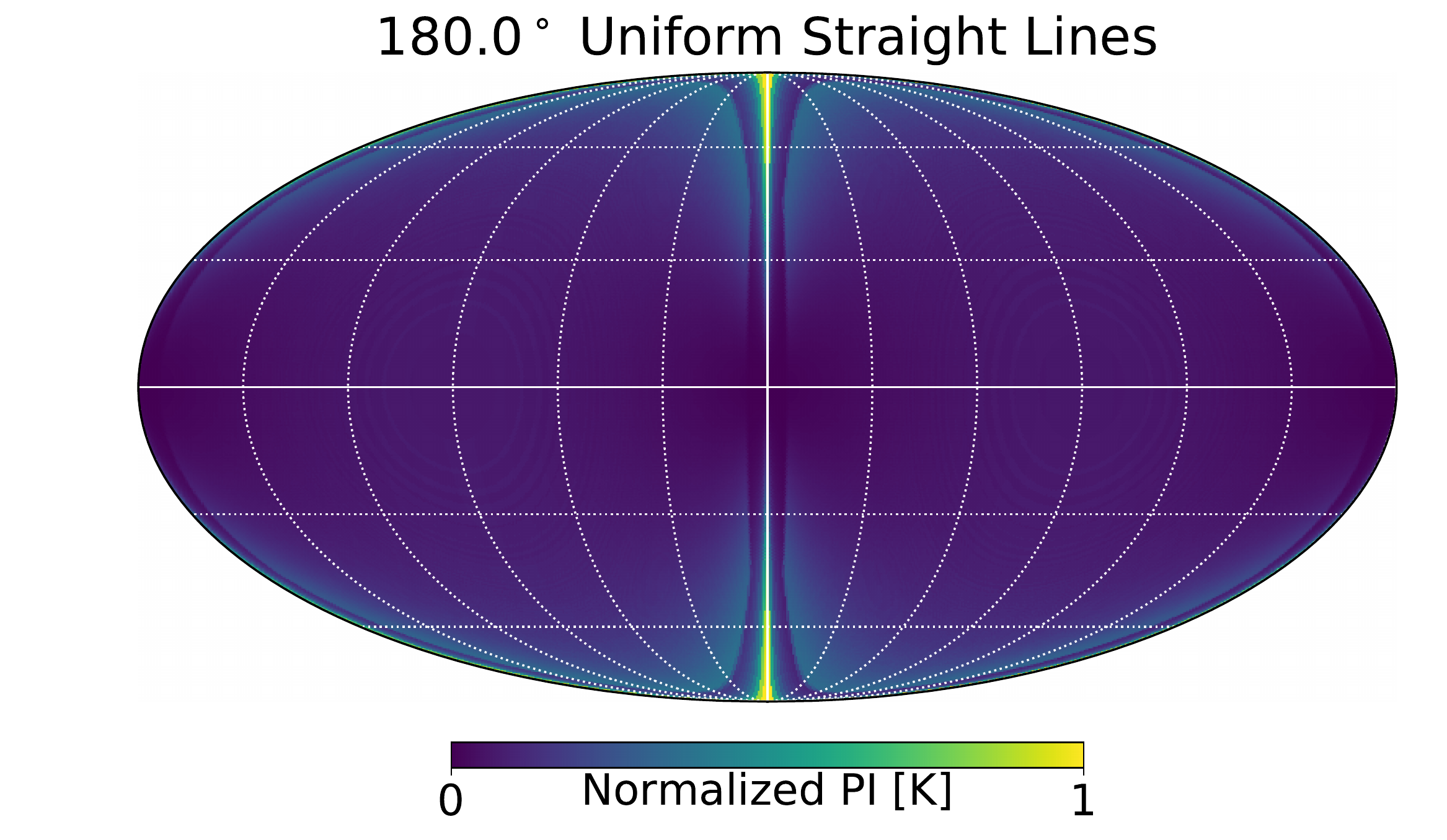}
\end{minipage}

\caption{ Left column: Top-down view of a selection of straight-line filaments oriented from $l_\alpha=10^\circ$ to $l_\alpha=90^\circ$ in steps of $20^\circ$ shown with the corresponding simulated synchrotron emission from these filaments. Right column: same as the left but for $l_\alpha=100^\circ$ to $l_\alpha=180^\circ$ in steps of $20^\circ$.}

\label{fig:appendixuniformfields}
\end{figure*}
\end{document}